\documentclass[twoside,11pt]{article}
\usepackage{jair, theapa, rawfonts}

\usepackage{graphicx}
\usepackage{xcolor}
\usepackage{enumitem}
\usepackage{bm}
\usepackage[hyphens]{url}
\usepackage{amsmath}
\newcommand{\indep}{\rotatebox[origin=c]{90}{$\models$}}
\newtheorem{definition}{Definition}
\usepackage{url}
\jairheading{71}{2021}{1137-1181}{03/2021}{08/2021}
\ShortHeadings{Socially Responsible AI Algorithms: Issues, Purposes, and Challenges}
{Cheng, Varshney, \& Liu}
\firstpageno{1}
\begin{document}

\title{Socially Responsible AI Algorithms:\\  Issues, Purposes, and Challenges}

\author{\name Lu Cheng \email lcheng35@asu.edu \\
       \addr Computer Science and Engineering, Arizona State University
       \AND
       \name Kush R. Varshney \email krvarshn@us.ibm.com \\
       \addr IBM Research -- Thomas J. Watson Research Center
       \AND
       \name Huan Liu \email huanliu@asu.edu \\
       \addr Computer Science and Engineering, Arizona State University}


\maketitle

\begin{abstract}
In the current era, people and society have grown increasingly reliant on artificial intelligence (AI) technologies. AI has the potential to drive us towards a future in which all of humanity flourishes. It also comes with substantial risks for oppression and calamity. Discussions about whether we should \textit{(re)trust} AI have repeatedly emerged in recent years and in many quarters, including industry, academia, healthcare, services, and so on. Technologists and AI researchers have a responsibility to develop trustworthy AI systems. They have responded with great effort to design more responsible AI algorithms. However, existing technical solutions are narrow in scope and have been primarily directed towards algorithms for scoring or classification tasks, with an emphasis on fairness and unwanted bias. To build long-lasting trust between AI and human beings, we argue that the key is to think beyond algorithmic fairness and connect major aspects of AI that potentially cause AI's indifferent behavior. In this survey, we provide a systematic framework of \textit{Socially Responsible AI Algorithms} that aims to examine the \textit{subjects} of AI indifference and the need for socially responsible AI algorithms, define the \textit{objectives}, and introduce the \textit{means} by which we may achieve these objectives. We further discuss how to leverage this framework to improve societal well-being through \textit{protection, information}, and \textit{prevention/mitigation}.
\end{abstract}

\section{Introduction}
Artificial intelligence (AI) has had and will continue to have a central role in countless aspects of life, livelihood, and liberty. AI is bringing forth a sea-change that is not only limited to technical domains, but is a truly sociotechnical phenomenon affecting healthcare, education, commerce, finance, and criminal justice, not to mention day-to-day life. AI offers both promise and perils. A report published by Martha Lane Fox's Doteveryone think tank \cite{people2019} reveals that 59\% of tech workers have worked on products they felt harmful to society, and more than 25\% of workers in AI who had such an experience quit their jobs as a result. This was particularly marked in relation to AI products. The rise of activism -- which has been regarded as one of the current few mechanisms to keep big tech companies in check \cite{Schwab_2021} -- against negative social impacts of big tech have brought Social Responsibility of AI into the spotlight of the media, the general public, and AI technologists and researchers \cite{abdalla2020grey}. Even researchers in universities and research institutes are trying hard to rectify the mistakes made by algorithms. Stanford's COVID-19 vaccine allocation algorithm, for example, prioritizes older employees over front-line workers \cite{Stanford}, turning much of our attention again to the transparency and fairness of AI. 

Research directed towards developing fair, transparent, accountable, and ethical AI algorithms has burgeoned with a focus on decision-making algorithms such as scoring or classification to mitigate unwanted bias and achieve fairness \cite{jagadish2019responsible}. However, this narrow subset of research risks blinding us to the challenges and opportunities that are presented by the full scope of AI. To identify potential higher-order effects on safety, privacy, and society at large, it is critical to think beyond algorithmic bias, to capture all the connections among different aspects related to AI algorithms. Therefore, this survey complements prior work through a holistic understanding of the relations between AI systems and humans. In this work, we begin by introducing an inclusive definition of \textit{Social Responsibility of AI}. Drawing on theories in business research, we then present a pyramid of Social Responsibility of AI that outlines four specific AI responsibilities in a hierarchy. This is adapted from the pyramid proposed for Corporate Social Responsibility (CSR) by \citeA{carroll1991pyramid}. In the second part of the survey, we review major aspects of AI algorithms and provide a systematic framework -- \textit{Socially Responsible AI Algorithms} (SRAs) -- that aims to understand the connections among these aspects. In particular, we examine the \textit{subjects} and \textit{causes} of socially indifferent AI algorithms\footnote{We define ``indifferent'' as the complement of responsible rather than ``irresponsible''.}, define the \textit{objectives}, and introduce the \textit{means} by which we can achieve SRAs. We further discuss how to leverage SRAs to improve daily life of human beings and address challenging societal issues through \textit{protecting, informing}, and \textit{preventing/mitigating}. We illustrate these ideas using recent studies on several emerging societal challenges. The survey concludes with open problems and challenges in SRAs. 

\noindent\textbf{Differences from Existing Surveys.} Some recent surveys focus on specific topics such as bias and fairness \cite{mehrabi2019survey,caton2020fairness}, interpretability/explainability \cite{carvalho2019machine,tjoa2019XAI}, and privacy-preservation \cite{beigi2020survey,dwork2008differential}. These surveys successfully draw great attention to the social responsibility of AI, leading to further developments in this important line of research. However, as indispensable components of socially responsible AI, these topics have been presented in their own self-contained ways. These works pave the way for looking at socially responsible AI holistically. Therefore, our survey aims to frame socially responsible AI with a more systematic view that goes beyond discussion of each independent line of research. We summarize our \textit{contributions} as follows:
\begin{itemize}[leftmargin=*]
    \item We formally define social responsibility of AI with three specified dimensions: \textit{principles, means}, and \textit{objectives}. We then propose the pyramid of social responsibility of AI, describing its four fundamental responsibilities: \textit{functional, legal, ethical}, and \textit{philanthropic} responsibilities. The pyramid embraces the entire range of AI responsibilities involving efforts from various disciplines. 
    \item We propose a systematic framework that discusses the essentials of socially responsible AI algorithms (SRAs) -- including its \textit{subjects, causes, means}, and \textit{objectives} -- and the roles of SRAs in \textit{protecting, informing} users, and \textit{preventing} them from negative impact of AI. This framework subsumes existing topics such as fairness and interpretability. 
    \item We look beyond prior research in socially responsible AI and identify an extensive list of open problems and challenges, ranging from understanding \textit{why} we need AI systems to showing the need to define new AI ethics principles and policies. We hope our discussions can spark future research on SRAs.  
\end{itemize}

\noindent\textbf{Intended Audience and Paper Organization.} This survey is intended for AI researchers, AI technologists, researchers, and practitioners from other disciplines who would like to contribute to making AI more socially responsible with their expertise. The rest of the survey is organized as follows: Section 2 introduces the definition and the pyramid of social responsibility of AI, and compares definitions of similar concepts. Section 3 discusses the framework of socially responsible algorithms and its essentials, followed by Section 4 that illustrates the roles of SRAs using several emerging societal issues as examples. Section 5 details the open problems and challenges that socially responsible AI currently confronts. The last section concludes the survey.
\section{Social Responsibility of AI}
Social Responsibility of AI includes efforts devoted to addressing both technical and societal issues. While similar concepts (e.g., ``Ethical AI'') repeatedly appear in the news, magazines, and scientific articles, ``Social Responsibility of AI'' has yet to be properly defined. In this section, we first attempt to provide an inclusive definition and then propose the \textit{Pyramid} of Social Responsibility of AI to outline the various responsibilities of AI in a hierarchy: functional responsibilities, legal responsibilities, ethical responsibilities, and philanthropic responsibilities. At last, we compare ``Socially Responsible AI'' with similar concepts.
\subsection{What is Social Responsibility of AI?}
\begin{definition}[Social Responsibility of AI]
Social Responsibility of AI refers to a human value-driven process where values such as Fairness, Transparency, Accountability, Reliability and Safety, Privacy and Security, and Inclusiveness are the principles; designing Socially Responsible AI Algorithms is the means; and addressing the social expectations of generating shared value -- enhancing both AI's ability and benefits to society -- is the main objective. 
\end{definition}
Here, we define three dimensions of Social Responsibility of AI: the \textit{principles} lay the foundations for ethical AI systems; the \textit{means} to reach the overarching goal of Social Responsibility of AI is to develop Socially Responsible AI Algorithms; and the \textit{objective} of Social Responsibility of AI is to improve both AI's capability and humanity with the second being the proactive goal.  
\subsection{The Pyramid of Social Responsibility of AI}
\begin{figure}
\center
  \includegraphics[width=.5\columnwidth]{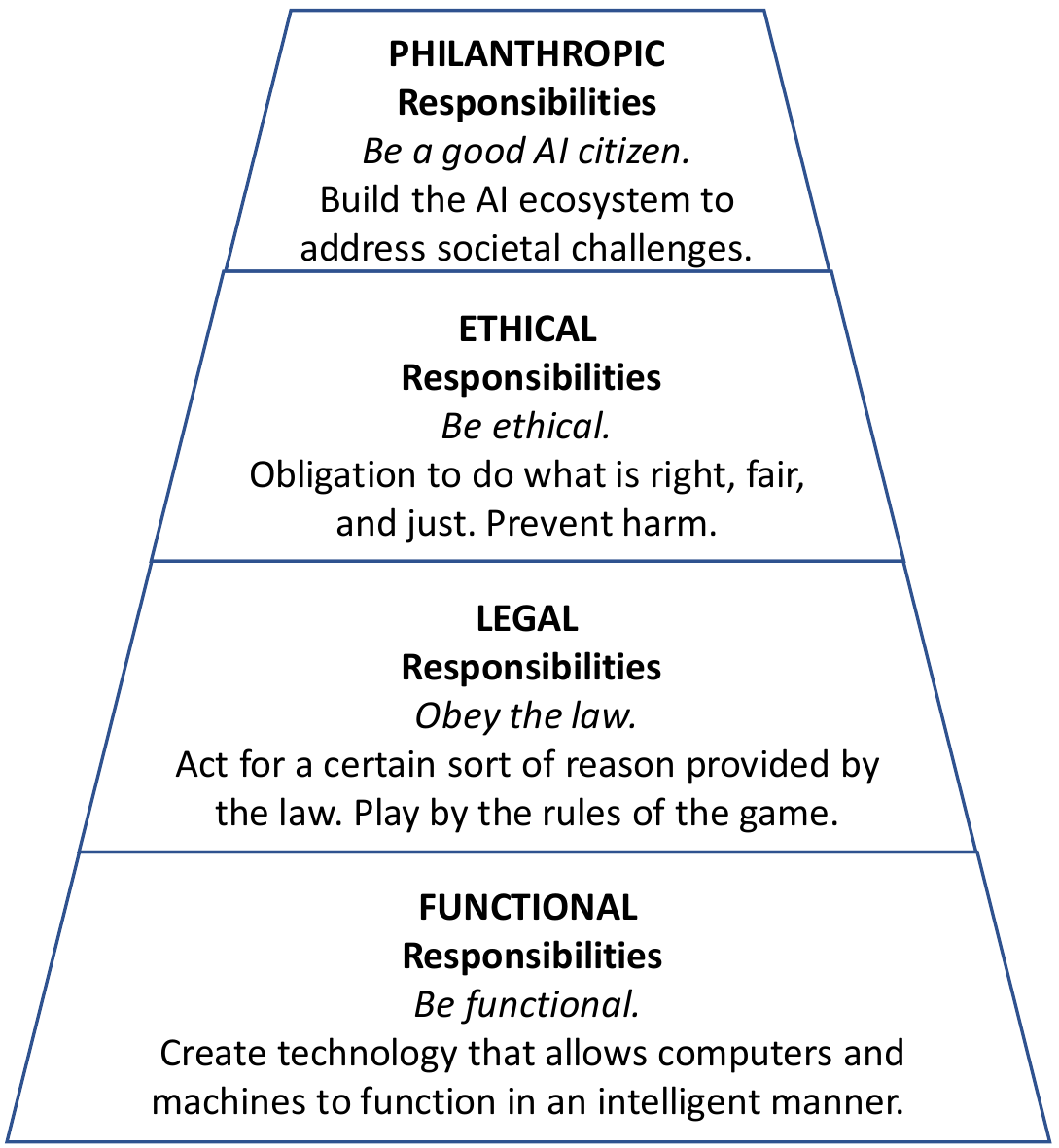}
  \caption{The pyramid of Social Responsibility of AI, adapted from the Pyramid of CSR by \citeA{carroll1991pyramid}.}
  \label{pyramid}
\end{figure}

Social Responsibility of AI should be framed in such a way that the entire range of AI responsibilities are embraced. Adapting Carroll's Pyramid of CSR \cite{carroll1991pyramid} in the AI context, we suggest four kinds of social responsibilities that constitute the Social Responsibility of AI: \textit{functional, legal, ethical,} and \textit{philanthropic} responsibilities, as shown in Figure \ref{pyramid}. By modularizing AI responsibilities, we hope to help AI technologists and researchers to reconcile these obligations and simultaneously fulfill all the components in the pyramid. All of these responsibilities have always existed, but functional responsibilities have been the main consideration until recently. Each type of responsibility requires close consideration. 

The pyramid portrays the four components of Social Responsibility of AI, beginning with the basic building block notion that the functional competence of AI undergirds all else. \textit{Functional responsibilities} require AI systems to perform in a manner consistent with profits maximization, operating efficiency, and other key performance indicators. Meanwhile, AI is expected to obey the law, which codifies the acceptable and unacceptable behaviors in our society. That is, \textit{legal responsibilities} require AI systems to perform in a manner consistent with expectations of government and law. All AI systems should at least meet the minimal legal requirements. At its most fundamental level, \textit{ethical responsibilities} are the obligation to do what is right, just, and fair, and to prevent or mitigate negative impact on stakeholders (e.g., users, the environment). To fulfill its ethical responsibilities, AI systems need to perform in a manner consistent with societal expectations and ethical norms, which cannot be compromised in order to achieve AI's functional responsibilities. Finally, in \textit{philanthropic responsibilities}, AI systems are expected to be \textit{good AI citizens} and to contribute to tackling societal challenges such as cancer and climate change. Particularly, it is important for AI systems to perform in a manner consistent with the philanthropic and charitable expectations of society to enhance people's quality of life. The distinguishing feature between ethical and philanthropic responsibilities is that the latter are not expected in an ethical sense. For example, while communities desire AI systems to be applied to humanitarian projects or purposes, they do not regard the AI systems as unethical if they do not provide such services. We explore the nature of Social Responsibility of AI by focusing on its components to help AI technologists to reconcile these obligations. Though these four components are depicted as separate concepts, they are not mutually exclusive. It is necessary for AI technologists and researchers to recognize that these obligations are in a constant but dynamic tension with one another.
\subsection{Comparisons of Similar Concepts}
Based on Definition 1 and the pyramid of socially responsibility of AI, we compare Socially Responsible AI with other similar concepts, as illustrated in Table \ref{comparisons}. The results show that Socially Responsible AI holds a systematic view that subsumes existing concepts and further considers the fundamental responsibilities of AI systems -- to be functional and legal, as well as their philanthropic responsibilities -- to be able to improve life quality of well beings and address challenging societal issues. In the rest of this survey, we focus our discussions on the \textit{ethical} (Section 3, essentials of SRAs) and \textit{philanthropic} (Section 4, roles of SRAs) responsibilities of AI given that both the functional and legal responsibilities are the usual focuses in AI research and development. An overview of SRAs research is illustrated in Figure \ref{outline}, which we will refer back to throughout the remainder of the survey. Importantly, in our view, the essentials of SRAs work toward ethical responsibilities, and their roles in society encompass both ethical and philanthropic responsibilities.
\begin{table}[t]
\begin{tabular}{|c|p{10cm}|}
\hline
Concepts       & \multicolumn{1}{c|}{Definitions}\\ \hline
\textbf{Robust AI}      & AI systems with the ability ``to cope with errors during execution and cope with erroneous input'' \cite{robust_2021}. \\ \hline
\textbf{Ethical AI}     & AI systems that do what is right, fair, and just. Prevent harm.\\ \hline
\textbf{Trustworthy AI} & AI systems that achieve their full potential if \textit{trust} can be established in the development, deployment, and use \cite{thiebes2020trustworthy}.  \\ \hline
\textbf{Fair AI}        & AI systems absent from ``any  prejudice  or  favoritism  toward an individual or a group based on their inherent or acquired characteristics'' \cite{mehrabi2019survey}.\\ \hline
\textbf{Safe AI}        & AI systems deployed in ways that do not harm humanity \cite{Safety2019}.\\ \hline
\textbf{Dependable AI}  & AI systems that focus on reliability, verifiability, explainability, and security \cite{singh2021trustworthy}. \\ \hline
\textbf{Human-centered AI} & AI systems that are ``continuously improving because of human input while providing an effective experience between human and robot''\footnotemark.\\ \hline
\end{tabular}
\caption{Definitions of concepts similar to Socially Responsible AI.}
\label{comparisons}
\end{table}
\footnotetext[2]{https://www.cognizant.com/glossary/human-centered-ai}
\section{Socially Responsible AI Algorithms (SRAs)}
\begin{figure}
\center
  \includegraphics[width=.8\textwidth]{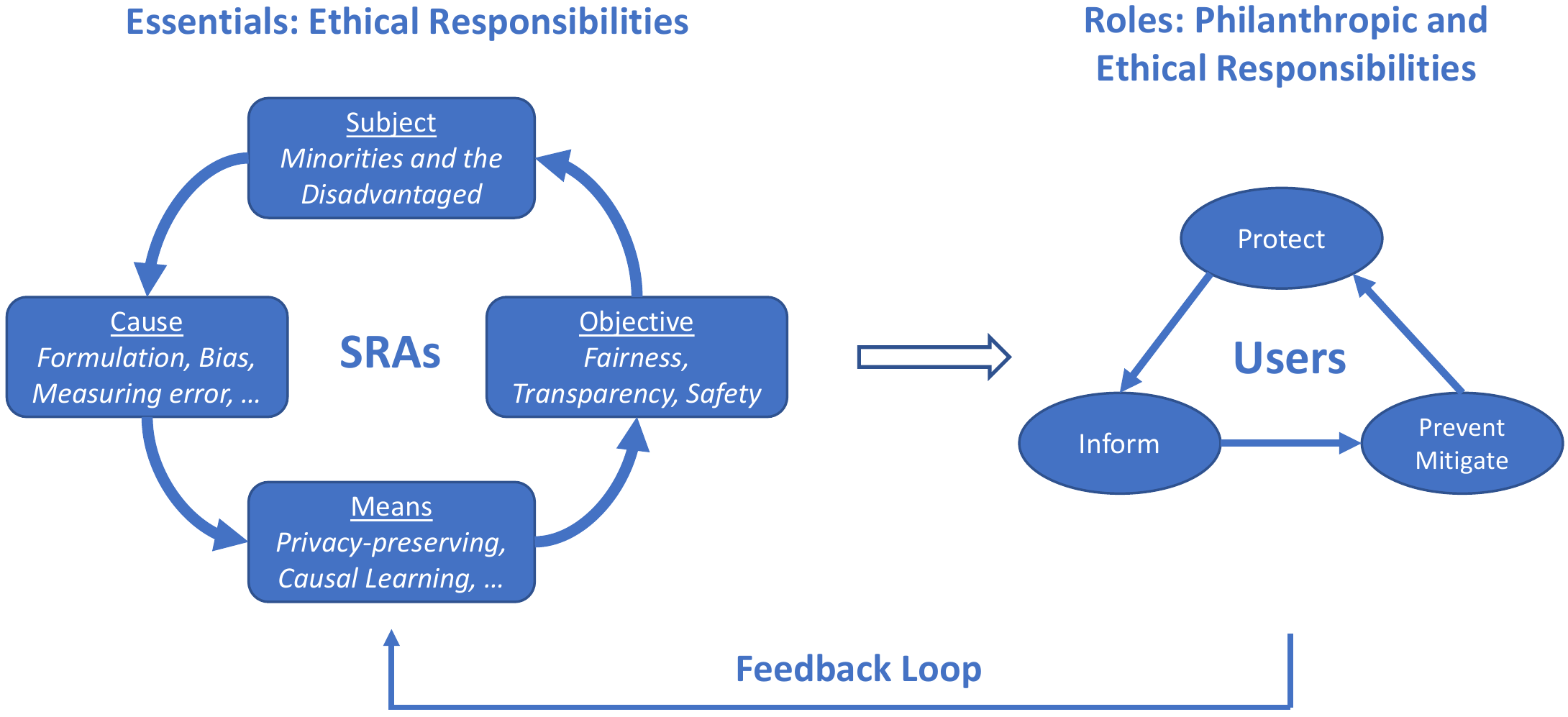}
  \caption{The framework of Socially Responsible AI Algorithms (SRAs). It consists of the essentials (i.e., the internal mechanisms) of SRAs (left), their roles (right), and feedback received from end users for helping SRAs gradually achieve the expected social values (bottom). The essentials of SRAs center on the ethical responsibilities of AI and the roles of SRAs require philanthropic responsibilities and ethical responsibilities.}
  \label{SRAs}
\end{figure}

The role of AI technologists and researchers carries a number of responsibilities. The most obvious is developing accurate, reliable, and trustworthy algorithms that can be depended on by their users. Yet, this has never been a trivial task. For example, due to the various types of human biases, e.g., confirmation bias, gender bias, and anchoring bias, AI technologists and researchers often inadvertently inject these same kinds of bias into the developed algorithms, especially when using machine learning techniques. For example, supervised machine learning is a common technique for learning and validating algorithms through manually annotated data, loss functions, and related evaluation metrics. Numerous uncertainties -- e.g., imbalanced data, ill-defined criteria for data annotation, over-simplified loss functions, and unexplainable results -- potentially lurk in this ``beautiful'' pipeline and will eventually lead to negative outcomes such as biases and discrimination. With the growing reliance on AI in almost any field in our society, we must bring upfront the vital question about \textit{how to develop Socially Responsible AI Algorithms}. While conclusive answers are yet to be found, we attempt to provide a systematic framework of SRAs (illustrated in Figure \ref{SRAs}) to discuss the components of AI's ethical responsibilities, the roles of SRAs in terms of AI's philanthropic and ethical responsibilities, and the feedback from users routed back as inputs to SRAs. We hope to broaden future discussions on this subject. In this regard, we define SRAs as follows:
\begin{figure}
\center
  \includegraphics[width=\textwidth]{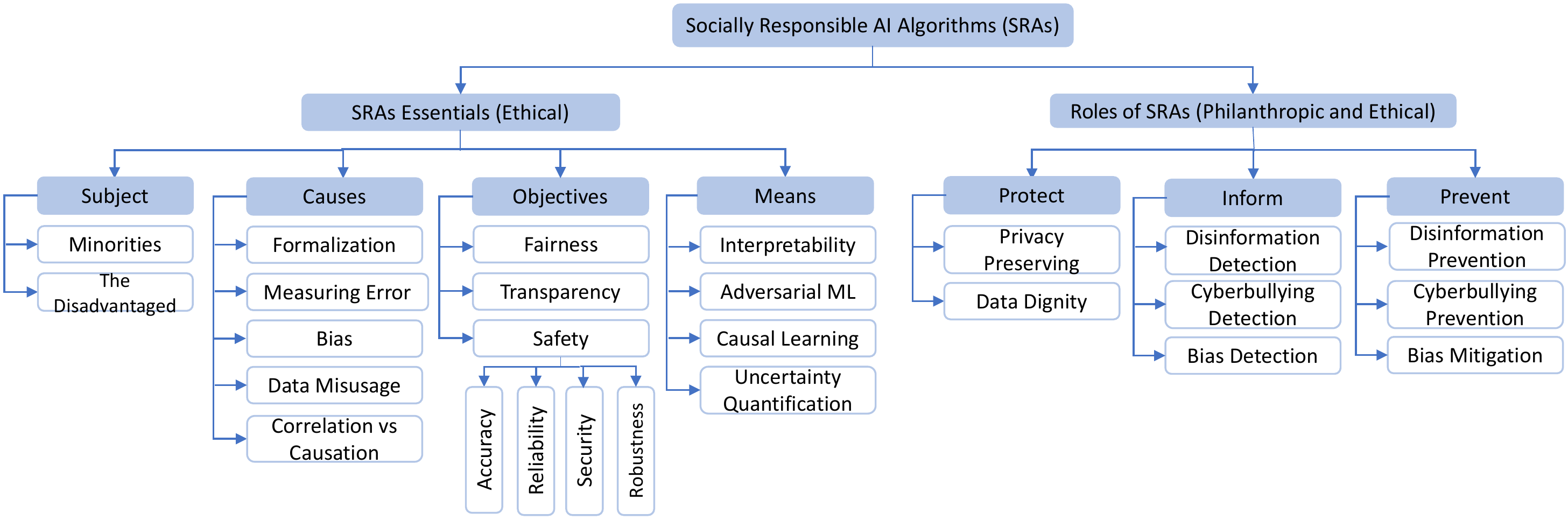}
  \caption{An overview of SRAs Research.}
  \label{outline}
\end{figure}
\begin{definition}[Socially Responsible AI Algorithms]\
Socially Responsible AI Algorithms are the intelligent algorithms that prioritize the needs of all stakeholders as the highest priority, especially the minoritized and disadvantaged users, in order to make just and trustworthy decisions. These obligations include protecting and informing users; preventing and mitigating negative impact; and maximizing the long-term beneficial impact. Socially Responsible AI Algorithms constantly receive feedback from users to continually accomplish the expected social values.  
\end{definition}
In this definition, we highlight that the functional (e.g., maximizing profits) and societal (e.g., transparency) objectives are integral parts of AI algorithms. SRAs aim to be socially responsible while still meeting and exceeding business objectives.
\subsection{Subjects of Socially Indifferent AI Algorithms}
Every human being can be a potential victim of socially indifferent AI algorithms. Mirroring society, the ones who suffer the most, both in frequency and severity, are minorities and disadvantaged groups such as black, indigenous and people of color (BIPOC), and females. For example, Google mislabeled an image of two black people as ``gorillas'' \cite{Guynn2015} and more frequently showed ads of high-paying jobs to males than females \cite{Carpenter2015}. Similar gender bias was also found in Facebook algorithms behind the job ads \cite{Horwitz_2021}. In domains with high-stakes decisions, e.g., financial services, healthcare, and criminal justice, it is not uncommon to identify instances where socially indifferent AI algorithms favor privileged groups. For example, the algorithm used in Correctional Offender Management Profiling for Alternative Sanctions (COMPAS) was found almost twice as likely to mislabel a black defendant as a future risk than a white defendant \cite{angwin2016machine}. Identifying the subjects of socially indifferent AI algorithms depends on the context. In another study, the journalistic organization ProPublica\footnote{https://www.propublica.org/} investigated algorithms that determine online prices for Princeton Review's tutoring classes. The results showed that people who lived in higher income areas were charged twice as much as the general public and than people living in a zip code with high population density. Asians were 1.8 times more likely to pay higher price, regardless of their income \cite{angwin2015tiger}. Analogously, these AI algorithms might put poor people who cannot afford internet service at disadvantage because they simply have never seen such data samples in the training process. 

When it comes to purpose-driven collection and use of data, each individual can be the subject of socially indifferent AI algorithms. Users' personal data are frequently collected and used without their consent. Such data includes granular details such as contact information, online browsing and session record, social media consumption, location and so on. While most of us are aware of our data being used, few have controls to where and how the data is used, and by whom. The misuse of data and lack of knowledge causes users to become the victims of privacy-leakage and distrust. 
\subsection{Causes of Socially Indifferent AI Algorithms}
There are many potential factors that can cause AI algorithms to be socially indifferent. Here, we list several causes that have been frequently discussed in literature \cite{mehrabi2019survey,getoor2019responsible}. They are formalization, measuring errors, bias, privacy, and correlation versus causation. 
\subsubsection{Formalization}
AI algorithms encompass data formalization, label formalization, formalization of loss function and evaluation metrics. We unconsciously make some frame of reference commitment to each of these formalizations. Firstly, the social and historical context are often left out when transforming raw data into numerical feature vectors. Therefore, AI algorithms are trained on pre-processed data with important contextual information missing. Secondly, data annotation can be problematic for a number of reasons. For example, what are the criteria? Who defines the criteria? Who are the annotators? How can it be ensured that they all follow the criteria? What we have for model training are only proxies of the true labels \cite{getoor2019responsible}. Ill-formulated loss functions can also result in socially indifferent AI algorithms. Many loss functions in the tasks are over-simplified to solely focus on maximizing profits and minimizing losses. The concerns of unethical optimization are recently discussed by \citeA{beale2019unethical}. Unknown to AI systems, certain strategies in the optimization space that are considered as unethical by stakeholder may be selected to satisfy the simplified task requirements. Lastly, the use of inappropriate benchmarks for evaluation may push algorithms away from the overarching goal of the task and fuel injustice.  
\subsubsection{Measuring Errors}
\begin{figure}
\center
  \includegraphics[width=.5\columnwidth]{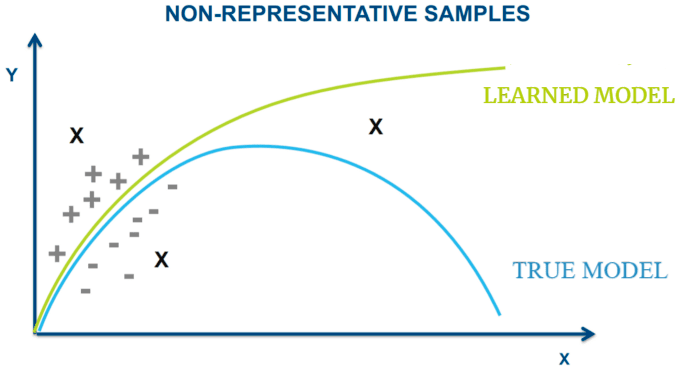}
  \caption{An example of measuring errors. The green line denotes the learned model and the blue one is the true model. `+' and `-' represent training data belonging to different classes; `X' represents testing data. Image taken from Getoor's slides for 2019 IEEE Big Data keynote \cite{getoor2019responsible} with permission.}
  \label{measureerror}
\end{figure}

Another cause of socially indifferent AI algorithms is the errors when measuring algorithm performance. When reporting results, researchers typically proclaim the proposed algorithms can achieve certain accuracy or F1 scores. However, this is based on assumptions that the training and test samples are representative of the target population and their distributions are similar enough. Yet, how often does the assumption hold in practice? As illustrated in Figure \ref{measureerror}, with non-representative samples, the learned model can achieve zero training error and perform well on the testing data at the initial stage. However, with more data being tested later, the model performance deteriorates because the learned model does not represent the true model.
\subsubsection{Bias}
Bias is one of the most discussed topics regarding responsible AI. We here focus on the data bias, automation bias, and algorithmic bias \cite{getoor2019responsible}.

\noindent\textbf{Data Bias.}
Data, especially big data, is often heterogeneous -- data with high variability of types and formats, e.g., text, image, and video. The availability of multiple data sources brings unprecedented opportunities as well as unequivocally presented challenges \cite{li2017feature}. For instance, high-dimensional data such as text is infamous for the danger of overfitting and the curse of dimensionality. Additionally, it is rather challenging to find subset of features that are predictive but uncorrelated. The required number samples for generalization also grows proportionally with feature dimension. One example is how the U.S. National Security Agency tried to use AI algorithms to identify potential terrorists. The Skynet project collected cellular network traffic in Pakistan and extracted 80 features for each cell phone user with only 7 known terrorists \cite{NSA}. The algorithm ended up identifying an Al Jazeera reporter covering Al Qaeda as a potential terrorist. Data heterogeneity is also against the well known $i.i.d.$ assumption in most learning algorithms \cite{li2017feature}. Therefore, training these algorithms on heterogeneous data can result in undesired results. Imbalanced subgroups is another source of data bias. As illustrated in \cite{mehrabi2019survey}, regression analysis based on the subgroups with balanced fitness level suggests positive correlation between BMI and daily pasta calorie intake whereas that based on less balanced data shows almost no relationship. 

\noindent\textbf{Automation Bias.} This type of bias refers to our preference to results suggested by automated decision-making systems while ignoring the contradictory information. Too much reliance on the automated systems without sparing additional thoughts in making final decisions, we might end up abdicating decision responsibility to AI algorithms.

\noindent\textbf{Algorithmic Bias.} Algorithmic bias regards biases added purely by the algorithm itself \cite{baeza2018bias}. Some algorithms are inadvertently taught prejudices and unethical biases by societal patterns hidden in the data. Typically, models fit better to features that frequently appear in the data. For example, an automatic AI recruiting tool will learn to make decisions for a given applicant of a software engineer position using observed patterns such as ``experience'', ``programming skills'', ``degree'', and ``past projects''. For a position where gender disparity is large, the algorithms mistakenly interpret this collective imbalance as a useful pattern in the data rather than undesirable noise that should have been discarded. Algorithmic bias is systematic and repeatable error in an AI system that creates discriminated outcome, e.g., privileging wealthy users over others. It can amplify, operationalize, and even legitimize institutional bias \cite{getoor2019responsible}. 
\subsubsection{Data Misuse}
Data is the fuel and new currency that has empowered tremendous progress in AI research. Search engines have to rely on data to craft precisely personalized recommendation that improves the online experience of consumers, including online shopping, book recommendation, entertainment, and so on. However, users' data are frequently misused without the consent and awareness of users. One example is the Facebook-Cambridge Analytical scandal \cite{FacebookCambridge2021} where millions of Facebook users' personal data was collected by Cambridge Analytica \cite{CambridgeAnalytica2021}, without their consent. In a more recent study \cite{cabanas2020does}, researchers show that Facebook allows advertisers to exploit its users' sensitive information for tailored ad campaigns. To make things worse, users often have no clue about where, how, and why their data is being used, and by whom. The lack of knowledge and choice over their data causes users to undervalue their personal data, and further creates issues such as privacy and distrust. 
\subsubsection{Correlation vs Causation}
\begin{figure}
\center
  \includegraphics[width=.4\columnwidth]{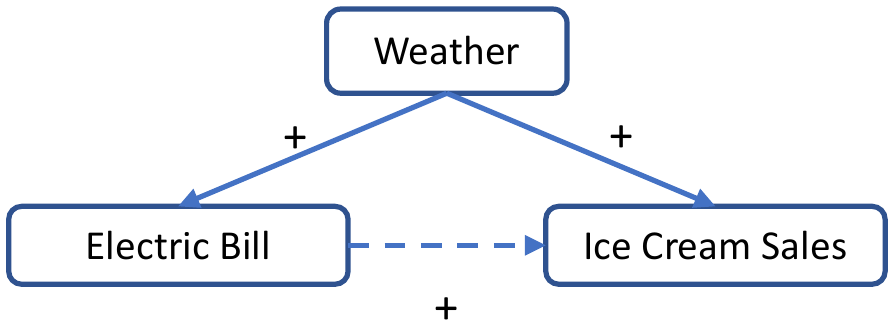}
  \caption{Confounders are common reasons for spurious correlation between two variables that are not causally connected.}
  \label{icecream}
\end{figure}
AI Algorithms can become socially indifferent when correlation is misinterpreted as causation. For example, in the diagram in Figure \ref{icecream}, we observe a strong correlation between the electric bill of an ice cream shop and ice cream sales. Apparently, high electric bill cannot \textit{cause} the ice cream sales to increase. Rather, weather is the common cause of electric bill and the sale, i.e., high temperature causes high electric bill and the increased ice cream sales. Weather -- the \textit{confounder} -- creates a spurious correlation between electric bill and ice cream sales. Causality is a generic relationship between the cause and the outcome \cite{guo2020survey}. While correlation helps with prediction, causation is important for decision making. One typical example is Simpson's Paradox \cite{blyth1972simpson}. It describes a phenomenon where a trend or association observed in subgroups maybe opposite to that observed when these subgroups are aggregated. For instance, in the study of analyzing the sex bias in graduation admissions at UC Berkeley \cite{bickel1975sex}, the admission rate was found higher in male applicants when using the entire data. However, when the admission data were separated and analyzed over the departments, female candidates had equal or even higher admission rate over male candidates. 
\subsection{Objectives of Socially Responsible AI Algorithms}
\begin{figure}
\center
  \includegraphics[width=.6\columnwidth]{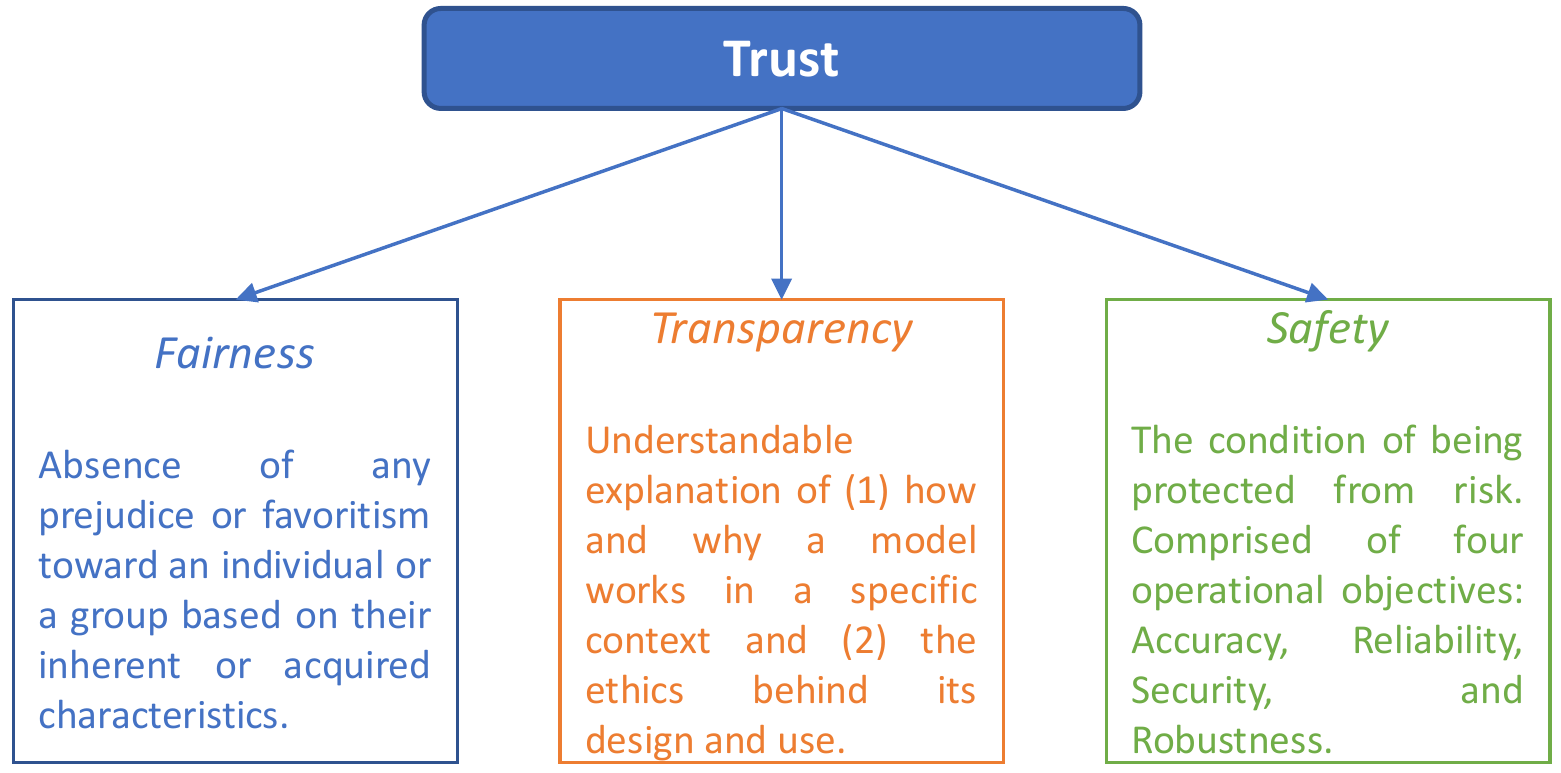}
  \caption{The objectives of Socially Responsible AI Algorithms.}
  \label{objective}
\end{figure}

Essentially, the goal is to (re)build \textit{trust} in AI. By definition, trust is the ``firm belief in the reliability, truth or ability of someone or something''\footnote{Definition from Oxford Languages.}. It is a high-level concept that needs to be specified by more concrete objectives. We here discuss the SRAs objectives that have been discussed comparatively frequently in literature. They are fairness, transparency, and safety as illustrated in Figure \ref{objective}.
\subsubsection{Fairness}
\label{fairness}
Fairness in AI has gained substantial attentions in both research and industry since 2010. For decades, researchers found it rather challenging to present a unified definition of fairness in part because fairness is a societal and ethical concept. This concept is mostly subjective, changes over social context, and evolves over time, making fairness a rather challenging goal to achieve in practice. Because SRAs is a decision-making process commensurate with social values, we here adopt a fairness definition in the context of decision-making:
\begin{definition}[Fairness]
``Fairness is the absence of any prejudice or favoritism toward an individual or a group based on their inherent or acquired characteristics'' \cite{mehrabi2019survey}.
\end{definition}

\noindent Note that even an ideally ``fair'' AI system defined in a specific context might still lead to biased decisions as the entire decision making process involves numerous elements such as policy makers and environment. While the concept of fairness is difficult to pin down, unfairness/bias/discrimination might be easier to identify. There are six types of discrimination \cite{mehrabi2019survey}. Direct discrimination results from protected attributes of individuals while indirect discrimination from seemingly neural and non-protected attributes. Systemic discrimination relates to policies that may show discrimination against subgroups of population. Statistical discrimination occurs when decision makers use average statistics to represent individuals. Depending whether the differences amongst different groups can be justified or not, we further have explainable and unexplainable discrimination. 

\subsubsection{Transparency}
Transparency is another important but quite ambiguous concept. This is partly because AI alone can be defined in more than 70 ways \cite{legg2007collection}. When we seek a transparent algorithm, we are asking for an \textit{understandable explanation} of how it works \cite{Yeo_2020}: What does the training set look like? Who collected the data? What is the algorithm doing? There are mainly three types of transparency with regard to human interpretability of AI algorithms \cite{weller2017challenges}: For a \textit{developer}, the goal of transparency is to understand how the algorithm works and get a sense of why; for a \textit{deployer} who owns and releases the algorithm to the public, the goal of transparency is to make the consumers to feel safe and comfortable to use the system; and what transparency means to a \textit{user} is understanding what the AI system is doing and why. We may further differentiate global transparency from local transparency, the former aims to explain the entire system whereas the latter explains a decision within a particular context. 

Yet, at the same time, disclosures about AI can pose potential risks: explanations can be hacked and releasing additional information may make AI more vulnerable to attacks. It is becoming clear that transparency is often beneficial but not universally good \cite{weller2017challenges}. The AI ``transparency paradox'' encourages different parties of AI systems to think more carefully about how to balance the transparency and the risks it poses. We can also see related discussions in recent work such as \cite{slack2020fooling}. The paper studied how the widely recognized interpretable algorithms LIME \cite{ribeiro2016should} and SHAP \cite{lundberg2017unified} could be hacked. As the authors illustrated, explanations can be purposefully manipulated, leading to a loss of trust not only in the model but also in its explanations \cite{slack2020fooling}. Consequently, while working towards the goal of transparency, we must also recognize that privacy and security are the indispensable conditions we need to satisfy. 
\subsubsection{Safety}
Because AI systems operate in a world with much uncertainty, volatility, and flux, another objective of SRAs is to be safe, accurate, and reliable \cite{varshney2017safety}. There are four operational objectives relevant to Safety: \textit{accuracy, reliability, security,} and \textit{robustness} \cite{leslie2019understanding}. In machine learning, \textit{accuracy} is typically measured by error rate or the fraction of instances for which the algorithm produces an incorrect output. As a standard performance metric, accuracy should be the fundamental component to establishing the approach to safe AI. It is necessary to specify a proper performance measure for evaluating any AI systems. For instance, when data for classification tasks is extremely imbalanced, precision, recall, and F1 scores are more appropriate than accuracy. The objective of \textit{reliability} is to ensure that AI systems behave as we anticipate. It is a measure of consistency and is important to establish confidence in the safety of AI systems. \textit{Security} encompasses the protection of information integrity, confidentiality, and continuous functionality to its users. Under harsh conditions (e.g., adversarial attack, perturbations, and implementation error), AI systems are expected to functions reliably and accurately, i.e., \textit{Robustness}.  
\subsection{Means Towards Socially Responsible AI Algorithms}
In this section, we review four primary machine learning techniques and statistical methods for achieving the goals of SRAs -- interpretability and explainability, adversarial machine learning, causal learning, and uncertainty quantification. Existing surveys have conducted comprehensive reviews on each of these techniques: e.g., interpretablity \cite{carvalho2019machine,tjoa2019XAI}, causal learning \cite{guo2020survey,yao2020causal}, adversarial machine learning \cite{chakraborty2018adversarial,akhtar2019brief}, and uncertainty quantification \cite{kabir2018neural}. We thereby focus on the basics and the most frequently discussed methods in each means.
\subsubsection{Interpretability and Explainability}
Interpretability and explanability are the keys to increasing transparency of AI algorithms. This is extremely important when we leverage these algorithms for high-stakes prediction applications, which deeply impact people's lives \cite{rudin2019stop}. Existing work in machine learning interpretability can be categorized according to different criteria. Depending on when the interpretability methods are applicable (before, during, or after building the machine learning model), we have pre-model (before), in-model (during), and post-model (after) interpretability. Pre-model techniques are only applicable to the data itself. It requires an in-depth understanding of the data before building the model, e.g., sparsity and dimensionality. Therefore, it is closely related to data interpretability \cite{carvalho2019machine}, in which classic descriptive statistics and data visualization methods are often used, including Principal Component Analysis \cite{wold1987principal} and t-SNE \cite{maaten2008visualizing}, and clustering methods such as $k$-means \cite{hartigan1979algorithm}. In-model interpretability asks for intrinsically interpretable AI algorithms (e.g., \citeR{yang2016hierarchical}), we can also refer to it as intrinsic interpretability. It can be achieved through imposition of constraints on the model such as causality, sparsity, or physical conditions from domain knowledge \cite{rudin2019stop}. In-model interpretability answers question \textit{how the model works} \cite{lipton2018mythos}. Decision trees, rule-based models, linear regression, attention network, and disentangled representation learning are in-model interpretability techniques. Post-model interpretability, or post-hoc interpretability (e.g., \citeR{mordvintsev2015inceptionism,ribeiro2016should}), is applied after model training. It answers the question \textit{what else can the model tell us} \cite{lipton2018mythos}. Post-model interpretability include local explanations \cite{ribeiro2016should}, saliency maps \cite{simonyan2013deep}, example-based explanations \cite{kim2016examples}, influence functions \cite{koh2017understanding}, feature visualization \cite{erhan2009visualizing}, and explaining by base interpretable models \cite{craven1996extracting}.

Another criterion to group current interpretability techniques is model-specific vs model-agnostic. Model-specific interpretation is based on internals of a specific model  \cite{molnar2020interpretable}. To illustrate, the coefficients of a linear regression model belong to model-specific interpretation. Model-agnostic methods do not have access to the model inner workings, rather, they are applied to any machine learning model after it has been trained. Essentially, the goal of interpretability is to help the user understand the decisions made by the machine learning models through the tool \textit{explanation}. There are pragmatic and non-pragmatic theories of explanation. The former indicates that explanation should be a \textit{good} answer that can be easily understood by the audience. The non-pragmatic theory emphasizes the \textit{correctness} of the answer to the why-question. Both need to have the following properties \cite{robnik2018perturbation}: expressive power, translucency, portability, and algorithmic complexity. 
\subsubsection{Adversarial Machine Learning}
\label{adversary}
Machine learning models, especially deep learning models, are vulnerable to crafted adversarial examples, which are imperceptible to human eyes but can easily fool deep neural networks (NN) in the testing/deploying stage \cite{yuan2019adversarial}. Adversarial examples have posed great concerns in the security and integrity of various applications. Adversarial machine learning, therefore, closely relates to the robustness of SRAs. 

The security of any machine learning model is measured with regard to the adversarial goals and capabilities \cite{chakraborty2018adversarial}. Identifying the threat surface \cite{papernot2016towards} of an AI system built on machine learning models is critical to understand where and how an adversary may subvert the system under attack. For example, the attack surface in a standard automated vehicle system can be defined with regard to the data processing pipeline. Typically, there are three types of attacks the attack surface can identify: evasion attack -- the adversary attempts to evade the system by manipulating malicious samples during testing phase, poisoning attack -- the adversary attempts to poison the training data by injecting carefully designed samples into the learning process, and exploratory attack -- it tries to collect as much information as possible about the learning algorithm of the underlying system and pattern in training data. Depending on the amount of information available to an adversary about the system, we can define different types of adversarial capabilities. In the training phase (i.e., training phase capabilities), there are three broad attack strategies: (1) data injection. The adversary can only augment new data to the training set; (2) data modification. The adversary has full access to the training data; and (3) logic corruption. The adversary can modify the learning algorithm. In the testing phase (i.e., testing phase capabilities), adversarial attacks focus on producing incorrect outputs. For white-box attack, an adversary has full knowledge about the model used for prediction: algorithm used in training, training data distribution, and the parameters of the fully trained model. The other type of attack is black-box attack, which, on the contrary, assumes no knowledge about the model and only uses historical information or information about the settings. The primary goal of black-box attack is to train a local model with the data distribution, i.e., non-adaptive attack, and with carefully selected dataset by querying the target model, i.e., adaptive attack.  

Exploratory attacks do not have access to the training data but aim to learn the current state by probing the learner. Commonly used techniques include model inversion attack \cite{fredrikson2014privacy,fredrikson2015model}, model extraction using APIs \cite{tramer2016stealing}, and inference attack \cite{ateniese2015hacking,shokri2017membership}. The popular attacks are evasion attacks where malicious inputs are craftily manipulated so as to fool the model to make false predictions. Poisoning attacks, however, modify the input during the training phase to obtain the desired results. Some of the well-known techniques are generative adversarial network (GAN) \cite{goodfellow2014generative}, adversarial examples generation (including training phase modification, e.g., \citeR{barreno2006can}, and testing phase modification, e.g., \citeR{papernot2016distillation}), GAN-based attack in collaborative deep learning \cite{hitaj2017deep}, and adversarial classification \cite{dalvi2004adversarial}.
\subsubsection{Causal Learning}
Causal inference and reasoning is a critical ingredient for AI to achieve human-level intelligence, an overarching goal of Socially Responsible AI. The momentum of integrating causality into responsible AI is growing, as witnessed by a number of works (e.g., \citeR{kusner2017counterfactual,xu2019achieving,holzinger2019causability}) studying SRAs through causal learning methods. 

\noindent\textbf{Basics of Causal Learning.} The two fundamental frameworks in causal learning are \textit{structural causal models} \cite{pearl2009causality} and \textit{potential outcome} \cite{rubin1974estimating}. Structural causal models rely on the causal graph, which is a special class of Bayesian network with edges denoting causal relationships. A more structured format is referred to as structural equations. One of the fundamental notions in structural causal models is the \textit{do}-calculus \cite{pearl2009causality}, an operation for \textit{intervention}. The difficulty to conduct causal study is the difference between the observational and interventional distribution, the latter describes what the distribution of outcome $Y$ is if we were to set covariates $X=x$. Potential outcome framework interprets causality as given the treatment and outcome, we can only observe one potential outcome. The counterfactuals -- potential outcome that would have been observed if the individual had received a different treatment -- however, can never be observed in reality. These two frameworks are the foundations of causal effect estimation (estimating effect of a treatment) and causal discovery (learning causal relations amongst different variables).

Many important concepts in causal inference have been adapted to AI such as intervention and counterfactual reasoning. Here, we introduce the causal concept most frequently used in SRAs -- propensity score, defined as ``conditional probability of assignment to a particular treatment given a vector of observed covariates'' \cite{rosenbaum1983central}. A popular propensity-based approach is Inverse Probability of Treatment Weighting \cite{hirano2003efficient}. To synthesize a randomized control trial \cite{rubin1974estimating}, it uses covariate balancing to weigh instances based on their propensity scores and the probability of an instance to receive the treatment. Let $t_i$ and $x_i$ be the treatment assignment and covariate of instance $i$, the weight $w_i$ is typically computed by the following formula:
\begin{equation}
w_i=\frac{t_i}{P(t_i|x_i)}+\frac{1-t_i}{1-P(t_i|x_i)},
\end{equation}
where $P(t_i|x_i)$ quantifies the propensity score. The weighted average of the observed outcomes for the treatment and control groups are defined as
\begin{equation}
    \hat{\tau}=\frac{1}{n_1}\sum_{i:t_i=1}w_iy_i-\frac{1}{n_0}\sum_{i:t_i=0}w_iy_i,
\end{equation}
where $n_1$ and $n_0$ denote the sizes of the treated and controlled groups.  

\noindent \textbf{Causal Learning for SRAs.} Firstly, it is becoming increasingly popular to use causal models to solve fairness-related issues. For example, the subject of causality and its importance to address fairness issue was discussed in \cite{loftus2018causal}. Causal models can also be used to discover and eliminate discrimination to make decisions that are irrespective of sensitive attributes, on individual-, group-, and system-level, see, e.g., \cite{zhang2016causal,zhang2017anti,nabi2018fair}. Secondly, bias alleviation is another field where causal learning methods are frequently discussed and affect many machine learning applications at large. The emerging research on debiasing recommender system \cite{wang2018position,wang2018deconfounded,joachims2017unbiased} can serve as one example. Due to the biased nature of user behavior data, recommender systems inevitably involve with various discrimination-related issues: recommending less career coaching services and high-paying jobs to women \cite{lambrecht2019algorithmic,datta2015automated}, recommending more male-authored books \cite{ekstrand2018exploring}, and minorities are less likely to become social influencers \cite{karimi2018homophily,stoica2018algorithmic}. Gender and ethnic biases were even found in a broader context, e.g., word embeddings trained on 100 years of text data \cite{garg2018word}. Causal approaches such as \cite{yang2020causal} aim to mitigate such bias in word embedding relations. 

Thirdly, causal learning methods also have had discernible achievements in transparency, especially the interpretability of black-box algorithms. Causality is particularly desired since these algorithms only capture correlations not real causes \cite{moraffah2020causal}. Further, it has been suggested that counterfactual explanations are the highest level of interpretability \cite{pearl2018theoretical}. For model-based interpretations, causal interpretability aims to explain the causal effect of a model component on the final decision \cite{chattopadhyay2019neural,parafita2019explaining,narendra2018explaining}. One example to differentiate it from traditional interpretability is only causal interpretability is able to answer question such as ``What is the effect of the $n$-th filter of the $m$-th layer of a neural network on the prediction of the model?''.  Counterfactual explanations is a type of example-based explanations, in which we look for data instances that can explain the underlying data distributions. Counterfactual explanations are human friendly, however, it is possible to have different true versions of explanations for the predicted results, i.e., the Rashomon effect \cite{molnar2020interpretable}. Studies such as \cite{wachter2017counterfactual,grath2018interpretable,liu2019generative} are proposed to address this issue. For detailed discussion on causal interpretability, please refer to \cite{moraffah2020causal}. Lastly, causal learning is inherently related to the robustness or adaptability of AI systems, which have been noted to lack the capability of reacting to new circumstances they are not trained for. Causal relationship, however, is expected to be invariant and robust across environments \cite{pearl2019seven,pearl2009causality}. This complements intensive earlier efforts toward ``transfer learning'', ``domain adaptation'', and ``lifelong learning'' \cite{chen2018lifelong}. Some current work seeking to extrapolate the relationship between AI robustness and causality includes the independent causal mechanism principle \cite{peters2017elements,scholkopf2019causality}, invariant prediction \cite{arjovsky2019invariant}, and disentangled causal mechanism \cite{suter2019robustly,bengio2019meta}.
\subsubsection{Uncertainty Quantification}
AI research continues to develop new state-of-the-art algorithms with superior performance and large-scaled datasets with high quality. Even using the best models and training data, it is still infeasible for AI systems to cover all the potential situations when deployed into real-world applications. As a matter of fact, AI systems always encounter new samples that are different from those used for training. The core question is how to leverage the strengths of these uncertainties. Recent research, e.g., \cite{bhatt2020uncertainty}, has advocated to measure, communicate, and use uncertainty as a form of transparency. There are also tools such as IBM's Uncertainty Quantification 360\footnote{http://uq360.mybluemix.net/overview} to provide AI practitioners access to related resources as common practices for AI transparency. Consequently, uncertainty quantification plays a crucial role in the optimization and decision-making process in SRAs.  
There are typically two kinds of uncertainties in risk analysis processes: first, the aleatory uncertainty describes the inherent randomness of systems. For example, an AI system can present different results even with the same set of inputs. The uncertainty arises from underlying random variations within the data. Second, the epistemic uncertainty represents the effect of an unknown phenomenon or an internal parameter. The primary reason leading to this type of uncertainty is the lack of observed data. As the variation among the data in aleatory uncertainty is often observable, we can well quantify the uncertainty and assess the risks. Quantification of epistemic uncertainty is more challenging because AI systems are forced to extrapolate over unseen situations \cite{staahl2020evaluation}. In the literature of uncertainty quantification, one of the most widely recognized techniques are prediction intervals (PI). For neural-network-based models, PI can be categorized into multi-step PI construction methods (e.g., Bayesian method) and direct PI construction methods (e.g., lower upper bound estimation). Here, we briefly discuss several methods in each category. Please refer to the survey \cite{kabir2018neural} for more details. 

\noindent\textbf{Multi-Step Prediction Intervals Construction Methods.} Delta method, Bayesian method, Mean-Variance Estimation method, and Bootstrap method are the four conventional multi-step methods reported in literature. Delta method constructs PIs through nonlinear regression using Tylor series expansion. Particularly, we linearize neural network models through optimization by minimizing the error-based loss function, sum square error. Under the assumption that uncertainty is from normal and homogeneous distribution, we then employ standard asymptotic theory to construct PIs. Delta method has been used in numerous case studies, e.g., \cite{lu2009prediction,ho2001neural}. Bayesian learning provides a natural framework for constructing PIs \cite{kasiviswanathan2016comparison,ungar1996estimating} as it optimizes the posterior distribution of parameters from the assumed prior distribution. Despite its high generalization power, Bayesian techniques are limited by large computational complexity due to the calculation of Hessian matrix. Bootstrap method is the most popular among the four conventional multi-step PI construction methods \cite{errouissi2015bootstrap,zio2006study,dybowski2001confidence}. It includes smooth, parametric, wild, pairs, residual, Gaussian process, and other types of bootstrap techniques. In NN-based pairs bootstrap algorithm, for example, the key is to generate bootstrapped pairs by uniform sampling with replacement from the original training data. The estimation is then conducted for a single bootstrapped dataset \cite{kabir2018neural}.

\begin{figure}
\center
  \includegraphics[width=.6\columnwidth]{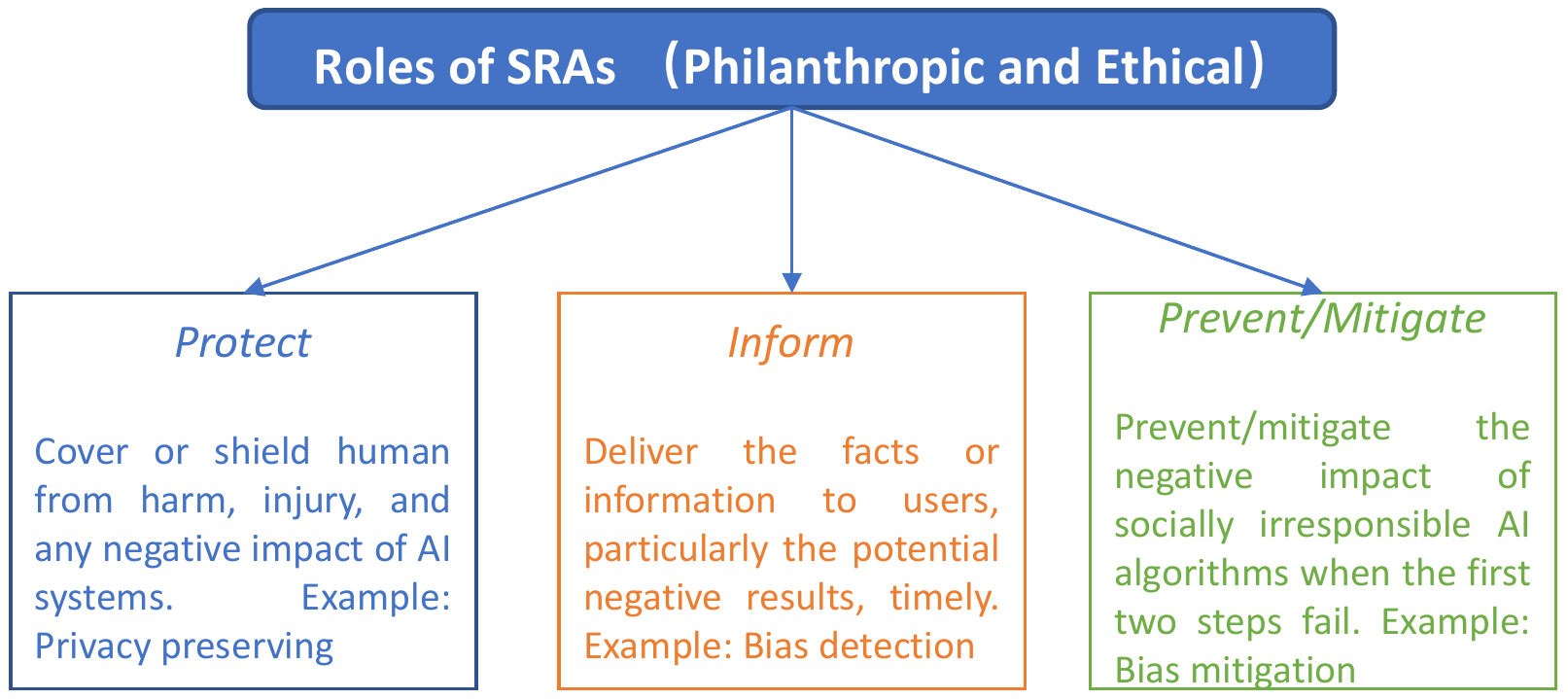}
  \caption{Illustration of what Socially Responsible AI Algorithms (SRAs) can do. It requires philanthropic responsibilities and ethical responsibilities.}
  \label{cando}
\end{figure}
\noindent\textbf{Direct Prediction Intervals Construction Methods.} This category of methods can tackle some of the limitations in previous methods, such as high demanding in computational power and stringent assumptions. When NN models are constructed through direct training without any assumptions, they can provide more adaptive and smarter PIs for any distribution of targets \cite{chu2015real}. Lower Upper Bound estimation method is such a technique that can be applied to arbitrary distribution of targets with more than one order reduced computation time. It directly calculates the lower and the upper bounds through trained NNs. Initially, Lower Upper Bound estimation NNs are optimized with the coverage width-based criterion, which presents several limitations. With all the benefits of the original Lower Upper Bound estimation method, the NN-based Direct Interval Forecasting method \cite{wan2013probabilistic} has much shorter computation time and narrower PIs credited to the improved cost function and the reduced average coverage error. Other approaches for improving the cost function of Lower Upper Bound estimation include the normalized root-mean-square width and particle swarm optimization \cite{quan2014particle}, optimal system by \cite{hosen2014improving}, the independent width and penalty factors \cite{khosravi2014constructing}, the deviation from mid-interval consideration \cite{marin2016prediction}, and the deviation information-based criterion \cite{zhang2014advanced}.  
\section{Roles of SRAs}
So far, we have introduced the essentials of SRAs to achieve the expected ethical responsibilities. But pragmatic questions regarding their intended use remain: How to operationalize SRAs? What can SRAs eventually do for societal well-being to address societal challenges? Both ethical and philanthropic responsibilities are indispensable ingredients of the answers. While the ultimate goal of SRAs is to \textit{do good} and \textit{be a good AI citizen}, their ethical responsibilities should be ensured first. When AI fails to fulfill its ethical responsibilities, its philanthropic benefits can be insignificant. For instance, despite the immense public good of COVID-19 vaccines, there has been great controversy about algorithms for their distribution, which have been shown to be inequitable \cite{covid2021}. Some argue that distribution algorithms should prioritize saving more lives and bringing the economy back more rapidly \cite{Vaccine_2020}; they support such an `unfair' allocation, but we would argue that that is not unfairness, but simply a difference of values and ethics. In our view, roles of SRAs are expected to encompass both ethical and philanthropic responsibilities. In this survey, we describe three dimensions that SRAs can help with to improve the quality of human life as illustrated in Figure \ref{cando}: Protect (e.g., protect users' personal information), Inform (e.g., fake news early detection), and Prevent/Mitigate (e.g., cyberbullying mitigation). We illustrate each dimension with research findings in several emerging societal issues. Particularly, for \textit{protecting} dimension, we focus on privacy preserving and data dignity; for \textit{informing} and \textit{preventing/mitigating} dimensions, we discuss three societal issues that raise growing concerns recently: disinformation, abusive language, and unwanted bias. Because there are many various forms of abusive language such as hate speech and profanity, and the body of work related to each form is vast and diverse, spanning multiple interconnected disciplines, this survey uses the form of cyberbullying as a representative for the illustrations. 
\subsection{Protecting}
The protecting dimension aims to cover or shield humans (especially the most vulnerable or at-risk) from harm, injury, and negative impact of AI systems, in order to intervene. This can be the protection of users' personal data and their interactions with AI systems. Two typical examples are privacy preserving and data dignity.
\subsubsection{Privacy-Preserving}
The capability of deep learning models has been greatly improved by the emerging powerful infrastructures such as clouds and collaborative learning for model training. The fuel of this power, however, comes from data, particularly sensitive data. This has raised growing privacy concerns such as illegitimate use of private data and the disclosure of sensitive data \cite{boulemtafes2020review,beigi2020survey}. Existing threats against privacy are typically from attacks such as the adversarial examples we discussed in Sec. \ref{adversary}. Specifically, there are direct information exposure (e.g., untrusted clouds), which is caused by direct intentional or unintentional data breaches, and indirect (inferred) information exposure (e.g., parameter inference), which is caused by direct access to the model or output. Existing privacy-preserving mechanisms can be classified into three categories, namely, private data aggregation methods, private training, and private inference \cite{mirshghallah2020privacy}. 

Data aggregation methods are either context-free or context-aware. A context-free approach such as differential privacy \cite{dwork2008differential}, is unaware of the context or what the data will be used for. Context-aware approach such as information-theoretic privacy \cite{schaefer2017information}, on the other hand, is aware of the context in which the data will be used. A na\"ive technique for privacy protection is to remove identifiers from data, such as name, address, and zip code. It has been used for protecting patients' information while processing their medical records, but the results are unsatisfying \cite{sweeney2002k,narayanan2008robust,homer2008resolving}. The k-Anonymity method can prevent information from re-identification by showing at least $k$ samples with exact same set of attributes for given combination of attributes that the adversary has access to \cite{sweeney2002k}. The most commonly used data aggregation method is differential privacy, which aims to estimate the effect of removing an individual from the dataset and keep the effect of the inclusion of one's data small. Some notable work includes the Laplace mechanism \cite{dwork2006calibrating}, differential privacy with Advanced Composition \cite{dwork2014algorithmic}, and local differential privacy \cite{kairouz2014extremal,erlingsson2014rappor}.

Information-theoretic privacy is a context-aware approach that explicitly models the dataset statistics. By contrast, context-free methods assume worse-case dataset statistics and adversaries. This line of research was studied by \citeA{diaz2019robustness}, \citeA{pinceti2019data}, and \citeA{varodayan2011smart}. The second type of privacy-preserving mechanism works during the training phase. Established work in private training is mostly used to guarantee differential privacy or semantic security and encryption \cite{goldwasser1984probabilistic}. The two most common methods for encryption are homomorphic encryption \cite{gentry2009fully} and secure multi-party computation \cite{makri2019epic}. The third type of privacy-preserving mechanism works during the inference phase. It aims at the trained systems that are deployed to offer inference-as-a-service \cite{mirshghallah2020privacy}. Most methods in private inference are similar to those in private training, except for the information-theoretic privacy \cite{malekzadeh2019mobile,malekzadeh2018protecting,malekzadeh2020privacy}. It is typically used to offer information-theoretic mathematical or empirical evidence of how these methods operate to improve privacy. There is also work using differential privacy \cite{wang2018not}, homomorphic encryption \cite{gilad2016cryptonets,chabanne2017privacy}, and secure multi-party computation \cite{liu2017oblivious}. 
\subsubsection{Data Dignity}
Beyond privacy preserving, what is more urgent to accomplish is data dignity. It allows users to have absolute control to how their data is being used and they are paid accordingly \cite{getoor2019responsible}. Data dignity encompasses the following aspects \cite{Safer2020}:
\begin{itemize}
    \item To help users objectively determine the benefits and risks associated with their digital presence and personal data.
    \item To let users control how their data will be used and the purpose of using the data.
    \item To allow users to negotiate the terms of using their data.
    \item To give users complete right and autonomy to be found, analyzed, or forgotten, apart from the fundamental right over their data. 
\end{itemize}
There are business models such as the Microsoft Data Bank designed to give users the control of their data and those shared by the Art of Research \cite{Hart2019} about how people can buy and sell their personal data. 
\subsection{Informing}
The informing dimension aims to deliver the facts or information to users, particularly the potential negative results, in a timely way. We illustrate it with a focus on the discussions of detecting disinformation, cyberbullying, and bias.  
\subsubsection{Disinformation Detection}
Disinformation is false information that is deliberately created and spread to deceive people, a social group, organization, or country \cite{pacepa2013disinformation}. The online information ecosystem is never short of disinformation and misinformation, and the growing concerns have been raised recently. Tackling disinformation is rather challenging mainly because (1) disinformation exists almost in all domains; (2) it is ever-changing with new problems, challenges, and threats emerging every day; (3) it entails the joint efforts 
of interdisciplinary research -- computer science, social science, politics, policy making, and psychology, cognitive science \cite{bhattacharjee2020disinformation}. Accurate and efficient identification of disinformation is the core to combat disinformation. Existing prominent approaches for disinformation detection primarily rely on news content, social context, user comments, fact-checking tools, and explainable and cross-domain detection. 

Early work on disinformation detection has been focused on hand-crafted features extracted from text, such as lexical and syntactic features \cite{feng2012syntactic,ott2011finding}. Apart from text, online platforms also provide abundant social information that can be leveraged to enrich the textual features, e.g., number of re-tweets and likes on Twitter. Informed by theories in social science and network science, another line of work exploits social network information to improve the detection performance. Common features are social context \cite{shu2019beyond}, user profile \cite{shu2018understanding}, user engagement \cite{shu2020fakenewsnet}, and relationships among news articles, readers, and publishers \cite{della2018automatic}. A unique function of online platforms is that they allow users to interact through comments. Recent work has shown that user comments can provide weak supervision signal for identifying the authenticity of news articles, which enables early detection of disinformation \cite{shu2020leveraging}. When the user comments are unavailable, it is possible to learn users' response to news articles and then generate user responses \cite{qian2018neural}. Fact-checking can be achieved manually or automatically. Manual fact-checking relies on domain experts or crowdsourced knowledge from users. Automatic fact-checking uses structure knowledge bases such as knowledge graph to verify the authenticity of news articles, see, e.g., \cite{ciampaglia2015computational}. Beyond within-domain detection, other tasks such as cross-domain detection \cite{janicka2019cross}, explanation \cite{shu2019defend}, and causal understanding of fake news dissemination \cite{cheng2021causal} have also been discussed in literature. 
\subsubsection{Cyberbullying Detection}
Cyberbullying differs from other forms of abusive language in that it is not an one-off incident but ``aggressively intentional acts carried out by a group or an individual using electronic forms of contact, \textit{repeatedly} or \textit{over time} against victims who cannot easily defend themselves''~\cite{smith2008cyberbullying}. The increasingly reported number of cyberbullying cases on social media and the resulting detrimental impact have raised great concerns in society. Cyberbullying detection is regularly figured as a binary classification problem. While it shares some similarities with document classification, it should be noted that cyberbullying identification is inherently more complicated than simply identifying oppressive content \cite{salawu2017approaches}. 

Distinct characteristics of cyberbullying such as \textit{power imbalance} and \textit{repetition} of aggressive acts are central to marking a message or a social media session \cite{cheng2020session} as cyberbullying. Several major challenges in cyberbullying detection have been discussed in literature such as the formulation of the unique bullying characteristics, e.g., repetition, data annotation, and severe class imbalance. Depending on the employed features, established work can be classified into four categories: content-based, sentiment-based, user-based, and network-based methods. Features extracted from social media content are lexical items such as keywords, Bag of Words, pronoun and punctuation. Empirical evaluations have shown that textual features are the most informative predictors for cyberbullying detection \cite{cheng2020unsupervised}. For instance, using number of offensive terms as content features is effective in detecting offensive and cursing behavior \cite{dinakar2012common,dadvar2013improving,kontostathis2013detecting}; Computing content similarity between tweets from different users can help capture users' personality traits and peer influence, two important factors of cyberbullying occurrences \cite{cheng2019pi}. Sentiment-based features typically include key-words, phrases and emojis, and they are often combined with content-based features \cite{dani2017sentiment}. A notable work \cite{xu2012learning} identified seven types of emotions in tweets such as anger, empathy, and fear. User-based features are typical characteristics of users, e.g., personality (e.g., hostility), demographics (e.g., age), and user activity (e.g., active users \cite{balakrishnan2015cyberbullying}). Hostility and neuroticism are found to be strongly related to cyberbullying behavior \cite{biel2011you,mishna2012risk}. Further, gender and age are indicative of cyberbullying in certain cases \cite{al2016cybercrime}. Network-based features measure the sociability of online users, e.g., number of friends, followers, and network embeddedness \cite{cheng2019xbully,huang2014cyber}. In addition, a number of methods seek to capture the temporal dynamics to characterize the repetition of cyberbullying, such as \cite{cheng2019hierarchical,chen2020henin,ge2021improving,cheng2021modeling}. 
\subsubsection{Bias Detection}
Compared to the well-defined notions of fairness, bias detection is much less studied and the solution is not as straightforward as it may seem \cite{fu2020ai}. The challenges arise from various perspectives. First, the data and algorithms used to make a decision are often not available to policy makers or enforcement agents. Second, algorithms are becoming increasingly complex and the uninterpretability limits an investigator's ability to identify systematic discrimination through analysis of algorithms. Rather, they have to examine the output from algorithms to check for anomalous results, increasing the difficulty and uncertainty of the task. 

Data exploratory analysis is a simple but effective tool to detect data bias. In this initial step of data analysis, we can use basic data statistics and visual exploration to understand what is in a dataset and the characteristics of the data. For algorithmic bias, one of the earliest methods is to compare the selection rate of different groups. Discrimination is highly possible if the selection rate for one group is sufficiently lower than that for other groups. For example, the US Equal Employment Opportunity Commission (EEOC) advocates the ``four-fifths rule'' or ``80\% rule'' \cite{feldman2015certifying} to identify a disparate impact. Suppose $Y$ denotes a binary class (e.g., hire or not), $A$ is the protected attribute (e.g., gender), a dataset presents disparate impact if 
\begin{equation}
 \frac{Pr(Y=1|A=0)}{Pr(Y=1|A=1)} \leq \tau =0.8.
\end{equation}
However, statistical disparity does not necessarily indicate discrimination. If one group has disproportionately more qualified members, we may expect the differences between groups in the results. 

A more frequently used approach is regression analysis \cite{ayres2010testing}, which is performed to examine the likelihood of favorable (or adverse) decisions across groups based on sensitive attributes. A significant, non-zero coefficient of the sensitive attributes given a correctly specified regression signals the presence of discrimination. However, we cannot guarantee to observe all the factors the decision maker considers. Therefore, instead of using rate at which decisions are made (e.g., the loan approval rates), bias detection can be based on the success rate of the decisions (e.g., the payback rate of the approved applicants \cite{becker1993nobel}), i.e., the \textit{outcome test}. Another less popular statistical approach for bias detection is benchmarking. The major challenge of benchmarking analysis is identifying the distribution of the sensitive attributes of the benchmark population where sensitive attributes are unlikely to influence the identification of being at-risk. Some solutions can be seen in \cite{mcconnell2001race,lange2001speed}. Recently, AI researchers have developed tools to automatically detect bias. For instance, drawing on techniques in natural language processing and moral foundation theories, the tool by \citeA{mokhberian2020moral} can understand structure and nuances of content consistently showing up on left-leaning and right-leaning news sites, aiming to help consumers better prepare for unfamiliar news source. In earlier efforts, an international research group launched a non-profit organization Project Implicit\footnote{https://implicit.harvard.edu/implicit} in 1998 aimed at detecting implicit social bias.      
\subsection{Preventing/Mitigating}
If both of the first two dimensions fail, we may rely on the last dimension to prevent/mitigate the negative impact of socially indifferent AI algorithms on the end-users. We continue the discussions about disinformation, cyberbullying, and bias, with a focus on the prevention and mitigation strategies.
\subsubsection{Disinformation Prevention/Mitigation}
Preventing the generation/spread of disinformation and mitigating its negative impact is an urgent task because disinformation typically spread faster \cite{vosoughi2018spread} than normal information due to the catchy news content and the ranking algorithms operating behind the online news platforms. To increase user engagement, social recommender systems are designed to recommend popular posts and trending content. Therefore, disinformation often gains more visibility. An effective approach for disinformation mitigation is to govern this visibility of news, e.g., recommendation and ranking based algorithms. Mitigation also relates to early detection. 

Network intervention can slow down the spread of disinformation by influencing the exposed users in a social network. For example, we can launch a counter-cascade that consists of fact-checked version of false news articles. This is commonly referred to as the influence limitation or minimization problem \cite{bhattacharjee2020disinformation}. Given a network with accessible counter-cascade, the goal is to find a (minimum) set of nodes in this network such that the effect of the original cascade can be minimized. A variety of approximation algorithms \cite{budak2011limiting,nguyen2012containment} have been proposed to solve the NP-hard problem and the variants. When applied to disinformation mitigation, they seek to inoculate as many nodes as possible in a short period of time. It is possible to extend the two cascades into tasks with multiple cascades, where we can further consider the different priorities of these cascades, i.e., each cascade influences the node in the network differently \cite{tong2018misinformation}. The second method for disinformation mitigation is content flagging: social media platforms allow users to `flag' or `report' a news content if they find it offensive, harmful, and/or false. Big social media companies such as Facebook hired professional moderators to manually investigate and/or remove these content. However, considering the millions of news generated/spread every minute \cite{Marr_2021}, it is impractical for these moderators to manually review all the news. The solution comes to the crowd wisdom -- users can choose to `flag' the content if it violates the community guidelines of the platform. Some platforms can further provide feedback for these users about if their fact-check is correct or not. User behavior is an effective predictor for disinformation detection \cite{cheng2021causal}, therefore, the third prevention method leverages the differences between user behaviors to identify susceptible or gullible users. For example, it is shown in \cite{rajabi2019user} that groups of vulnerable Twitter users can be identified in fake news consumption. Other studies \cite{guess2019less} also suggest that older people are more likely to spread disinformation.
\subsubsection{Cyberbullying Prevention/Mitigation}
In contrast to the large amount of work in cyberbullying detection, efforts for its prevention and mitigation have been a few. Some research suggests that prevention/mitigation strategies are defined at different levels \cite{smith2008cyberbullying}. At technological level, we can consider providing parental control service, firewall blocking service, online services rules, text-message control, and mobile parental control, e.g., KnowBullying and BullyBlocker \cite{silva2016bullyblocker}. Another effective tool is psychological approach, such as talking and listening to cyber-victims, providing counseling services, encouraging victims to make new relations and join social clubs. At education level, we are responsible to educate end-users, help improve their technical and cognitive skills. At administrative level, it is important for organizations and government to develop policies to regulate using free service and enhance workplace environment. Therefore, the goal of cyberbullying prevention/mitigation can only be accomplished with interdisciplinary collaborations, e.g., psychology, public health, computer science, and other behavioral and social sciences \cite{kraft2009effectiveness}. One example is that computer and social scientists attempted to understand behavior of users in realistic environments by designing social media site for experimentation such as controlled study and post-study survey \cite{difranzo2018upstanding,ashktorab2017designing}. 

Existing solutions to preventing cyberbullying can report/control/warn about message content (e.g., \citeR{dinakar2012common,vishwamitra2017mcdefender}), provide support for victims (e.g., \citeR{vishwamitra2017mcdefender}), and educate both victims and bullies (e.g., \citeR{dinakar2012common}). A variety of anti-bully apps are also available to promote well-being of users. For example, NoMoreBullyingMe App provides online meditation techniques to support victims; ``Honestly'' App \cite{Br_2015} encourages users to share positive responses with each other (e.g., sing a song). However, current cyberbullying prevention strategies often do not work as desired because of the complexity and nuance with which adolescents bully others online \cite{ashktorab2016designing}.  
\subsubsection{Bias Mitigation}
Prior approaches for bias mitigation focus on either designing fair machine learning algorithms or theorizing on the social and ethical aspects of machine learning discrimination \cite{caton2020fairness}. From the technical aspect, approaches to fairness typically can be categorized into pre-processing (prior to modelling), in-processing (at the point of modelling), and post-processing (after modelling). One condition to use pre-processing approaches is that the algorithm is allowed to modify the training data \cite{bellamy2019ai}. We can then transform the data to remove the discrimination \cite{d2017conscientious}. In-processing approaches eliminate bias by modifying algorithms during the training process \cite{d2017conscientious}. We can either incorporate fairness notion into the objective function or impose fairness constraint \cite{bellamy2019ai,berk2017convex}. When neither training data nor model can be modified, we can use post-processing approaches to reassign the predicted labels based on a defined function and a holdout set which was not used in the model training phase \cite{bellamy2019ai,berk2017convex}. Most of these approaches are built on the notion of protected or sensitive variables that define the (un)privileged groups. Commonly used protected variables are age, gender, marital status, race, and disabilities. A shared characteristic of these groups is they are disproportionately (less) more likely to be positively classified. Fairness measures are important to quantify fairness in the development of fairness approaches. However, creating generalized notions of fairness quantification is a challenging task \cite{caton2020fairness}. Depending on the protected target, fairness metrics are usually designed for individual fairness (e.g., every one is treated equally), group fairness (e.g., different groups such as women vs men are treated equally), or subgroup fairness. Drawing on theories in causal inference, individual fairness also includes counterfactual fairness which describes that a decision is fair towards an individual if the result was same when s/he had taken a different sensitive attribute \cite{kusner2017counterfactual}.

Recent years have witnessed immense progress of fair machine learning -- a variety of methods have been proposed to address bias and discrimination over different applications. We focus on two mainstream methods: fair classification and fair regression. A review of machine learning fairness can be referred to \cite{mehrabi2019survey,caton2020fairness}. 

\noindent \textit{(1) Fair Classification.} For a (binary) classifier with sensitive variable $S$, the target variable $Y$, and the classification score $R$, general fairness desiderata have three ``non-discrimination'' criteria: \textit{Independence}, i.e., $R\indep S$; \textit{Separation}, i.e., $R\indep S|Y$; and \textit{Sufficiency}, i.e., $Y\indep S | R$. Fair machine learning algorithms need to adopt/create specific fairness definitions that fit into context \cite{goel2018non,kamishima2012fairness,krasanakis2018adaptive,menon2018cost,calders2010three}. Common methods in fair classification include blinding \cite{calmon2017optimized,hardt2016equality}, causal methods \cite{galhotra2017fairness,johnson2016impartial}, transformation \cite{feldman2015certifying,dwork2012fairness,wei2020optimized}, sampling and subgroup analysis \cite{celis2016fair,dwork2018decoupled}, adversarial learning \cite{feng2019learning,xu2019achieving,sattigeri2019fairness}, reweighing \cite{jiang2020identifying,calders2010three}, and regularization and constraint optimization \cite{kamishima2012fairness,bechavod2017penalizing,cheng2021mitigating}. 

\noindent\textit{(2) Fair Regression.} The goal of fair regression is to jointly minimize the difference between true and predicted values and ensure fairness. It follows the general formulation of fair classification but with continuous rather than binary/categorical target variable. Accordingly, the fairness definition, metrics, and the basic algorithms are adapted from classification to regression. For example, it is suggested using statistical parity and bounded-group-loss metrics to measure fairness in regression \cite{agarwal2019fair}. Bias in linear regression is considered as the effects of a sensitive attribute on the target variable through the mean difference between groups and AUC metrics \cite{calders2013controlling}. One commonly used approach in fair regression is regularization, e.g., \cite{kamishima2012fairness,berk2017convex}.  

Apart from fair machine learning, algorithm operators are encouraged to share enough details about how research is carried out to allow others to replicate it. This is a leap for mitigating bias as it helps end-users with different technical background to understand how the algorithm works before making any decision. It is also suggested that AI technologists and researchers develop a bias impact statement as a self-regulatory practice. It can help probe and avert any potential biases that are injected into or resultant from algorithmic decision \cite{lee2019algorithmic}. Some example questions in the statement are ``What will the automated decision do?'', ``How will potential bias be detected?'', and ``What are the operator incentives''. In algorithm design, researchers are also responsible to encourage the role of diversity within the team, training data, and the level of cultural sensitivity. The ``diversity-in-design'' mechanism aims to take deliberate and transparent actions to address the upfront cultural biases and stereotypes. Furthermore, we might also consider updating nondiscrimination and other civil rights laws to interpret and redress online disparate impacts \cite{lee2019algorithmic}. An example of such consideration is to unambiguously define the thresholds and parameters for the disparate treatments of protected groups before the algorithm design. 
\section{Open Problems and Challenges}
\begin{figure}
\center
  \includegraphics[width=.7\columnwidth]{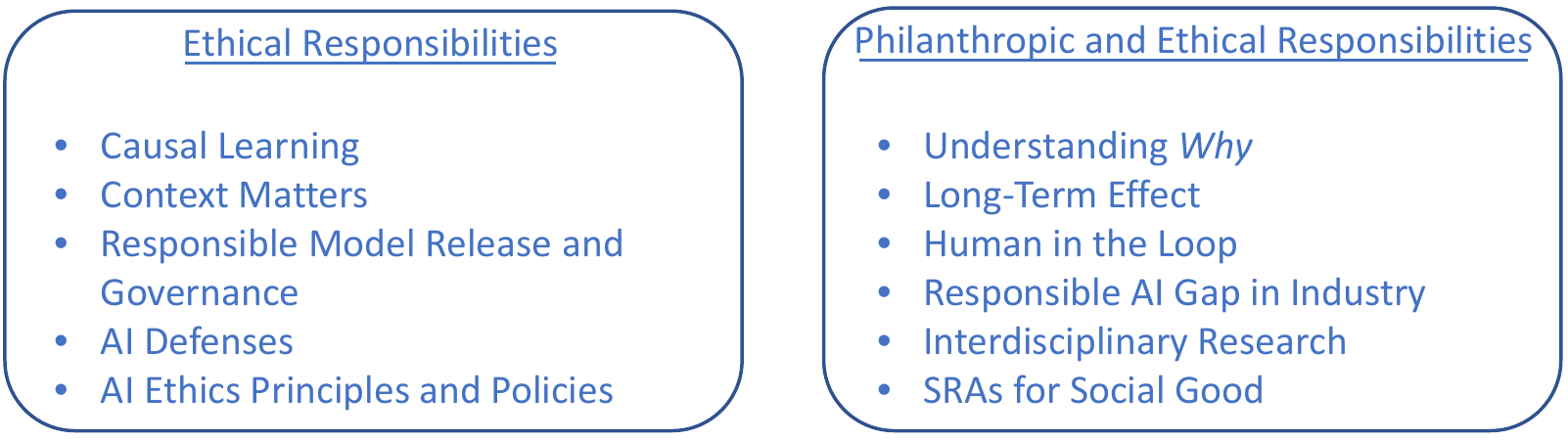}
  \caption{Primary challenges and open problems we confront in developing SRAs. Some challenges relate to SRAs' internal mechanisms that fulfill AI's ethical responsibilities whilst others relate to SRAs' roles to which both ethical and philanthropic responsibilities are the keys.}
  \label{challenges}
\end{figure}

This survey reveals that the current understanding of SRAs is insufficient and future efforts are in great need. Here, we describe several primary challenges, as summarized in Figure \ref{challenges}, in an attempt to broaden the discussions on future directions and potential solutions. 

\noindent\textbf{Causal Learning.} The correlation fallacy causes AI algorithms to meet with fundamental obstacles in order to commit social responsibility. These obstacles are robustness, explainability, and cause-effect connections \cite{pearl2019seven}. The era of big data has changed the ways of learning causality, and meanwhile, causal learning becomes an indispensable ingredient for AI systems to achieve human-level intelligence. There are a number of benefits to incorporate causality in the next-generation of AI. For example, teaching AI algorithms to understand ``why'' can help them transfer their knowledge to different but similar domains. Early efforts in SRAs attempted to employ causal learning concept and methods such as intervention, counterfactual, \textit{do}-calculus, propensity scoring to address fairness (e.g., counterfactual fairness) and interpretability (causal interpretability) issues. They have shown prominent results in these tasks.

\noindent\textbf{Context Matters.} Context is the core to SRAs due to its inherently elaborate nature, e.g., the ``Transparency Paradox''. Understanding and quantifying the relationships among the various principles (some are tradeoffs and some are not), e.g., fairness, transparency, and safety, have to be placed in specific context. One such context is the \textit{social context}. Existing SRAs (e.g., fair machine learning), once introduced into a new social context, may render current technical interventions ineffective, inaccurate, and even dangerously misguided \cite{selbst2019fairness}. A recent study \cite{suhr2020does} found that while fair ranking algorithms such as Det-Greedy \cite{geyik2019fairness} help increase the exposure of minority candidates, their effectiveness is limited by the job contexts in which employers have a preference to particular genders. How to properly integrate social context into SRAs is still an open problem. \textit{Algorithmic context} (e.g., supervised learning, unsupervised learning, and reinforcement learning) is also extremely important when designing SRAs for the given data. A typical example is the feedback loop problem in predictive policing \cite{ensign2018runaway}. A subtle algorithmic choice can have huge ramifications on the results. Consequently, we need to understand the algorithmic context to make the right algorithmic choices when designing socially responsible AI systems. Designing context-aware SRAs is the key to achieving Social Responsibility of AI.

\noindent\textbf{Responsible Model Release and Governance.} Nontransparent model reporting is one of the main causes of AI indifferent behaviors. As a critical step to clarify the intended use cases of AI systems and the contexts for which they are well suited, responsible model release and governance has been receiving growing attentions from both industry and academia. One role of SRAs is to bring together the tools, solutions, practices, and people to govern the built AI systems across its life cycle \cite{IBM_2020}. At this early stage, some research results suggested that released models be accompanied by documentation detailing various characteristics of the systems, e.g., what it does, how it works, and why it matters. For example, the AI FactSheets \cite{arnold2019factsheets} advocates to use a factsheet completed and voluntarily released by AI developers to increase the transparency of their services. A similar concept is model cards \cite{mitchell2019model}, short documents that provide benchmarked evaluation for the trained AI models in a variety of conditions, e.g., different cultural or demographic groups. Typically, a model card should include the model details, intended use, evaluation metrics, training/evaluation data, ethical considerations, and caveats and recommendations. To help increase transparency, manage risk, and build trust in AI, AI technologists and researchers are responsible to address various challenges faced in creating useful AI release documentation \cite{hind2020experiences} and develop effective AI governance tools.

\noindent\textbf{AI Defenses.} Developing AI systems that outwit malicious AI is still at an early stage \cite{Knight2020}. Since we have not fully understood how AI systems work, they are not only vulnerable to attack but also likely to fail in surprising ways \cite{yuan2019adversarial,chakraborty2018adversarial}. As a result, it is critical and urgent to work on designing systems that are provably robust to help ensure that the AI systems are not vulnerable to adversaries. At least two capabilities an ``AI firewall'' needs to be equipped with: one capability is to probe an AI algorithm for weaknesses (e.g., perturb the input of an AI system to make it misbehave) and the other one is to automatically intercept potentially problematic inputs. Some big tech companies have started building their own AI defenses to identify the weak spots, e.g., the ``red team'' in Facebook, the software framework released by Microsoft, Nvidia, IBM, and 9 other companies. AI defenses reflect the fundamental weakness in modern AI and make AI systems more robust and intelligent.

\noindent\textbf{AI Ethics Principles and Policies.} Current AI principles and policies for ethical practice have at least two common criticisms: (1) they are too vaguely formulated to prove to be helpful in guiding practice; and (2) they are defined primarily by AI researchers and powerful people with mainstream populations in mind \cite{young2019toward}. For the first criticism, to help operationalize AI principles in practice and organizations confront inevitable value trade-offs, it has been suggested to redefine AI principles based on philosophical theories in applied ethics \cite{canca2020operationalizing}. Particularly, it categorizes published AI principles (e.g., fairness, accountability, and transparency) into three widely used \textit{core principles} in applied ethics: autonomy, beneficence (avoiding harm and doing good), and justice. The core principles ``invoke those values that theories in moral and political philosophy argue to be intrinsically valuable, meaning their value is not derived from something else'' \cite{canca2020operationalizing}. Existing AI principles are \textit{instrumental principles} that ``build on concepts whose values are derived from their instrumental effect in protecting and promoting intrinsic values'' \cite{canca2020operationalizing}. Operationazable AI principles help effectively put ethical AI in practice and reduce the responsible AI Gap in companies. To address the second criticism, we need to best elicit the inputs and values of diverse voices from the Subjects of SRAs, i.e., the minority and disadvantaged groups, and incorporate their perspectives into the tech policy document design process. If we align values of AI systems through a panel of people (who are compensated for doing this), they too can influence the system behavior, and not just the powerful people or AI researchers.

\noindent\textbf{Understanding \textit{Why}.} Many AI systems are designed and developed without fully understanding \textit{why}: What do we wish the AI system do? This is often the reason that these systems fail to represent the goals of the real tasks, a primary source of AI risks. The problem can become more challenging when the AI system is animated through a number of lines of code that lack nuance, creating a machine that does not align with our true intentions. As the first step, understanding \textit{why} clearly defines our social expectation of AI systems and paves way for more specific questions such as ``What is the problem? Who will define it? and what are the right people to include?''. Answering \textit{why} helps us effectively abolish the development of socially indifferent AI systems in the first place and also helps understand the kinds of deception an AI system may learn by itself.

\noindent\textbf{Long-Term Effect.} SRAs include social concepts such as fairness that can evolve over time along with the constant changes of human values and social dynamics. This raises the concerns about the commitment SRAs need to fulfill in the long term. For example, despite the various types of fairness definitions, once introduced into the dimension of time, the number of fairness definitions may be explosive. In addition, current fairness criteria may be considered as unfair in the future. Fairness criteria are essentially designed to promote long-term well-being. However, even a static fairness notion can fail to protect the target groups when there is a feedback loop in the overall system \cite{liu2018delayed}. How to build AI systems that can commit long-term responsibility is extremely challenging and rarely studied thus far. Initial results of long-term fairness \cite{liu2018delayed,hu2018short} highlight the importance of measurement and temporal modeling in the evaluation of fairness criteria. 

\noindent\textbf{Humans in the Loop.} While existing techniques in SRAs have indeed made significant progress towards responsible AI systems, their usefulness can be limited in some settings where the decisions made are actually poorer for every individual. For issues of fairness in prediction, for example, many findings (e.g., \citeR{pfohl2020empirical}) have shown the concerns about the fairness-performance trade-off: the imposition of fairness comes at a cost to model performance. Predictions are less reliable and moreover, different notions of fairness can make approaches to fairness conflict with one another. Having human in the loop matters when it comes to contextualizing the objectives of SRAs, especially for high-stake decisions. For instance, there are situations where the cut-off values of fairness for two subgroups are different, and humans can help calibrate the differences.    

\noindent\textbf{Responsible AI Gap in Industry.} The far-reaching effect of reputational damage and employee disengagement result from AI misbehavior has forced company executives to begin understanding the risks of poorly designed AI systems and the importance of SRAs. While seeing many potential benefits of developing responsible AI systems such as increasing market share and long-term profitability, companies lack the knowledge of how to cross the ``Responsible AI Gap'' between principles and tangible actions \cite{six2020}. This is partly because companies view responsible AI solely as risk-avoidance mechanism and overlook its financial rewards. To capture the benefits of responsible AI in companies' day-to-day business, companies need to go far beyond SRAs and examine every aspect of the end-to-end AI systems. A recent article \cite{six2020} suggested six basic steps to bridge the gulf between responsible AI and the reality: \textit{Empower responsible AI leadership, Develop principles, policies, and training, Establish human and AI governance, Conduct Responsible AI reviews, Integrate tools and methods, and Build and test a response plan}. Even though the gap might be huge, small efforts built over time can let SRAs achieve a transformational impact on the businesses.

\noindent\textbf{Interdisciplinary Research.} Current public dialog on SRAs has been focused on a narrow subset of fields, blinding us to the opportunities presented by interdisciplinary research. It is necessary to work with researchers from different disciplines whose contributions are sorely needed, e.g., psychologist, social scientist, educators, and humanities. Non-profit organizations are both the beneficiaries and benefactors of SRAs. In partnering with non-profits and social enterprises will not only unleash AI's potential for benefiting societal well-being, but also let AI technologists and researchers have the opportunity to encounter the real problems we are currently facing. A better understanding of what problems need to be solved helps identify SRAs that need to be created. Moreover, as big tech companies bankroll more work of academic researchers, much of ethics-based research gets concentrated in the hands of a few companies that can afford it \cite{Rivero2020}. This is problematic because we are over reliant on the same companies that are producing socially indifferent AI systems. We need interdisciplinary and decentralized research to create SRAs and simultaneously achieve the four levels in the pyramid of Social Responsibility of AI. 

\noindent\textbf{SRAs for Social Good.} The last challenge regards the intended use of SRAs. When SRAs are leveraged to uplift humanity, a trust in AI is further enhanced. There has been a burgeoning of AI-for-social-good movement that produces AI algorithms to help reduce poverty, hunger, inequality, injustice, climate change, ill health, and other causes of human suffering \cite{varshney2019trustworthy}. Compared to deploying cutting-edge AI systems to solve these critical issues, a more urgent question to examine is ``What makes an AI project good'' in order to prevent the detrimental consequences of AI. In addition to Protecting, Informing, and Preventing, social good applications also relate closely to \textit{Fundraise} and \textit{Greenlight} \cite{social2020}. Applying SRAs to target solicitations for donations largely helps with fundraising for non-profits, charitable organizations, and universities. Greenlight describes how SRAs can help allocate grants and other types of resources by predicting the success rates of project proposals. It plays an important role in improving execution effectiveness of organizations. Developing social good applications that leverage power of SRAs to benefit society is an equally endeavor for AI technologists and researchers. 
\section{Conclusion}
This survey examines multiple dimensions of research in Social Responsibility of AI, seeking to broaden the current discussions primarily focused on decision-making algorithms that perform scoring and classification tasks. We argue that having a full scope of AI to capture the connections among all the major dimensions is the key to Socially Responsible AI Algorithms (SRAs). This work starts with an inclusive definition of Social Responsibility of AI, highlighting the principles (e.g., Fairness, Inclusiveness), means (e.g., SRAs), and objective (e.g., improving humanity). To better frame the Social Responsibility of AI, we also introduce the pyramid with four-level responsibilities of AI systems: functional responsibilities, legal responsibilities, ethical responsibilities, and philanthropic responsibilities. We then focus our discussions on how to achieve Social Responsibility of AI via the proposed framework SRAs. In the definition of SRAs, we emphasize that the functional and societal aspects are integral parts of AI algorithms. Given that both the functional and legal responsibilities are the usual focuses in AI research and development, we particularly investigate the essentials to achieve AI's ethical responsibilities: the subjects, causes, objectives, and means. For the intended use (i.e., roles) of SRAs, we discuss the need of philanthropic and ethical responsibilities for AI systems to protect and inform users, and prevent/mitigate the negative impact. We conclude with several open problems and major challenges in SRAs. At this pivotal moment in the development of AI, it is of vital importance to discuss AI ethics and specify Social Responsibility of AI. Drawing from the theory of moral license \cite{list2018corporate} -- when humans are good, we give ourselves moral license to be bad -- we argue that simply asking AI to do good is insufficient and inefficient, and more can be done for AI technologists and researchers to develop socially responsible AI systems. We hope this work can propel future research in various fields to tackle together the challenges and steer a course towards a beneficial AI future.
\section*{Acknowledgements}
This material is based upon work supported by, or in part by, the U.S. Army Research Laboratory (ARL), the U.S. Army Research Office (ARO), the Office of Naval Research (ONR) under contract/grant numbers W911NF2110030, W911NF2020124, and N00014-21-1-4002, as well as by the National Science Foundation (NSF) grants 1909555 and 2036127. We thank Dr. Lise Getoor and Dr. Hosagrahar V. Jagadish for their invaluable suggestions.
\vskip 0.2in
\bibliography{sample}

\begin{thebibliography}{}

\bibitem[\protect\BCAY{Abdalla\ \BBA\ Abdalla}{Abdalla\ \BBA\
  Abdalla}{2020}]{abdalla2020grey}
Abdalla, M.\BBACOMMA\  \BBA\ Abdalla, M. \BBOP2020\BBCP.
\newblock \BBOQ The grey hoodie project: Big tobacco, big tech, and the threat
  on academic integrity\BBCQ.

\bibitem[\protect\BCAY{Agarwal, Dud{\'\i}k,\ \BBA\ Wu}{Agarwal
  et~al.}{2019}]{agarwal2019fair}
Agarwal, A., Dud{\'\i}k, M., \BBA\ Wu, Z.~S. \BBOP2019\BBCP.
\newblock \BBOQ Fair regression: Quantitative definitions and reduction-based
  algorithms\BBCQ.

\bibitem[\protect\BCAY{Akhtar\ \BBA\ Dasgupta}{Akhtar\ \BBA\
  Dasgupta}{2019}]{akhtar2019brief}
Akhtar, Z.\BBACOMMA\  \BBA\ Dasgupta, D. \BBOP2019\BBCP.
\newblock \BBOQ A brief survey of adversarial machine learning and defense
  strategies\BBCQ.

\bibitem[\protect\BCAY{Al-garadi, Varathan,\ \BBA\ Ravana}{Al-garadi
  et~al.}{2016}]{al2016cybercrime}
Al-garadi, M.~A., Varathan, K.~D., \BBA\ Ravana, S.~D. \BBOP2016\BBCP.
\newblock \BBOQ Cybercrime detection in online communications: The experimental
  case of cyberbullying detection in the twitter network\BBCQ\
\newblock {\Bem Computers in Human Behavior}, {\Bem 63}, 433--443.

\bibitem[\protect\BCAY{Angwin\ \BBA\ Larson}{Angwin\ \BBA\
  Larson}{2015}]{angwin2015tiger}
Angwin, J.\BBACOMMA\  \BBA\ Larson, J. \BBOP2015\BBCP.
\newblock \BBOQ The tiger mom tax: Asians are nearly twice as likely to get a
  higher price from princeton review\BBCQ\
\newblock {\Bem Retrieved September}, {\Bem 1}, 2015.

\bibitem[\protect\BCAY{Angwin, Larson, Mattu,\ \BBA\ Kirchner}{Angwin
  et~al.}{2016}]{angwin2016machine}
Angwin, J., Larson, J., Mattu, S., \BBA\ Kirchner, L. \BBOP2016\BBCP.
\newblock \BBOQ Machine bias\BBCQ\
\newblock {\Bem ProPublica, May}, {\Bem 23}, 2016.

\bibitem[\protect\BCAY{Arjovsky, Bottou, Gulrajani,\ \BBA\ Lopez-Paz}{Arjovsky
  et~al.}{2019}]{arjovsky2019invariant}
Arjovsky, M., Bottou, L., Gulrajani, I., \BBA\ Lopez-Paz, D. \BBOP2019\BBCP.
\newblock \BBOQ Invariant risk minimization\BBCQ.

\bibitem[\protect\BCAY{Arnold, Bellamy, Hind, Houde, Mehta, Mojsilovi{\'c},
  Nair, Ramamurthy, Olteanu, Piorkowski, et~al.}{Arnold
  et~al.}{2019}]{arnold2019factsheets}
Arnold, M., Bellamy, R.~K., Hind, M., Houde, S., Mehta, S., Mojsilovi{\'c}, A.,
  Nair, R., Ramamurthy, K.~N., Olteanu, A., Piorkowski, D., et~al.
  \BBOP2019\BBCP.
\newblock \BBOQ Factsheets: Increasing trust in {AI} services through
  supplier's declarations of conformity\BBCQ\
\newblock {\Bem IBM Journal of Research and Development}, {\Bem 63\/}(4/5),
  6--1.

\bibitem[\protect\BCAY{Ashktorab}{Ashktorab}{2017}]{ashktorab2017designing}
Ashktorab, Z. \BBOP2017\BBCP.
\newblock {\Bem Designing Cyberbullying Prevention and Mitigation Tools}.
\newblock Ph.D.\ thesis.

\bibitem[\protect\BCAY{Ashktorab\ \BBA\ Vitak}{Ashktorab\ \BBA\
  Vitak}{2016}]{ashktorab2016designing}
Ashktorab, Z.\BBACOMMA\  \BBA\ Vitak, J. \BBOP2016\BBCP.
\newblock \BBOQ Designing cyberbullying mitigation and prevention solutions
  through participatory design with teenagers\BBCQ\
\newblock In {\Bem CHI}, \BPGS\ 3895--3905.

\bibitem[\protect\BCAY{Asimov}{Asimov}{2020}]{Stanford}
Asimov, N. \BBOP2020\BBCP\
\newblock
  \url{https://www.govtech.com/em/safety/Stanfords-Vaccine-Algorithm-Left-Frontline-Workers-at-Back-of-Line-.html}.

\bibitem[\protect\BCAY{Ateniese, Mancini, Spognardi, Villani, Vitali,\ \BBA\
  Felici}{Ateniese et~al.}{2015}]{ateniese2015hacking}
Ateniese, G., Mancini, L.~V., Spognardi, A., Villani, A., Vitali, D., \BBA\
  Felici, G. \BBOP2015\BBCP.
\newblock \BBOQ Hacking smart machines with smarter ones: How to extract
  meaningful data from machine learning classifiers\BBCQ\
\newblock {\Bem International Journal of Security and Networks}, {\Bem
  10\/}(3), 137--150.

\bibitem[\protect\BCAY{Ayres}{Ayres}{2010}]{ayres2010testing}
Ayres, I. \BBOP2010\BBCP.
\newblock \BBOQ Testing for discrimination and the problem of” included
  variable bias”,”\BBCQ\
\newblock \BTR, mimeo, Yale Law School.

\bibitem[\protect\BCAY{Baeza-Yates}{Baeza-Yates}{2018}]{baeza2018bias}
Baeza-Yates, R. \BBOP2018\BBCP.
\newblock \BBOQ Bias on the web\BBCQ\
\newblock {\Bem Communications of the ACM}, {\Bem 61\/}(6), 54--61.

\bibitem[\protect\BCAY{Balakrishnan}{Balakrishnan}{2015}]{balakrishnan2015cyberbullying}
Balakrishnan, V. \BBOP2015\BBCP.
\newblock \BBOQ Cyberbullying among young adults in malaysia: The roles of
  gender, age and internet frequency\BBCQ\
\newblock {\Bem Computers in Human Behavior}, {\Bem 46}, 149--157.

\bibitem[\protect\BCAY{Barreno, Nelson, Sears, Joseph,\ \BBA\ Tygar}{Barreno
  et~al.}{2006}]{barreno2006can}
Barreno, M., Nelson, B., Sears, R., Joseph, A.~D., \BBA\ Tygar, J.~D.
  \BBOP2006\BBCP.
\newblock \BBOQ Can machine learning be secure?\BBCQ\
\newblock In {\Bem ASIACCS}, \BPGS\ 16--25.

\bibitem[\protect\BCAY{Beale, Battey, Davison,\ \BBA\ MacKay}{Beale
  et~al.}{2019}]{beale2019unethical}
Beale, N., Battey, H., Davison, A.~C., \BBA\ MacKay, R.~S. \BBOP2019\BBCP.
\newblock \BBOQ An unethical optimization principle\BBCQ.

\bibitem[\protect\BCAY{Bechavod\ \BBA\ Ligett}{Bechavod\ \BBA\
  Ligett}{2017}]{bechavod2017penalizing}
Bechavod, Y.\BBACOMMA\  \BBA\ Ligett, K. \BBOP2017\BBCP.
\newblock \BBOQ Penalizing unfairness in binary classification\BBCQ.

\bibitem[\protect\BCAY{Becker}{Becker}{1993}]{becker1993nobel}
Becker, G.~S. \BBOP1993\BBCP.
\newblock \BBOQ Nobel lecture: The economic way of looking at behavior\BBCQ\
\newblock {\Bem Journal of Political Economy}, {\Bem 101\/}(3), 385--409.

\bibitem[\protect\BCAY{Beigi\ \BBA\ Liu}{Beigi\ \BBA\
  Liu}{2020}]{beigi2020survey}
Beigi, G.\BBACOMMA\  \BBA\ Liu, H. \BBOP2020\BBCP.
\newblock \BBOQ A survey on privacy in social media: Identification,
  mitigation, and applications\BBCQ\
\newblock {\Bem ACM Transactions on Data Science}, {\Bem 1\/}(1), 1--38.

\bibitem[\protect\BCAY{Bellamy, Dey, Hind, Hoffman, Houde, Kannan, Lohia,
  Martino, Mehta, Mojsilovi{\'c}, et~al.}{Bellamy et~al.}{2019}]{bellamy2019ai}
Bellamy, R.~K., Dey, K., Hind, M., Hoffman, S.~C., Houde, S., Kannan, K.,
  Lohia, P., Martino, J., Mehta, S., Mojsilovi{\'c}, A., et~al. \BBOP2019\BBCP.
\newblock \BBOQ {AI} fairness 360: An extensible toolkit for detecting and
  mitigating algorithmic bias\BBCQ\
\newblock {\Bem IBM Journal of Research and Development}, {\Bem 63\/}(4/5),
  4--1.

\bibitem[\protect\BCAY{Bengio, Deleu, Rahaman, Ke, Lachapelle, Bilaniuk,
  Goyal,\ \BBA\ Pal}{Bengio et~al.}{2019}]{bengio2019meta}
Bengio, Y., Deleu, T., Rahaman, N., Ke, R., Lachapelle, S., Bilaniuk, O.,
  Goyal, A., \BBA\ Pal, C. \BBOP2019\BBCP.
\newblock \BBOQ A meta-transfer objective for learning to disentangle causal
  mechanisms\BBCQ.

\bibitem[\protect\BCAY{Berk, Heidari, Jabbari, Joseph, Kearns, Morgenstern,
  Neel,\ \BBA\ Roth}{Berk et~al.}{2017}]{berk2017convex}
Berk, R., Heidari, H., Jabbari, S., Joseph, M., Kearns, M., Morgenstern, J.,
  Neel, S., \BBA\ Roth, A. \BBOP2017\BBCP.
\newblock \BBOQ A convex framework for fair regression\BBCQ.

\bibitem[\protect\BCAY{Bhatt, Antor{\'a}n, Zhang, Liao, Sattigeri, Fogliato,
  Melancon, Krishnan, Stanley, Tickoo, et~al.}{Bhatt
  et~al.}{2020}]{bhatt2020uncertainty}
Bhatt, U., Antor{\'a}n, J., Zhang, Y., Liao, Q.~V., Sattigeri, P., Fogliato,
  R., Melancon, G.~G., Krishnan, R., Stanley, J., Tickoo, O., et~al.
  \BBOP2020\BBCP.
\newblock \BBOQ Uncertainty as a form of transparency: Measuring,
  communicating, and using uncertainty\BBCQ.

\bibitem[\protect\BCAY{Bhattacharjee, Shu, Gao,\ \BBA\ Liu}{Bhattacharjee
  et~al.}{2020}]{bhattacharjee2020disinformation}
Bhattacharjee, A., Shu, K., Gao, M., \BBA\ Liu, H. \BBOP2020\BBCP.
\newblock \BBOQ Disinformation in the online information ecosystem: Detection,
  mitigation and challenges\BBCQ.

\bibitem[\protect\BCAY{Bickel, Hammel,\ \BBA\ O'Connell}{Bickel
  et~al.}{1975}]{bickel1975sex}
Bickel, P.~J., Hammel, E.~A., \BBA\ O'Connell, J.~W. \BBOP1975\BBCP.
\newblock \BBOQ Sex bias in graduate admissions: Data from {B}erkeley\BBCQ\
\newblock {\Bem Science}, {\Bem 187\/}(4175), 398--404.

\bibitem[\protect\BCAY{Biel, Aran,\ \BBA\ Gatica-Perez}{Biel
  et~al.}{2011}]{biel2011you}
Biel, J.-I., Aran, O., \BBA\ Gatica-Perez, D. \BBOP2011\BBCP.
\newblock \BBOQ You are known by how you vlog: Personality impressions and
  nonverbal behavior in youtube.\BBCQ\
\newblock In {\Bem ICWSM}, \BPGS\ 446--449. Citeseer.

\bibitem[\protect\BCAY{Blyth}{Blyth}{1972}]{blyth1972simpson}
Blyth, C.~R. \BBOP1972\BBCP.
\newblock \BBOQ On {S}impson's paradox and the sure-thing principle\BBCQ\
\newblock {\Bem Journal of the American Statistical Association}, {\Bem
  67\/}(338), 364--366.

\bibitem[\protect\BCAY{Boulemtafes, Derhab,\ \BBA\ Challal}{Boulemtafes
  et~al.}{2020}]{boulemtafes2020review}
Boulemtafes, A., Derhab, A., \BBA\ Challal, Y. \BBOP2020\BBCP.
\newblock \BBOQ A review of privacy-preserving techniques for deep
  learning\BBCQ\
\newblock {\Bem Neurocomputing}, {\Bem 384}, 21--45.

\bibitem[\protect\BCAY{Branswell}{Branswell}{2021}]{covid2021}
Branswell, H. \BBOP2021\BBCP\
\newblock
  \url{https://www.statnews.com/2021/03/08/a-pandemic-expert-weighs-in-on-the-long-road-ahead-for-covid-19-vaccine-distribution/}.

\bibitem[\protect\BCAY{Budak, Agrawal,\ \BBA\ El~Abbadi}{Budak
  et~al.}{2011}]{budak2011limiting}
Budak, C., Agrawal, D., \BBA\ El~Abbadi, A. \BBOP2011\BBCP.
\newblock \BBOQ Limiting the spread of misinformation in social networks\BBCQ\
\newblock In {\Bem WWW}, \BPGS\ 665--674.

\bibitem[\protect\BCAY{Caba{\~n}as, Cuevas, Arrate,\ \BBA\ Cuevas}{Caba{\~n}as
  et~al.}{2020}]{cabanas2020does}
Caba{\~n}as, J.~G., Cuevas, {\'A}., Arrate, A., \BBA\ Cuevas, R.
  \BBOP2020\BBCP.
\newblock \BBOQ Does facebook use sensitive data for advertising
  purposes?\BBCQ\
\newblock {\Bem Communications of the ACM}, {\Bem 64\/}(1), 62--69.

\bibitem[\protect\BCAY{Calders, Karim, Kamiran, Ali,\ \BBA\ Zhang}{Calders
  et~al.}{2013}]{calders2013controlling}
Calders, T., Karim, A., Kamiran, F., Ali, W., \BBA\ Zhang, X. \BBOP2013\BBCP.
\newblock \BBOQ Controlling attribute effect in linear regression\BBCQ\
\newblock In {\Bem ICDM}, \BPGS\ 71--80. IEEE.

\bibitem[\protect\BCAY{Calders\ \BBA\ Verwer}{Calders\ \BBA\
  Verwer}{2010}]{calders2010three}
Calders, T.\BBACOMMA\  \BBA\ Verwer, S. \BBOP2010\BBCP.
\newblock \BBOQ Three naive {B}ayes approaches for discrimination-free
  classification\BBCQ\
\newblock {\Bem Data Mining and Knowledge Discovery}, {\Bem 21\/}(2), 277--292.

\bibitem[\protect\BCAY{Calmon, Wei, Vinzamuri, Ramamurthy,\ \BBA\
  Varshney}{Calmon et~al.}{2017}]{calmon2017optimized}
Calmon, F., Wei, D., Vinzamuri, B., Ramamurthy, K.~N., \BBA\ Varshney, K.~R.
  \BBOP2017\BBCP.
\newblock \BBOQ Optimized pre-processing for discrimination prevention\BBCQ\
\newblock In {\Bem NeurIPS}, \BPGS\ 3992--4001.

\bibitem[\protect\BCAY{Canca}{Canca}{2020}]{canca2020operationalizing}
Canca, C. \BBOP2020\BBCP.
\newblock \BBOQ Operationalizing {AI} ethics principles\BBCQ\
\newblock {\Bem Communications of the ACM}, {\Bem 63\/}(12), 18--21.

\bibitem[\protect\BCAY{Carpenter}{Carpenter}{2015}]{Carpenter2015}
Carpenter, J. \BBOP2015\BBCP\
\newblock
  \url{https://www.washingtonpost.com/news/the-intersect/wp/2015/07/06/googles-algorithm-shows-prestigious-job-ads-to-men-but-not-to-women-heres-why-that-should-worry-you/}.

\bibitem[\protect\BCAY{Carroll et~al.}{Carroll
  et~al.}{1991}]{carroll1991pyramid}
Carroll, A.~B.\BBACOMMA\  et~al. \BBOP1991\BBCP.
\newblock \BBOQ The pyramid of corporate social responsibility: Toward the
  moral management of organizational stakeholders\BBCQ\
\newblock {\Bem Business Horizons}, {\Bem 34\/}(4), 39--48.

\bibitem[\protect\BCAY{Carvalho, Pereira,\ \BBA\ Cardoso}{Carvalho
  et~al.}{2019}]{carvalho2019machine}
Carvalho, D.~V., Pereira, E.~M., \BBA\ Cardoso, J.~S. \BBOP2019\BBCP.
\newblock \BBOQ Machine learning interpretability: A survey on methods and
  metrics\BBCQ\
\newblock {\Bem Electronics}, {\Bem 8\/}(8), 832.

\bibitem[\protect\BCAY{Caton\ \BBA\ Haas}{Caton\ \BBA\
  Haas}{2020}]{caton2020fairness}
Caton, S.\BBACOMMA\  \BBA\ Haas, C. \BBOP2020\BBCP.
\newblock \BBOQ Fairness in machine learning: A survey\BBCQ.

\bibitem[\protect\BCAY{Celis, Deshpande, Kathuria,\ \BBA\ Vishnoi}{Celis
  et~al.}{2016}]{celis2016fair}
Celis, L.~E., Deshpande, A., Kathuria, T., \BBA\ Vishnoi, N.~K. \BBOP2016\BBCP.
\newblock \BBOQ How to be fair and diverse?\BBCQ.

\bibitem[\protect\BCAY{Chabanne, de~Wargny, Milgram, Morel,\ \BBA\
  Prouff}{Chabanne et~al.}{2017}]{chabanne2017privacy}
Chabanne, H., de~Wargny, A., Milgram, J., Morel, C., \BBA\ Prouff, E.
  \BBOP2017\BBCP.
\newblock \BBOQ Privacy-preserving classification on deep neural network.\BBCQ\
\newblock {\Bem IACR Cryptol. ePrint Arch.}, {\Bem 2017}, 35.

\bibitem[\protect\BCAY{Chakraborty, Alam, Dey, Chattopadhyay,\ \BBA\
  Mukhopadhyay}{Chakraborty et~al.}{2018}]{chakraborty2018adversarial}
Chakraborty, A., Alam, M., Dey, V., Chattopadhyay, A., \BBA\ Mukhopadhyay, D.
  \BBOP2018\BBCP.
\newblock \BBOQ Adversarial attacks and defences: A survey\BBCQ.

\bibitem[\protect\BCAY{Chattopadhyay, Manupriya, Sarkar,\ \BBA\
  Balasubramanian}{Chattopadhyay et~al.}{2019}]{chattopadhyay2019neural}
Chattopadhyay, A., Manupriya, P., Sarkar, A., \BBA\ Balasubramanian, V.~N.
  \BBOP2019\BBCP.
\newblock \BBOQ Neural network attributions: A causal perspective\BBCQ.

\bibitem[\protect\BCAY{Chen\ \BBA\ Li}{Chen\ \BBA\ Li}{2020}]{chen2020henin}
Chen, H.-Y.\BBACOMMA\  \BBA\ Li, C.-T. \BBOP2020\BBCP.
\newblock \BBOQ Henin: Learning heterogeneous neural interaction networks for
  explainable cyberbullying detection on social media\BBCQ.

\bibitem[\protect\BCAY{Chen\ \BBA\ Liu}{Chen\ \BBA\
  Liu}{2018}]{chen2018lifelong}
Chen, Z.\BBACOMMA\  \BBA\ Liu, B. \BBOP2018\BBCP.
\newblock \BBOQ Lifelong machine learning\BBCQ\
\newblock {\Bem Synthesis Lectures on Artificial Intelligence and Machine
  Learning}, {\Bem 12\/}(3), 1--207.

\bibitem[\protect\BCAY{Cheng, Guo, Shu,\ \BBA\ Liu}{Cheng
  et~al.}{2021}]{cheng2021causal}
Cheng, L., Guo, R., Shu, K., \BBA\ Liu, H. \BBOP2021\BBCP.
\newblock \BBOQ Causal understanding of fake news dissemination on social
  media\BBCQ\
\newblock In {\Bem KDD}.

\bibitem[\protect\BCAY{Cheng, Guo, Silva, Hall,\ \BBA\ Liu}{Cheng
  et~al.}{2019}]{cheng2019hierarchical}
Cheng, L., Guo, R., Silva, Y., Hall, D., \BBA\ Liu, H. \BBOP2019\BBCP.
\newblock \BBOQ Hierarchical attention networks for cyberbullying detection on
  the instagram social network\BBCQ\
\newblock In {\Bem SDM}, \BPGS\ 235--243. SIAM.

\bibitem[\protect\BCAY{Cheng, Guo, Silva, Hall,\ \BBA\ Liu}{Cheng
  et~al.}{2021}]{cheng2021modeling}
Cheng, L., Guo, R., Silva, Y.~N., Hall, D., \BBA\ Liu, H. \BBOP2021\BBCP.
\newblock \BBOQ Modeling temporal patterns of cyberbullying detection with
  hierarchical attention networks\BBCQ\
\newblock {\Bem ACM/IMS Transactions on Data Science}, {\Bem 2\/}(2), 1--23.

\bibitem[\protect\BCAY{Cheng, Li, Silva, Hall,\ \BBA\ Liu}{Cheng
  et~al.}{2019a}]{cheng2019pi}
Cheng, L., Li, J., Silva, Y.~N., Hall, D.~L., \BBA\ Liu, H. \BBOP2019a\BBCP.
\newblock \BBOQ {PI}-bully: Personalized cyberbullying detection with peer
  influence.\BBCQ\
\newblock In {\Bem IJCAI}, \BPGS\ 5829--5835.

\bibitem[\protect\BCAY{Cheng, Li, Silva, Hall,\ \BBA\ Liu}{Cheng
  et~al.}{2019b}]{cheng2019xbully}
Cheng, L., Li, J., Silva, Y.~N., Hall, D.~L., \BBA\ Liu, H. \BBOP2019b\BBCP.
\newblock \BBOQ Xbully: Cyberbullying detection within a multi-modal
  context\BBCQ\
\newblock In {\Bem WSDM}, \BPGS\ 339--347.

\bibitem[\protect\BCAY{Cheng, Mosallanezhad, Silva, Hall,\ \BBA\ Liu}{Cheng
  et~al.}{2021}]{cheng2021mitigating}
Cheng, L., Mosallanezhad, A., Silva, Y.~N., Hall, D.~L., \BBA\ Liu, H.
  \BBOP2021\BBCP.
\newblock \BBOQ Mitigating bias in session-based cyberbullying detection: A
  non-compromising approach\BBCQ\
\newblock In {\Bem Proceedings of ACL}.

\bibitem[\protect\BCAY{Cheng, Shu, Wu, Silva, Hall,\ \BBA\ Liu}{Cheng
  et~al.}{2020a}]{cheng2020unsupervised}
Cheng, L., Shu, K., Wu, S., Silva, Y.~N., Hall, D.~L., \BBA\ Liu, H.
  \BBOP2020a\BBCP.
\newblock \BBOQ Unsupervised cyberbullying detection via time-informed gaussian
  mixture model\BBCQ\
\newblock In {\Bem CIKM}.

\bibitem[\protect\BCAY{Cheng, Silva, Hall,\ \BBA\ Liu}{Cheng
  et~al.}{2020b}]{cheng2020session}
Cheng, L., Silva, Y., Hall, D., \BBA\ Liu, H. \BBOP2020b\BBCP.
\newblock \BBOQ Session-based cyberbullying detection: Problems and
  challenges\BBCQ.

\bibitem[\protect\BCAY{Chu, Li, Pedro,\ \BBA\ Coimbra}{Chu
  et~al.}{2015}]{chu2015real}
Chu, Y., Li, M., Pedro, H.~T., \BBA\ Coimbra, C.~F. \BBOP2015\BBCP.
\newblock \BBOQ Real-time prediction intervals for intra-hour dni
  forecasts\BBCQ\
\newblock {\Bem Renewable Energy}, {\Bem 83}, 234--244.

\bibitem[\protect\BCAY{Ciampaglia, Shiralkar, Rocha, Bollen, Menczer,\ \BBA\
  Flammini}{Ciampaglia et~al.}{2015}]{ciampaglia2015computational}
Ciampaglia, G.~L., Shiralkar, P., Rocha, L.~M., Bollen, J., Menczer, F., \BBA\
  Flammini, A. \BBOP2015\BBCP.
\newblock \BBOQ Computational fact checking from knowledge networks\BBCQ\
\newblock {\Bem PloS One}, {\Bem 10\/}(6), e0128193.

\bibitem[\protect\BCAY{Cowen}{Cowen}{2020}]{Vaccine_2020}
Cowen, T. \BBOP2020\BBCP\
\newblock
  \url{https://www.bloomberg.com/opinion/articles/2020-11-23/vaccine-distribution-shouldn-t-be-fair}.

\bibitem[\protect\BCAY{Craven\ \BBA\ Shavlik}{Craven\ \BBA\
  Shavlik}{1996}]{craven1996extracting}
Craven, M.\BBACOMMA\  \BBA\ Shavlik, J.~W. \BBOP1996\BBCP.
\newblock \BBOQ Extracting tree-structured representations of trained
  networks\BBCQ\
\newblock In {\Bem NeurIPS}, \BPGS\ 24--30.

\bibitem[\protect\BCAY{Dadvar, Trieschnigg, Ordelman,\ \BBA\ de~Jong}{Dadvar
  et~al.}{2013}]{dadvar2013improving}
Dadvar, M., Trieschnigg, D., Ordelman, R., \BBA\ de~Jong, F. \BBOP2013\BBCP.
\newblock \BBOQ Improving cyberbullying detection with user context\BBCQ\
\newblock In {\Bem ECIR}, \BPGS\ 693--696. Springer.

\bibitem[\protect\BCAY{d'Alessandro, O'Neil,\ \BBA\ LaGatta}{d'Alessandro
  et~al.}{2017}]{d2017conscientious}
d'Alessandro, B., O'Neil, C., \BBA\ LaGatta, T. \BBOP2017\BBCP.
\newblock \BBOQ Conscientious classification: A data scientist's guide to
  discrimination-aware classification\BBCQ\
\newblock {\Bem Big Data}, {\Bem 5\/}(2), 120--134.

\bibitem[\protect\BCAY{Dalvi, Domingos, Sanghai,\ \BBA\ Verma}{Dalvi
  et~al.}{2004}]{dalvi2004adversarial}
Dalvi, N., Domingos, P., Sanghai, S., \BBA\ Verma, D. \BBOP2004\BBCP.
\newblock \BBOQ Adversarial classification\BBCQ\
\newblock In {\Bem KDD}, \BPGS\ 99--108.

\bibitem[\protect\BCAY{Dani, Li,\ \BBA\ Liu}{Dani
  et~al.}{2017}]{dani2017sentiment}
Dani, H., Li, J., \BBA\ Liu, H. \BBOP2017\BBCP.
\newblock \BBOQ Sentiment informed cyberbullying detection in social
  media\BBCQ\
\newblock In {\Bem ECML PKDD}, \BPGS\ 52--67. Springer.

\bibitem[\protect\BCAY{Datta, Tschantz,\ \BBA\ Datta}{Datta
  et~al.}{2015}]{datta2015automated}
Datta, A., Tschantz, M.~C., \BBA\ Datta, A. \BBOP2015\BBCP.
\newblock \BBOQ Automated experiments on ad privacy settings: A tale of
  opacity, choice, and discrimination\BBCQ\
\newblock {\Bem PoPETs}, {\Bem 2015\/}(1), 92--112.

\bibitem[\protect\BCAY{Della~Vedova, Tacchini, Moret, Ballarin, DiPierro,\
  \BBA\ de~Alfaro}{Della~Vedova et~al.}{2018}]{della2018automatic}
Della~Vedova, M.~L., Tacchini, E., Moret, S., Ballarin, G., DiPierro, M., \BBA\
  de~Alfaro, L. \BBOP2018\BBCP.
\newblock \BBOQ Automatic online fake news detection combining content and
  social signals\BBCQ\
\newblock In {\Bem 2018 FRUCT}, \BPGS\ 272--279. IEEE.

\bibitem[\protect\BCAY{Diaz, Wang, Calmon,\ \BBA\ Sankar}{Diaz
  et~al.}{2019}]{diaz2019robustness}
Diaz, M., Wang, H., Calmon, F.~P., \BBA\ Sankar, L. \BBOP2019\BBCP.
\newblock \BBOQ On the robustness of information-theoretic privacy measures and
  mechanisms\BBCQ\
\newblock {\Bem IEEE Transactions on Information Theory}, {\Bem 66\/}(4),
  1949--1978.

\bibitem[\protect\BCAY{DiFranzo, Taylor, Kazerooni, Wherry,\ \BBA\
  Bazarova}{DiFranzo et~al.}{2018}]{difranzo2018upstanding}
DiFranzo, D., Taylor, S.~H., Kazerooni, F., Wherry, O.~D., \BBA\ Bazarova,
  N.~N. \BBOP2018\BBCP.
\newblock \BBOQ Upstanding by design: Bystander intervention in
  cyberbullying\BBCQ\
\newblock In {\Bem CHI}, \BPGS\ 1--12.

\bibitem[\protect\BCAY{Dinakar, Jones, Havasi, Lieberman,\ \BBA\
  Picard}{Dinakar et~al.}{2012}]{dinakar2012common}
Dinakar, K., Jones, B., Havasi, C., Lieberman, H., \BBA\ Picard, R.
  \BBOP2012\BBCP.
\newblock \BBOQ Common sense reasoning for detection, prevention, and
  mitigation of cyberbullying\BBCQ\
\newblock {\Bem ACM TiiS}, {\Bem 2\/}(3), 1--30.

\bibitem[\protect\BCAY{Dobrin}{Dobrin}{2020}]{IBM_2020}
Dobrin, S. \BBOP2020\BBCP\
\newblock
  \url{https://www.ibm.com/blogs/watson/2020/12/how-ibm-is-advancing-ai-governance-to-help-clients-build-trust-and-transparency/}.

\bibitem[\protect\BCAY{Dwork}{Dwork}{2008}]{dwork2008differential}
Dwork, C. \BBOP2008\BBCP.
\newblock \BBOQ Differential privacy: A survey of results\BBCQ\
\newblock In {\Bem TAMC}, \BPGS\ 1--19. Springer.

\bibitem[\protect\BCAY{Dwork, Hardt, Pitassi, Reingold,\ \BBA\ Zemel}{Dwork
  et~al.}{2012}]{dwork2012fairness}
Dwork, C., Hardt, M., Pitassi, T., Reingold, O., \BBA\ Zemel, R.
  \BBOP2012\BBCP.
\newblock \BBOQ Fairness through awareness\BBCQ\
\newblock In {\Bem ITCS}, \BPGS\ 214--226.

\bibitem[\protect\BCAY{Dwork, Immorlica, Kalai,\ \BBA\ Leiserson}{Dwork
  et~al.}{2018}]{dwork2018decoupled}
Dwork, C., Immorlica, N., Kalai, A.~T., \BBA\ Leiserson, M. \BBOP2018\BBCP.
\newblock \BBOQ Decoupled classifiers for group-fair and efficient machine
  learning\BBCQ\
\newblock In {\Bem FAT*}, \BPGS\ 119--133.

\bibitem[\protect\BCAY{Dwork, McSherry, Nissim,\ \BBA\ Smith}{Dwork
  et~al.}{2006}]{dwork2006calibrating}
Dwork, C., McSherry, F., Nissim, K., \BBA\ Smith, A. \BBOP2006\BBCP.
\newblock \BBOQ Calibrating noise to sensitivity in private data analysis\BBCQ\
\newblock In {\Bem Theory of Cryptography Conference}, \BPGS\ 265--284.
  Springer.

\bibitem[\protect\BCAY{Dwork, Roth, et~al.}{Dwork
  et~al.}{2014}]{dwork2014algorithmic}
Dwork, C., Roth, A., et~al. \BBOP2014\BBCP.
\newblock \BBOQ The algorithmic foundations of differential privacy.\BBCQ\
\newblock {\Bem FnT-TCS}, {\Bem 9\/}(3-4), 211--407.

\bibitem[\protect\BCAY{Dybowski\ \BBA\ Roberts}{Dybowski\ \BBA\
  Roberts}{2001}]{dybowski2001confidence}
Dybowski, R.\BBACOMMA\  \BBA\ Roberts, S.~J. \BBOP2001\BBCP.
\newblock \BBOQ Confidence intervals and prediction intervals for feed-forward
  neural networks\BBCQ.

\bibitem[\protect\BCAY{Ekstrand, Tian, Kazi, Mehrpouyan,\ \BBA\
  Kluver}{Ekstrand et~al.}{2018}]{ekstrand2018exploring}
Ekstrand, M.~D., Tian, M., Kazi, M. R.~I., Mehrpouyan, H., \BBA\ Kluver, D.
  \BBOP2018\BBCP.
\newblock \BBOQ Exploring author gender in book rating and recommendation\BBCQ\
\newblock In {\Bem RecSys}, \BPGS\ 242--250.

\bibitem[\protect\BCAY{Ensign, Friedler, Neville, Scheidegger,\ \BBA\
  Venkatasubramanian}{Ensign et~al.}{2018}]{ensign2018runaway}
Ensign, D., Friedler, S.~A., Neville, S., Scheidegger, C., \BBA\
  Venkatasubramanian, S. \BBOP2018\BBCP.
\newblock \BBOQ Runaway feedback loops in predictive policing\BBCQ\
\newblock In {\Bem FAT*}, \BPGS\ 160--171.

\bibitem[\protect\BCAY{Erhan, Bengio, Courville,\ \BBA\ Vincent}{Erhan
  et~al.}{2009}]{erhan2009visualizing}
Erhan, D., Bengio, Y., Courville, A., \BBA\ Vincent, P. \BBOP2009\BBCP.
\newblock \BBOQ Visualizing higher-layer features of a deep network\BBCQ\
\newblock {\Bem University of Montreal}, {\Bem 1341\/}(3), 1.

\bibitem[\protect\BCAY{Erlingsson, Pihur,\ \BBA\ Korolova}{Erlingsson
  et~al.}{2014}]{erlingsson2014rappor}
Erlingsson, {\'U}., Pihur, V., \BBA\ Korolova, A. \BBOP2014\BBCP.
\newblock \BBOQ Rappor: Randomized aggregatable privacy-preserving ordinal
  response\BBCQ\
\newblock In {\Bem CCS}, \BPGS\ 1054--1067.

\bibitem[\protect\BCAY{Errouissi, Cardenas-Barrera, Meng, Castillo-Guerra,
  Gong,\ \BBA\ Chang}{Errouissi et~al.}{2015}]{errouissi2015bootstrap}
Errouissi, R., Cardenas-Barrera, J., Meng, J., Castillo-Guerra, E., Gong, X.,
  \BBA\ Chang, L. \BBOP2015\BBCP.
\newblock \BBOQ Bootstrap prediction interval estimation for wind speed
  forecasting\BBCQ\
\newblock In {\Bem 2015 IEEE ECCE}, \BPGS\ 1919--1924. IEEE.

\bibitem[\protect\BCAY{Feige}{Feige}{2019}]{Safety2019}
Feige, I. \BBOP2019\BBCP\
\newblock \url{https://faculty.ai/blog/what-is-ai-safety/}.

\bibitem[\protect\BCAY{Feldman, Friedler, Moeller, Scheidegger,\ \BBA\
  Venkatasubramanian}{Feldman et~al.}{2015}]{feldman2015certifying}
Feldman, M., Friedler, S.~A., Moeller, J., Scheidegger, C., \BBA\
  Venkatasubramanian, S. \BBOP2015\BBCP.
\newblock \BBOQ Certifying and removing disparate impact\BBCQ\
\newblock In {\Bem KDD}, \BPGS\ 259--268.

\bibitem[\protect\BCAY{Feng, Yang, Lyu, Tan, Sun,\ \BBA\ Wang}{Feng
  et~al.}{2019}]{feng2019learning}
Feng, R., Yang, Y., Lyu, Y., Tan, C., Sun, Y., \BBA\ Wang, C. \BBOP2019\BBCP.
\newblock \BBOQ Learning fair representations via an adversarial
  framework\BBCQ.

\bibitem[\protect\BCAY{Feng, Banerjee,\ \BBA\ Choi}{Feng
  et~al.}{2012}]{feng2012syntactic}
Feng, S., Banerjee, R., \BBA\ Choi, Y. \BBOP2012\BBCP.
\newblock \BBOQ Syntactic stylometry for deception detection\BBCQ\
\newblock In {\Bem ACL}, \BPGS\ 171--175.

\bibitem[\protect\BCAY{Fredrikson, Jha,\ \BBA\ Ristenpart}{Fredrikson
  et~al.}{2015}]{fredrikson2015model}
Fredrikson, M., Jha, S., \BBA\ Ristenpart, T. \BBOP2015\BBCP.
\newblock \BBOQ Model inversion attacks that exploit confidence information and
  basic countermeasures\BBCQ\
\newblock In {\Bem CCS}, \BPGS\ 1322--1333.

\bibitem[\protect\BCAY{Fredrikson, Lantz, Jha, Lin, Page,\ \BBA\
  Ristenpart}{Fredrikson et~al.}{2014}]{fredrikson2014privacy}
Fredrikson, M., Lantz, E., Jha, S., Lin, S., Page, D., \BBA\ Ristenpart, T.
  \BBOP2014\BBCP.
\newblock \BBOQ Privacy in pharmacogenetics: An end-to-end case study of
  personalized warfarin dosing\BBCQ\
\newblock In {\Bem USENIX Security}, \BPGS\ 17--32.

\bibitem[\protect\BCAY{Fu, Huang,\ \BBA\ Singh}{Fu et~al.}{2020}]{fu2020ai}
Fu, R., Huang, Y., \BBA\ Singh, P.~V. \BBOP2020\BBCP.
\newblock \BBOQ {AI} and algorithmic bias: Source, detection, mitigation and
  implications\BBCQ.

\bibitem[\protect\BCAY{Galhotra, Brun,\ \BBA\ Meliou}{Galhotra
  et~al.}{2017}]{galhotra2017fairness}
Galhotra, S., Brun, Y., \BBA\ Meliou, A. \BBOP2017\BBCP.
\newblock \BBOQ Fairness testing: testing software for discrimination\BBCQ\
\newblock In {\Bem ESEC/FSE}, \BPGS\ 498--510.

\bibitem[\protect\BCAY{Garg, Schiebinger, Jurafsky,\ \BBA\ Zou}{Garg
  et~al.}{2018}]{garg2018word}
Garg, N., Schiebinger, L., Jurafsky, D., \BBA\ Zou, J. \BBOP2018\BBCP.
\newblock \BBOQ Word embeddings quantify 100 years of gender and ethnic
  stereotypes\BBCQ\
\newblock {\Bem PNAS}, {\Bem 115\/}(16), E3635--E3644.

\bibitem[\protect\BCAY{Ge, Cheng,\ \BBA\ Liu}{Ge
  et~al.}{2021}]{ge2021improving}
Ge, S., Cheng, L., \BBA\ Liu, H. \BBOP2021\BBCP.
\newblock \BBOQ Improving cyberbullying detection with user interaction\BBCQ\
\newblock In {\Bem Proceedings of the Web Conference 2021}, \BPGS\ 496--506.

\bibitem[\protect\BCAY{Gentry}{Gentry}{2009}]{gentry2009fully}
Gentry, C. \BBOP2009\BBCP.
\newblock \BBOQ Fully homomorphic encryption using ideal lattices\BBCQ\
\newblock In {\Bem STOC}, \BPGS\ 169--178.

\bibitem[\protect\BCAY{Gershgorn}{Gershgorn}{2019}]{NSA}
Gershgorn, D. \BBOP2019\BBCP\
\newblock
  \url{https://www.popsci.com/nsas-skynet-might-not-be-able-to-tell-what-makes-terrorist/}.

\bibitem[\protect\BCAY{Getoor}{Getoor}{2019}]{getoor2019responsible}
Getoor, L. \BBOP2019\BBCP.
\newblock \BBOQ Responsible data science\BBCQ\
\newblock In {\Bem Big Data}, \BPGS\ 1--1. IEEE.

\bibitem[\protect\BCAY{Geyik, Ambler,\ \BBA\ Kenthapadi}{Geyik
  et~al.}{2019}]{geyik2019fairness}
Geyik, S.~C., Ambler, S., \BBA\ Kenthapadi, K. \BBOP2019\BBCP.
\newblock \BBOQ Fairness-aware ranking in search \& recommendation systems with
  application to linkedin talent search\BBCQ\
\newblock In {\Bem KDD}, \BPGS\ 2221--2231.

\bibitem[\protect\BCAY{Gilad-Bachrach, Dowlin, Laine, Lauter, Naehrig,\ \BBA\
  Wernsing}{Gilad-Bachrach et~al.}{2016}]{gilad2016cryptonets}
Gilad-Bachrach, R., Dowlin, N., Laine, K., Lauter, K., Naehrig, M., \BBA\
  Wernsing, J. \BBOP2016\BBCP.
\newblock \BBOQ Cryptonets: Applying neural networks to encrypted data with
  high throughput and accuracy\BBCQ\
\newblock In {\Bem ICML}, \BPGS\ 201--210.

\bibitem[\protect\BCAY{Goel, Yaghini,\ \BBA\ Faltings}{Goel
  et~al.}{2018}]{goel2018non}
Goel, N., Yaghini, M., \BBA\ Faltings, B. \BBOP2018\BBCP.
\newblock \BBOQ Non-discriminatory machine learning through convex fairness
  criteria\BBCQ\
\newblock In {\Bem AIES}, \BPGS\ 116--116.

\bibitem[\protect\BCAY{Goldwasser\ \BBA\ Micali}{Goldwasser\ \BBA\
  Micali}{1984}]{goldwasser1984probabilistic}
Goldwasser, S.\BBACOMMA\  \BBA\ Micali, S. \BBOP1984\BBCP.
\newblock \BBOQ Probabilistic encryption\BBCQ\
\newblock {\Bem JCSS}, {\Bem 28\/}(2), 270--299.

\bibitem[\protect\BCAY{Goodfellow, Pouget-Abadie, Mirza, Xu, Warde-Farley,
  Ozair, Courville,\ \BBA\ Bengio}{Goodfellow
  et~al.}{2014}]{goodfellow2014generative}
Goodfellow, I., Pouget-Abadie, J., Mirza, M., Xu, B., Warde-Farley, D., Ozair,
  S., Courville, A., \BBA\ Bengio, Y. \BBOP2014\BBCP.
\newblock \BBOQ Generative adversarial nets\BBCQ\
\newblock In {\Bem NeurIPS}, \BPGS\ 2672--2680.

\bibitem[\protect\BCAY{Grath, Costabello, Van, Sweeney, Kamiab, Shen,\ \BBA\
  Lecue}{Grath et~al.}{2018}]{grath2018interpretable}
Grath, R.~M., Costabello, L., Van, C.~L., Sweeney, P., Kamiab, F., Shen, Z.,
  \BBA\ Lecue, F. \BBOP2018\BBCP.
\newblock \BBOQ Interpretable credit application predictions with
  counterfactual explanations\BBCQ.

\bibitem[\protect\BCAY{Grover}{Grover}{2020}]{Safer2020}
Grover, V. \BBOP2020\BBCP\
\newblock
  \url{https://www.martechadvisor.com/articles/data-governance/data-dignity-for-better-data-privacy/}.

\bibitem[\protect\BCAY{Guess, Nagler,\ \BBA\ Tucker}{Guess
  et~al.}{2019}]{guess2019less}
Guess, A., Nagler, J., \BBA\ Tucker, J. \BBOP2019\BBCP.
\newblock \BBOQ Less than you think: Prevalence and predictors of fake news
  dissemination on facebook\BBCQ\
\newblock {\Bem Science Advances}, {\Bem 5\/}(1), eaau4586.

\bibitem[\protect\BCAY{Guo, Cheng, Li, Hahn,\ \BBA\ Liu}{Guo
  et~al.}{2020}]{guo2020survey}
Guo, R., Cheng, L., Li, J., Hahn, P.~R., \BBA\ Liu, H. \BBOP2020\BBCP.
\newblock \BBOQ A survey of learning causality with data: Problems and
  methods\BBCQ\
\newblock {\Bem CSUR}, {\Bem 53\/}(4), 1--37.

\bibitem[\protect\BCAY{Guynn}{Guynn}{2015}]{Guynn2015}
Guynn, J. \BBOP2015\BBCP\
\newblock
  \url{https://www.usatoday.com/story/tech/2015/07/01/google-apologizes-after-photos-identify-black-people-as-gorillas/29567465/}.

\bibitem[\protect\BCAY{Hardt, Price,\ \BBA\ Srebro}{Hardt
  et~al.}{2016}]{hardt2016equality}
Hardt, M., Price, E., \BBA\ Srebro, N. \BBOP2016\BBCP.
\newblock \BBOQ Equality of opportunity in supervised learning\BBCQ\
\newblock In {\Bem NeurIPS}, \BPGS\ 3315--3323.

\bibitem[\protect\BCAY{Hart}{Hart}{2019}]{Hart2019}
Hart, V. \BBOP2019\BBCP.
\newblock \BBOQ Data dignity at radicalxchange - the art of research\BBCQ\
\newblock \url{https://theartofresearch.org/data-dignity-at-radicalxchange/}.

\bibitem[\protect\BCAY{Hartigan\ \BBA\ Wong}{Hartigan\ \BBA\
  Wong}{1979}]{hartigan1979algorithm}
Hartigan, J.~A.\BBACOMMA\  \BBA\ Wong, M.~A. \BBOP1979\BBCP.
\newblock \BBOQ Algorithm as 136: A k-means clustering algorithm\BBCQ\
\newblock {\Bem Journal of the Royal Statistical Society. Series C (applied
  statistics)}, {\Bem 28\/}(1), 100--108.

\bibitem[\protect\BCAY{Hind, Houde, Martino, Mojsilovic, Piorkowski, Richards,\
  \BBA\ Varshney}{Hind et~al.}{2020}]{hind2020experiences}
Hind, M., Houde, S., Martino, J., Mojsilovic, A., Piorkowski, D., Richards, J.,
  \BBA\ Varshney, K.~R. \BBOP2020\BBCP.
\newblock \BBOQ Experiences with improving the transparency of {AI} models and
  services\BBCQ\
\newblock In {\Bem Extended Abstracts of the 2020 CHI Conference on Human
  Factors in Computing Systems}, \BPGS\ 1--8.

\bibitem[\protect\BCAY{Hirano, Imbens,\ \BBA\ Ridder}{Hirano
  et~al.}{2003}]{hirano2003efficient}
Hirano, K., Imbens, G.~W., \BBA\ Ridder, G. \BBOP2003\BBCP.
\newblock \BBOQ Efficient estimation of average treatment effects using the
  estimated propensity score\BBCQ\
\newblock {\Bem Econometrica}, {\Bem 71\/}(4), 1161--1189.

\bibitem[\protect\BCAY{Hitaj, Ateniese,\ \BBA\ Perez-Cruz}{Hitaj
  et~al.}{2017}]{hitaj2017deep}
Hitaj, B., Ateniese, G., \BBA\ Perez-Cruz, F. \BBOP2017\BBCP.
\newblock \BBOQ Deep models under the gan: information leakage from
  collaborative deep learning\BBCQ\
\newblock In {\Bem CCS}, \BPGS\ 603--618.

\bibitem[\protect\BCAY{Ho, Xie, Tang, Xu,\ \BBA\ Goh}{Ho
  et~al.}{2001}]{ho2001neural}
Ho, S., Xie, M., Tang, L., Xu, K., \BBA\ Goh, T. \BBOP2001\BBCP.
\newblock \BBOQ Neural network modeling with confidence bounds: a case study on
  the solder paste deposition process\BBCQ\
\newblock {\Bem IEEE Transactions on Electronics Packaging Manufacturing},
  {\Bem 24\/}(4), 323--332.

\bibitem[\protect\BCAY{Holzinger, Langs, Denk, Zatloukal,\ \BBA\
  M{\"u}ller}{Holzinger et~al.}{2019}]{holzinger2019causability}
Holzinger, A., Langs, G., Denk, H., Zatloukal, K., \BBA\ M{\"u}ller, H.
  \BBOP2019\BBCP.
\newblock \BBOQ Causability and explainability of artificial intelligence in
  medicine\BBCQ\
\newblock {\Bem Wiley Interdisciplinary Reviews: Data Mining and Knowledge
  Discovery}, {\Bem 9\/}(4), e1312.

\bibitem[\protect\BCAY{Homer, Szelinger, Redman, Duggan, Tembe, Muehling,
  Pearson, Stephan, Nelson,\ \BBA\ Craig}{Homer
  et~al.}{2008}]{homer2008resolving}
Homer, N., Szelinger, S., Redman, M., Duggan, D., Tembe, W., Muehling, J.,
  Pearson, J.~V., Stephan, D.~A., Nelson, S.~F., \BBA\ Craig, D.~W.
  \BBOP2008\BBCP.
\newblock \BBOQ Resolving individuals contributing trace amounts of dna to
  highly complex mixtures using high-density snp genotyping microarrays\BBCQ\
\newblock {\Bem PLoS Genet}, {\Bem 4\/}(8), e1000167.

\bibitem[\protect\BCAY{Horwitz}{Horwitz}{2021}]{Horwitz_2021}
Horwitz, J. \BBOP2021\BBCP.
\newblock \BBOQ Facebook algorithm shows gender bias in job ads, study
  finds\BBCQ\
\newblock
  \url{https://www.wsj.com/articles/facebook-shows-men-and-women-different-job-ads-study-finds-11617969600}.

\bibitem[\protect\BCAY{Hosen, Khosravi, Nahavandi,\ \BBA\ Creighton}{Hosen
  et~al.}{2014}]{hosen2014improving}
Hosen, M.~A., Khosravi, A., Nahavandi, S., \BBA\ Creighton, D. \BBOP2014\BBCP.
\newblock \BBOQ Improving the quality of prediction intervals through optimal
  aggregation\BBCQ\
\newblock {\Bem IEEE Transactions on Industrial Electronics}, {\Bem 62\/}(7),
  4420--4429.

\bibitem[\protect\BCAY{Hu\ \BBA\ Chen}{Hu\ \BBA\ Chen}{2018}]{hu2018short}
Hu, L.\BBACOMMA\  \BBA\ Chen, Y. \BBOP2018\BBCP.
\newblock \BBOQ A short-term intervention for long-term fairness in the labor
  market\BBCQ\
\newblock In {\Bem The Web Conference}, \BPGS\ 1389--1398.

\bibitem[\protect\BCAY{Huang, Singh,\ \BBA\ Atrey}{Huang
  et~al.}{2014}]{huang2014cyber}
Huang, Q., Singh, V.~K., \BBA\ Atrey, P.~K. \BBOP2014\BBCP.
\newblock \BBOQ Cyberbullying detection using social and textual analysis\BBCQ\
\newblock In {\Bem SAM}, \BPGS\ 3--6.

\bibitem[\protect\BCAY{Jagadish}{Jagadish}{2019}]{jagadish2019responsible}
Jagadish, V.~H. \BBOP2019\BBCP.
\newblock \BBOQ Responsible data science\BBCQ\
\newblock In {\Bem 2019 WSDM}, \BPGS\ 1--1. ACM.

\bibitem[\protect\BCAY{Janicka, Pszona,\ \BBA\ Wawer}{Janicka
  et~al.}{2019}]{janicka2019cross}
Janicka, M., Pszona, M., \BBA\ Wawer, A. \BBOP2019\BBCP.
\newblock \BBOQ Cross-domain failures of fake news detection\BBCQ\
\newblock {\Bem Computaci{\'o}n y Sistemas}, {\Bem 23\/}(3).

\bibitem[\protect\BCAY{Jiang\ \BBA\ Nachum}{Jiang\ \BBA\
  Nachum}{2020}]{jiang2020identifying}
Jiang, H.\BBACOMMA\  \BBA\ Nachum, O. \BBOP2020\BBCP.
\newblock \BBOQ Identifying and correcting label bias in machine learning\BBCQ\
\newblock In {\Bem AISTATS}, \BPGS\ 702--712.

\bibitem[\protect\BCAY{Joachims, Swaminathan,\ \BBA\ Schnabel}{Joachims
  et~al.}{2017}]{joachims2017unbiased}
Joachims, T., Swaminathan, A., \BBA\ Schnabel, T. \BBOP2017\BBCP.
\newblock \BBOQ Unbiased learning-to-rank with biased feedback\BBCQ\
\newblock In {\Bem WSDM}, \BPGS\ 781--789.

\bibitem[\protect\BCAY{Johnson, Foster,\ \BBA\ Stine}{Johnson
  et~al.}{2016}]{johnson2016impartial}
Johnson, K.~D., Foster, D.~P., \BBA\ Stine, R.~A. \BBOP2016\BBCP.
\newblock \BBOQ Impartial predictive modeling: Ensuring fairness in arbitrary
  models\BBCQ.

\bibitem[\protect\BCAY{Kabir, Khosravi, Hosen,\ \BBA\ Nahavandi}{Kabir
  et~al.}{2018}]{kabir2018neural}
Kabir, H.~D., Khosravi, A., Hosen, M.~A., \BBA\ Nahavandi, S. \BBOP2018\BBCP.
\newblock \BBOQ Neural network-based uncertainty quantification: A survey of
  methodologies and applications\BBCQ\
\newblock {\Bem IEEE Access}, {\Bem 6}, 36218--36234.

\bibitem[\protect\BCAY{Kairouz, Oh,\ \BBA\ Viswanath}{Kairouz
  et~al.}{2014}]{kairouz2014extremal}
Kairouz, P., Oh, S., \BBA\ Viswanath, P. \BBOP2014\BBCP.
\newblock \BBOQ Extremal mechanisms for local differential privacy\BBCQ\
\newblock In {\Bem NeurIPS}, \BPGS\ 2879--2887.

\bibitem[\protect\BCAY{Kamishima, Akaho, Asoh,\ \BBA\ Sakuma}{Kamishima
  et~al.}{2012}]{kamishima2012fairness}
Kamishima, T., Akaho, S., Asoh, H., \BBA\ Sakuma, J. \BBOP2012\BBCP.
\newblock \BBOQ Fairness-aware classifier with prejudice remover
  regularizer\BBCQ\
\newblock In {\Bem ECML PKDD}, \BPGS\ 35--50. Springer.

\bibitem[\protect\BCAY{Karimi, G{\'e}nois, Wagner, Singer,\ \BBA\
  Strohmaier}{Karimi et~al.}{2018}]{karimi2018homophily}
Karimi, F., G{\'e}nois, M., Wagner, C., Singer, P., \BBA\ Strohmaier, M.
  \BBOP2018\BBCP.
\newblock \BBOQ Homophily influences ranking of minorities in social
  networks\BBCQ\
\newblock {\Bem Scientific Reports}, {\Bem 8\/}(1), 1--12.

\bibitem[\protect\BCAY{Kasiviswanathan\ \BBA\ Sudheer}{Kasiviswanathan\ \BBA\
  Sudheer}{2016}]{kasiviswanathan2016comparison}
Kasiviswanathan, K.\BBACOMMA\  \BBA\ Sudheer, K. \BBOP2016\BBCP.
\newblock \BBOQ Comparison of methods used for quantifying prediction interval
  in artificial neural network hydrologic models\BBCQ\
\newblock {\Bem Modeling Earth Systems and Environment}, {\Bem 2\/}(1), 22.

\bibitem[\protect\BCAY{Khosravi, Nahavandi, Srinivasan,\ \BBA\
  Khosravi}{Khosravi et~al.}{2014}]{khosravi2014constructing}
Khosravi, A., Nahavandi, S., Srinivasan, D., \BBA\ Khosravi, R. \BBOP2014\BBCP.
\newblock \BBOQ Constructing optimal prediction intervals by using neural
  networks and bootstrap method\BBCQ\
\newblock {\Bem IEEE Transactions on Neural Networks and Learning Systems},
  {\Bem 26\/}(8), 1810--1815.

\bibitem[\protect\BCAY{Kim, Khanna,\ \BBA\ Koyejo}{Kim
  et~al.}{2016}]{kim2016examples}
Kim, B., Khanna, R., \BBA\ Koyejo, O.~O. \BBOP2016\BBCP.
\newblock \BBOQ Examples are not enough, learn to criticize! criticism for
  interpretability\BBCQ\
\newblock In {\Bem NeurIPS}, \BPGS\ 2280--2288.

\bibitem[\protect\BCAY{Knight}{Knight}{2020}]{Knight2020}
Knight, W. \BBOP2020\BBCP.
\newblock \BBOQ This company uses {AI} to outwit malicious {AI}\BBCQ\
\newblock
  \url{https://www.wired.com/story/company-uses-ai-outwit-malicious-ai/}.

\bibitem[\protect\BCAY{Koh\ \BBA\ Liang}{Koh\ \BBA\
  Liang}{2017}]{koh2017understanding}
Koh, P.~W.\BBACOMMA\  \BBA\ Liang, P. \BBOP2017\BBCP.
\newblock \BBOQ Understanding black-box predictions via influence
  functions\BBCQ.

\bibitem[\protect\BCAY{Kontostathis, Reynolds, Garron,\ \BBA\
  Edwards}{Kontostathis et~al.}{2013}]{kontostathis2013detecting}
Kontostathis, A., Reynolds, K., Garron, A., \BBA\ Edwards, L. \BBOP2013\BBCP.
\newblock \BBOQ Detecting cyberbullying: query terms and techniques\BBCQ\
\newblock In {\Bem Web Science}, \BPGS\ 195--204.

\bibitem[\protect\BCAY{Kraft\ \BBA\ Wang}{Kraft\ \BBA\
  Wang}{2009}]{kraft2009effectiveness}
Kraft, E.~M.\BBACOMMA\  \BBA\ Wang, J. \BBOP2009\BBCP.
\newblock \BBOQ Effectiveness of cyber bullying prevention strategies: A study
  on students' perspectives.\BBCQ\
\newblock {\Bem International Journal of Cyber Criminology}, {\Bem 3\/}(2).

\bibitem[\protect\BCAY{Krasanakis, Spyromitros-Xioufis, Papadopoulos,\ \BBA\
  Kompatsiaris}{Krasanakis et~al.}{2018}]{krasanakis2018adaptive}
Krasanakis, E., Spyromitros-Xioufis, E., Papadopoulos, S., \BBA\ Kompatsiaris,
  Y. \BBOP2018\BBCP.
\newblock \BBOQ Adaptive sensitive reweighting to mitigate bias in
  fairness-aware classification\BBCQ\
\newblock In {\Bem The Web Conference}, \BPGS\ 853--862.

\bibitem[\protect\BCAY{Kusner, Loftus, Russell,\ \BBA\ Silva}{Kusner
  et~al.}{2017}]{kusner2017counterfactual}
Kusner, M.~J., Loftus, J., Russell, C., \BBA\ Silva, R. \BBOP2017\BBCP.
\newblock \BBOQ Counterfactual fairness\BBCQ\
\newblock In {\Bem NeurIPS}, \BPGS\ 4066--4076.

\bibitem[\protect\BCAY{Lambrecht\ \BBA\ Tucker}{Lambrecht\ \BBA\
  Tucker}{2019}]{lambrecht2019algorithmic}
Lambrecht, A.\BBACOMMA\  \BBA\ Tucker, C. \BBOP2019\BBCP.
\newblock \BBOQ Algorithmic bias? an empirical study of apparent gender-based
  discrimination in the display of stem career ads\BBCQ\
\newblock {\Bem Management Science}, {\Bem 65\/}(7), 2966--2981.

\bibitem[\protect\BCAY{Lange, Blackman,\ \BBA\ Johnson}{Lange
  et~al.}{2001}]{lange2001speed}
Lange, J.~E., Blackman, K.~O., \BBA\ Johnson, M.~B. \BBOP2001\BBCP.
\newblock \BBOQ Speed violation survey of the new jersey turnpike: Final
  report\BBCQ.

\bibitem[\protect\BCAY{Lee, Resnick,\ \BBA\ Barton}{Lee
  et~al.}{2019}]{lee2019algorithmic}
Lee, N.~T., Resnick, P., \BBA\ Barton, G. \BBOP2019\BBCP.
\newblock \BBOQ Algorithmic bias detection and mitigation: Best practices and
  policies to reduce consumer harms\BBCQ.

\bibitem[\protect\BCAY{Legg, Hutter, et~al.}{Legg
  et~al.}{2007}]{legg2007collection}
Legg, S., Hutter, M., et~al. \BBOP2007\BBCP.
\newblock \BBOQ A collection of definitions of intelligence\BBCQ\
\newblock {\Bem FAIA}, {\Bem 157}, 17.

\bibitem[\protect\BCAY{Leslie}{Leslie}{2019}]{leslie2019understanding}
Leslie, D. \BBOP2019\BBCP.
\newblock \BBOQ Understanding artificial intelligence ethics and safety\BBCQ.

\bibitem[\protect\BCAY{Li, Cheng, Wang, Morstatter, Trevino, Tang,\ \BBA\
  Liu}{Li et~al.}{2017}]{li2017feature}
Li, J., Cheng, K., Wang, S., Morstatter, F., Trevino, R.~P., Tang, J., \BBA\
  Liu, H. \BBOP2017\BBCP.
\newblock \BBOQ Feature selection: A data perspective\BBCQ\
\newblock {\Bem CSUR}, {\Bem 50\/}(6), 1--45.

\bibitem[\protect\BCAY{Lipton}{Lipton}{2018}]{lipton2018mythos}
Lipton, Z.~C. \BBOP2018\BBCP.
\newblock \BBOQ The mythos of model interpretability\BBCQ\
\newblock {\Bem Queue}, {\Bem 16\/}(3), 31--57.

\bibitem[\protect\BCAY{List\ \BBA\ Momeni}{List\ \BBA\
  Momeni}{2018}]{list2018corporate}
List, J.~A.\BBACOMMA\  \BBA\ Momeni, F. \BBOP2018\BBCP.
\newblock \BBOQ When corporate social responsibility backfires: Theory and
  evidence from a natural field experiment\BBCQ\
\newblock \BTR, National Bureau of Economic Research.

\bibitem[\protect\BCAY{Liu, Juuti, Lu,\ \BBA\ Asokan}{Liu
  et~al.}{2017}]{liu2017oblivious}
Liu, J., Juuti, M., Lu, Y., \BBA\ Asokan, N. \BBOP2017\BBCP.
\newblock \BBOQ Oblivious neural network predictions via minionn
  transformations\BBCQ\
\newblock In {\Bem CCS}, \BPGS\ 619--631.

\bibitem[\protect\BCAY{Liu, Dean, Rolf, Simchowitz,\ \BBA\ Hardt}{Liu
  et~al.}{2018}]{liu2018delayed}
Liu, L.~T., Dean, S., Rolf, E., Simchowitz, M., \BBA\ Hardt, M. \BBOP2018\BBCP.
\newblock \BBOQ Delayed impact of fair machine learning\BBCQ.

\bibitem[\protect\BCAY{Liu, Kailkhura, Loveland,\ \BBA\ Han}{Liu
  et~al.}{2019}]{liu2019generative}
Liu, S., Kailkhura, B., Loveland, D., \BBA\ Han, Y. \BBOP2019\BBCP.
\newblock \BBOQ Generative counterfactual introspection for explainable deep
  learning\BBCQ.

\bibitem[\protect\BCAY{Loftus, Russell, Kusner,\ \BBA\ Silva}{Loftus
  et~al.}{2018}]{loftus2018causal}
Loftus, J.~R., Russell, C., Kusner, M.~J., \BBA\ Silva, R. \BBOP2018\BBCP.
\newblock \BBOQ Causal reasoning for algorithmic fairness\BBCQ.

\bibitem[\protect\BCAY{Lu\ \BBA\ Viljanen}{Lu\ \BBA\
  Viljanen}{2009}]{lu2009prediction}
Lu, T.\BBACOMMA\  \BBA\ Viljanen, M. \BBOP2009\BBCP.
\newblock \BBOQ Prediction of indoor temperature and relative humidity using
  neural network models: model comparison\BBCQ\
\newblock {\Bem Neural Computing and Applications}, {\Bem 18\/}(4), 345.

\bibitem[\protect\BCAY{Lundberg\ \BBA\ Lee}{Lundberg\ \BBA\
  Lee}{2017}]{lundberg2017unified}
Lundberg, S.~M.\BBACOMMA\  \BBA\ Lee, S.-I. \BBOP2017\BBCP.
\newblock \BBOQ A unified approach to interpreting model predictions\BBCQ\
\newblock In {\Bem NeurIPS}, \BPGS\ 4765--4774.

\bibitem[\protect\BCAY{Maaten\ \BBA\ Hinton}{Maaten\ \BBA\
  Hinton}{2008}]{maaten2008visualizing}
Maaten, L. v.~d.\BBACOMMA\  \BBA\ Hinton, G. \BBOP2008\BBCP.
\newblock \BBOQ Visualizing data using t-sne\BBCQ\
\newblock {\Bem JMLR}, {\Bem 9\/}(Nov), 2579--2605.

\bibitem[\protect\BCAY{Makri, Rotaru, Smart,\ \BBA\ Vercauteren}{Makri
  et~al.}{2019}]{makri2019epic}
Makri, E., Rotaru, D., Smart, N.~P., \BBA\ Vercauteren, F. \BBOP2019\BBCP.
\newblock \BBOQ {EPIC}: efficient private image classification (or: Learning
  from the masters)\BBCQ\
\newblock In {\Bem CT-RSA}, \BPGS\ 473--492. Springer.

\bibitem[\protect\BCAY{Malekzadeh, Clegg, Cavallaro,\ \BBA\ Haddadi}{Malekzadeh
  et~al.}{2018}]{malekzadeh2018protecting}
Malekzadeh, M., Clegg, R.~G., Cavallaro, A., \BBA\ Haddadi, H. \BBOP2018\BBCP.
\newblock \BBOQ Protecting sensory data against sensitive inferences\BBCQ\
\newblock In {\Bem Proceedings of the 1st Workshop on Privacy by Design in
  Distributed Systems}, \BPGS\ 1--6.

\bibitem[\protect\BCAY{Malekzadeh, Clegg, Cavallaro,\ \BBA\ Haddadi}{Malekzadeh
  et~al.}{2019}]{malekzadeh2019mobile}
Malekzadeh, M., Clegg, R.~G., Cavallaro, A., \BBA\ Haddadi, H. \BBOP2019\BBCP.
\newblock \BBOQ Mobile sensor data anonymization\BBCQ.

\bibitem[\protect\BCAY{Malekzadeh, Clegg, Cavallaro,\ \BBA\ Haddadi}{Malekzadeh
  et~al.}{2020}]{malekzadeh2020privacy}
Malekzadeh, M., Clegg, R.~G., Cavallaro, A., \BBA\ Haddadi, H. \BBOP2020\BBCP.
\newblock \BBOQ Privacy and utility preserving sensor-data
  transformations\BBCQ.

\bibitem[\protect\BCAY{Mar{\'\i}n, Valencia,\ \BBA\ S{\'a}ez}{Mar{\'\i}n
  et~al.}{2016}]{marin2016prediction}
Mar{\'\i}n, L.~G., Valencia, F., \BBA\ S{\'a}ez, D. \BBOP2016\BBCP.
\newblock \BBOQ Prediction interval based on type-2 fuzzy systems for wind
  power generation and loads in microgrid control design\BBCQ\
\newblock In {\Bem 2016 FUZZ-IEEE}, \BPGS\ 328--335. IEEE.

\bibitem[\protect\BCAY{Marr}{Marr}{2021}]{Marr_2021}
Marr, B. \BBOP2021\BBCP\
\newblock
  \url{https://bernardmarr.com/how-much-data-do-we-create-every-day-the-mind-blowing-stats-everyone-should-read/}.

\bibitem[\protect\BCAY{McConnell\ \BBA\ Scheidegger}{McConnell\ \BBA\
  Scheidegger}{2001}]{mcconnell2001race}
McConnell, E.~H.\BBACOMMA\  \BBA\ Scheidegger, A.~R. \BBOP2001\BBCP.
\newblock \BBOQ Race and speeding citations: Comparing speeding citations
  issued by air traffic officers with those issued by ground traffic
  officers\BBCQ\
\newblock In {\Bem ACJS, Washington, DC}.

\bibitem[\protect\BCAY{Mehrabi, Morstatter, Saxena, Lerman,\ \BBA\
  Galstyan}{Mehrabi et~al.}{2019}]{mehrabi2019survey}
Mehrabi, N., Morstatter, F., Saxena, N., Lerman, K., \BBA\ Galstyan, A.
  \BBOP2019\BBCP.
\newblock \BBOQ A survey on bias and fairness in machine learning\BBCQ.

\bibitem[\protect\BCAY{Menon\ \BBA\ Williamson}{Menon\ \BBA\
  Williamson}{2018}]{menon2018cost}
Menon, A.~K.\BBACOMMA\  \BBA\ Williamson, R.~C. \BBOP2018\BBCP.
\newblock \BBOQ The cost of fairness in binary classification\BBCQ\
\newblock In {\Bem FAT*}, \BPGS\ 107--118.

\bibitem[\protect\BCAY{Miller~C}{Miller~C}{2019}]{people2019}
Miller~C, C.~R. \BBOP2019\BBCP.
\newblock \BBOQ People, power and technology: The tech workers’ view\BBCQ\
\newblock
  \url{https://doteveryone.org.uk/wp-content/uploads/2019/04/PeoplePowerTech_Doteveryone_May2019.pdf}.

\bibitem[\protect\BCAY{Mills, Baltassis, Santinelli, Carlisi, Duranton,\ \BBA\
  Gallego}{Mills et~al.}{2020}]{six2020}
Mills, S., Baltassis, E., Santinelli, M., Carlisi, C., Duranton, S., \BBA\
  Gallego, A. \BBOP2020\BBCP.
\newblock \BBOQ Six steps to bridge the responsible {AI} gap\BBCQ\
\newblock
  \url{https://www.bcg.com/publications/2020/six-steps-for-socially-responsible-artificial-intelligence}.

\bibitem[\protect\BCAY{Mirshghallah, Taram, Vepakomma, Singh, Raskar,\ \BBA\
  Esmaeilzadeh}{Mirshghallah et~al.}{2020}]{mirshghallah2020privacy}
Mirshghallah, F., Taram, M., Vepakomma, P., Singh, A., Raskar, R., \BBA\
  Esmaeilzadeh, H. \BBOP2020\BBCP.
\newblock \BBOQ Privacy in deep learning: A survey\BBCQ.

\bibitem[\protect\BCAY{Mishna, Khoury-Kassabri, Gadalla,\ \BBA\ Daciuk}{Mishna
  et~al.}{2012}]{mishna2012risk}
Mishna, F., Khoury-Kassabri, M., Gadalla, T., \BBA\ Daciuk, J. \BBOP2012\BBCP.
\newblock \BBOQ Risk factors for involvement in cyber bullying: Victims,
  bullies and bully--victims\BBCQ\
\newblock {\Bem Children and Youth Services Review}, {\Bem 34\/}(1), 63--70.

\bibitem[\protect\BCAY{Mitchell, Wu, Zaldivar, Barnes, Vasserman, Hutchinson,
  Spitzer, Raji,\ \BBA\ Gebru}{Mitchell et~al.}{2019}]{mitchell2019model}
Mitchell, M., Wu, S., Zaldivar, A., Barnes, P., Vasserman, L., Hutchinson, B.,
  Spitzer, E., Raji, I.~D., \BBA\ Gebru, T. \BBOP2019\BBCP.
\newblock \BBOQ Model cards for model reporting\BBCQ\
\newblock In {\Bem FAT*}, \BPGS\ 220--229.

\bibitem[\protect\BCAY{Mokhberian, Abeliuk, Cummings,\ \BBA\ Lerman}{Mokhberian
  et~al.}{2020}]{mokhberian2020moral}
Mokhberian, N., Abeliuk, A., Cummings, P., \BBA\ Lerman, K. \BBOP2020\BBCP.
\newblock \BBOQ Moral framing and ideological bias of news\BBCQ\
\newblock In {\Bem SocInfo}, \BPGS\ 206--219. Springer.

\bibitem[\protect\BCAY{Molnar}{Molnar}{2020}]{molnar2020interpretable}
Molnar, C. \BBOP2020\BBCP.
\newblock {\Bem Interpretable Machine Learning}.
\newblock Lulu. com.

\bibitem[\protect\BCAY{Moraffah, Karami, Guo, Raglin,\ \BBA\ Liu}{Moraffah
  et~al.}{2020}]{moraffah2020causal}
Moraffah, R., Karami, M., Guo, R., Raglin, A., \BBA\ Liu, H. \BBOP2020\BBCP.
\newblock \BBOQ Causal interpretability for machine learning-problems, methods
  and evaluation\BBCQ\
\newblock {\Bem SIGKDD Explorations}, {\Bem 22\/}(1), 18--33.

\bibitem[\protect\BCAY{Mordvintsev, Olah,\ \BBA\ Tyka}{Mordvintsev
  et~al.}{2015}]{mordvintsev2015inceptionism}
Mordvintsev, A., Olah, C., \BBA\ Tyka, M. \BBOP2015\BBCP.
\newblock \BBOQ Inceptionism: Going deeper into neural networks\BBCQ.

\bibitem[\protect\BCAY{Nabi\ \BBA\ Shpitser}{Nabi\ \BBA\
  Shpitser}{2018}]{nabi2018fair}
Nabi, R.\BBACOMMA\  \BBA\ Shpitser, I. \BBOP2018\BBCP.
\newblock \BBOQ Fair inference on outcomes\BBCQ\
\newblock In {\Bem AAAI}, \lowercase{\BVOL}\ 2018, \BPG\ 1931. NIH Public
  Access.

\bibitem[\protect\BCAY{Narayanan\ \BBA\ Shmatikov}{Narayanan\ \BBA\
  Shmatikov}{2008}]{narayanan2008robust}
Narayanan, A.\BBACOMMA\  \BBA\ Shmatikov, V. \BBOP2008\BBCP.
\newblock \BBOQ Robust de-anonymization of large sparse datasets\BBCQ\
\newblock In {\Bem 2008 IEEE sp}, \BPGS\ 111--125. IEEE.

\bibitem[\protect\BCAY{Narendra, Sankaran, Vijaykeerthy,\ \BBA\ Mani}{Narendra
  et~al.}{2018}]{narendra2018explaining}
Narendra, T., Sankaran, A., Vijaykeerthy, D., \BBA\ Mani, S. \BBOP2018\BBCP.
\newblock \BBOQ Explaining deep learning models using causal inference\BBCQ.

\bibitem[\protect\BCAY{Nguyen, Yan, Thai,\ \BBA\ Eidenbenz}{Nguyen
  et~al.}{2012}]{nguyen2012containment}
Nguyen, N.~P., Yan, G., Thai, M.~T., \BBA\ Eidenbenz, S. \BBOP2012\BBCP.
\newblock \BBOQ Containment of misinformation spread in online social
  networks\BBCQ\
\newblock In {\Bem Web Science}, \BPGS\ 213--222.

\bibitem[\protect\BCAY{Ott, Choi, Cardie,\ \BBA\ Hancock}{Ott
  et~al.}{2011}]{ott2011finding}
Ott, M., Choi, Y., Cardie, C., \BBA\ Hancock, J.~T. \BBOP2011\BBCP.
\newblock \BBOQ Finding deceptive opinion spam by any stretch of the
  imagination\BBCQ.

\bibitem[\protect\BCAY{Pacepa\ \BBA\ Rychlak}{Pacepa\ \BBA\
  Rychlak}{2013}]{pacepa2013disinformation}
Pacepa, I.~M.\BBACOMMA\  \BBA\ Rychlak, R.~J. \BBOP2013\BBCP.
\newblock {\Bem Disinformation: Former Spy Chief Reveals Secret Strategy for
  Undermining Freedom, Attacking Religion, and Promoting Terrorism}.
\newblock Wnd Books.

\bibitem[\protect\BCAY{Papernot, McDaniel, Sinha,\ \BBA\ Wellman}{Papernot
  et~al.}{2016a}]{papernot2016towards}
Papernot, N., McDaniel, P., Sinha, A., \BBA\ Wellman, M. \BBOP2016a\BBCP.
\newblock \BBOQ Towards the science of security and privacy in machine
  learning\BBCQ.

\bibitem[\protect\BCAY{Papernot, McDaniel, Wu, Jha,\ \BBA\ Swami}{Papernot
  et~al.}{2016b}]{papernot2016distillation}
Papernot, N., McDaniel, P., Wu, X., Jha, S., \BBA\ Swami, A. \BBOP2016b\BBCP.
\newblock \BBOQ Distillation as a defense to adversarial perturbations against
  deep neural networks\BBCQ\
\newblock In {\Bem IEEE Symposium on SP}, \BPGS\ 582--597. IEEE.

\bibitem[\protect\BCAY{Parafita\ \BBA\ Vitri{\`a}}{Parafita\ \BBA\
  Vitri{\`a}}{2019}]{parafita2019explaining}
Parafita, {\'A}.\BBACOMMA\  \BBA\ Vitri{\`a}, J. \BBOP2019\BBCP.
\newblock \BBOQ Explaining visual models by causal attribution\BBCQ.

\bibitem[\protect\BCAY{Pearl}{Pearl}{2009}]{pearl2009causality}
Pearl, J. \BBOP2009\BBCP.
\newblock {\Bem Causality}.
\newblock Cambridge university press.

\bibitem[\protect\BCAY{Pearl}{Pearl}{2018}]{pearl2018theoretical}
Pearl, J. \BBOP2018\BBCP.
\newblock \BBOQ Theoretical impediments to machine learning with seven sparks
  from the causal revolution\BBCQ.

\bibitem[\protect\BCAY{Pearl}{Pearl}{2019}]{pearl2019seven}
Pearl, J. \BBOP2019\BBCP.
\newblock \BBOQ The seven tools of causal inference, with reflections on
  machine learning\BBCQ\
\newblock {\Bem Communications of the ACM}, {\Bem 62\/}(3), 54--60.

\bibitem[\protect\BCAY{Peters, Janzing,\ \BBA\ Sch{\"o}lkopf}{Peters
  et~al.}{2017}]{peters2017elements}
Peters, J., Janzing, D., \BBA\ Sch{\"o}lkopf, B. \BBOP2017\BBCP.
\newblock {\Bem Elements of causal inference}.
\newblock The MIT Press.

\bibitem[\protect\BCAY{Pfohl, Foryciarz,\ \BBA\ Shah}{Pfohl
  et~al.}{2020}]{pfohl2020empirical}
Pfohl, S.~R., Foryciarz, A., \BBA\ Shah, N.~H. \BBOP2020\BBCP.
\newblock \BBOQ An empirical characterization of fair machine learning for
  clinical risk prediction\BBCQ.

\bibitem[\protect\BCAY{Pinceti, Kosut,\ \BBA\ Sankar}{Pinceti
  et~al.}{2019}]{pinceti2019data}
Pinceti, A., Kosut, O., \BBA\ Sankar, L. \BBOP2019\BBCP.
\newblock \BBOQ Data-driven generation of synthetic load datasets preserving
  spatio-temporal features\BBCQ\
\newblock In {\Bem 2019 IEEE PESGM}, \BPGS\ 1--5. IEEE.

\bibitem[\protect\BCAY{Qian, Gong, Sharma,\ \BBA\ Liu}{Qian
  et~al.}{2018}]{qian2018neural}
Qian, F., Gong, C., Sharma, K., \BBA\ Liu, Y. \BBOP2018\BBCP.
\newblock \BBOQ Neural user response generator: Fake news detection with
  collective user intelligence.\BBCQ\
\newblock In {\Bem IJCAI}, \lowercase{\BVOL}~18, \BPGS\ 3834--3840.

\bibitem[\protect\BCAY{Quan, Srinivasan,\ \BBA\ Khosravi}{Quan
  et~al.}{2014}]{quan2014particle}
Quan, H., Srinivasan, D., \BBA\ Khosravi, A. \BBOP2014\BBCP.
\newblock \BBOQ Particle swarm optimization for construction of neural
  network-based prediction intervals\BBCQ\
\newblock {\Bem Neurocomputing}, {\Bem 127}, 172--180.

\bibitem[\protect\BCAY{Rajabi, Shehu,\ \BBA\ Purohit}{Rajabi
  et~al.}{2019}]{rajabi2019user}
Rajabi, Z., Shehu, A., \BBA\ Purohit, H. \BBOP2019\BBCP.
\newblock \BBOQ User behavior modelling for fake information mitigation on
  social web\BBCQ\
\newblock In {\Bem SBP-BRiMS}, \BPGS\ 234--244. Springer.

\bibitem[\protect\BCAY{Ribeiro, Singh,\ \BBA\ Guestrin}{Ribeiro
  et~al.}{2016}]{ribeiro2016should}
Ribeiro, M.~T., Singh, S., \BBA\ Guestrin, C. \BBOP2016\BBCP.
\newblock \BBOQ ``{W}hy should {I} trust you?'' explaining the predictions of
  any classifier\BBCQ\
\newblock In {\Bem KDD}, \BPGS\ 1135--1144.

\bibitem[\protect\BCAY{Rivero}{Rivero}{2020}]{Rivero2020}
Rivero, N. \BBOP2020\BBCP.
\newblock \BBOQ Google showed us the danger of letting corporations lead {AI}
  research\BBCQ\
\newblock
  \url{https://qz.com/1945293/the-dangers-of-letting-google-lead-ai-research/}.

\bibitem[\protect\BCAY{Robnik-{\v{S}}ikonja\ \BBA\
  Bohanec}{Robnik-{\v{S}}ikonja\ \BBA\ Bohanec}{2018}]{robnik2018perturbation}
Robnik-{\v{S}}ikonja, M.\BBACOMMA\  \BBA\ Bohanec, M. \BBOP2018\BBCP.
\newblock \BBOQ Perturbation-based explanations of prediction models\BBCQ\
\newblock In {\Bem Human and Machine Learning}, \BPGS\ 159--175. Springer.

\bibitem[\protect\BCAY{Rosenbaum\ \BBA\ Rubin}{Rosenbaum\ \BBA\
  Rubin}{1983}]{rosenbaum1983central}
Rosenbaum, P.~R.\BBACOMMA\  \BBA\ Rubin, D.~B. \BBOP1983\BBCP.
\newblock \BBOQ The central role of the propensity score in observational
  studies for causal effects\BBCQ\
\newblock {\Bem Biometrika}, {\Bem 70\/}(1), 41--55.

\bibitem[\protect\BCAY{Rubin}{Rubin}{1974}]{rubin1974estimating}
Rubin, D.~B. \BBOP1974\BBCP.
\newblock \BBOQ Estimating causal effects of treatments in randomized and
  nonrandomized studies\BBCQ\
\newblock {\Bem Journal of Educational Psychology}, {\Bem 66\/}(5), 688.

\bibitem[\protect\BCAY{Rudin}{Rudin}{2019}]{rudin2019stop}
Rudin, C. \BBOP2019\BBCP.
\newblock \BBOQ Stop explaining black box machine learning models for high
  stakes decisions and use interpretable models instead\BBCQ\
\newblock {\Bem Nature Machine Intelligence}, {\Bem 1\/}(5), 206--215.

\bibitem[\protect\BCAY{Salawu, He,\ \BBA\ Lumsden}{Salawu
  et~al.}{2017}]{salawu2017approaches}
Salawu, S., He, Y., \BBA\ Lumsden, J. \BBOP2017\BBCP.
\newblock \BBOQ Approaches to automated detection of cyberbullying: A
  survey\BBCQ.

\bibitem[\protect\BCAY{Sattigeri, Hoffman, Chenthamarakshan,\ \BBA\
  Varshney}{Sattigeri et~al.}{2019}]{sattigeri2019fairness}
Sattigeri, P., Hoffman, S.~C., Chenthamarakshan, V., \BBA\ Varshney, K.~R.
  \BBOP2019\BBCP.
\newblock \BBOQ Fairness gan: Generating datasets with fairness properties
  using a generative adversarial network\BBCQ\
\newblock {\Bem IBM Journal of Research and Development}, {\Bem 63\/}(4/5),
  3--1.

\bibitem[\protect\BCAY{Schaefer, Boche, Khisti,\ \BBA\ Poor}{Schaefer
  et~al.}{2017}]{schaefer2017information}
Schaefer, R.~F., Boche, H., Khisti, A., \BBA\ Poor, H.~V. \BBOP2017\BBCP.
\newblock {\Bem Information Theoretic Security and Privacy of Information
  Systems}.
\newblock Cambridge University Press.

\bibitem[\protect\BCAY{Sch{\"o}lkopf}{Sch{\"o}lkopf}{2019}]{scholkopf2019causality}
Sch{\"o}lkopf, B. \BBOP2019\BBCP.
\newblock \BBOQ Causality for machine learning\BBCQ.

\bibitem[\protect\BCAY{Schwab}{Schwab}{2021}]{Schwab_2021}
Schwab, K. \BBOP2021\BBCP.
\newblock \BBOQ ‘this is bigger than just timnit’: How google tried to
  silence a critic and ignited a movement\BBCQ\
\newblock
  \url{https://www.fastcompany.com/90608471/timnit-gebru-google-ai-ethics-equitable-tech-movement}.

\bibitem[\protect\BCAY{Selbst, Boyd, Friedler, Venkatasubramanian,\ \BBA\
  Vertesi}{Selbst et~al.}{2019}]{selbst2019fairness}
Selbst, A.~D., Boyd, D., Friedler, S.~A., Venkatasubramanian, S., \BBA\
  Vertesi, J. \BBOP2019\BBCP.
\newblock \BBOQ Fairness and abstraction in sociotechnical systems\BBCQ\
\newblock In {\Bem FAT*}, \BPGS\ 59--68.

\bibitem[\protect\BCAY{Shaul}{Shaul}{2015}]{Br_2015}
Shaul, B. \BBOP2015\BBCP.
\newblock \BBOQ Honestly looks to combat cyberbullying on ios, android\BBCQ\
\newblock
  \url{https://www.adweek.com/performance-marketing/honestly-looks-to-combat-cyberbullying-on-ios-android/}.

\bibitem[\protect\BCAY{Shokri, Stronati, Song,\ \BBA\ Shmatikov}{Shokri
  et~al.}{2017}]{shokri2017membership}
Shokri, R., Stronati, M., Song, C., \BBA\ Shmatikov, V. \BBOP2017\BBCP.
\newblock \BBOQ Membership inference attacks against machine learning
  models\BBCQ\
\newblock In {\Bem 2017 IEEE Symposium on SP}, \BPGS\ 3--18. IEEE.

\bibitem[\protect\BCAY{Shu, Cui, Wang, Lee,\ \BBA\ Liu}{Shu
  et~al.}{2019}]{shu2019defend}
Shu, K., Cui, L., Wang, S., Lee, D., \BBA\ Liu, H. \BBOP2019\BBCP.
\newblock \BBOQ d{EFEND}: Explainable fake news detection\BBCQ\
\newblock In {\Bem KDD}, \BPGS\ 395--405.

\bibitem[\protect\BCAY{Shu, Mahudeswaran, Wang, Lee,\ \BBA\ Liu}{Shu
  et~al.}{2020}]{shu2020fakenewsnet}
Shu, K., Mahudeswaran, D., Wang, S., Lee, D., \BBA\ Liu, H. \BBOP2020\BBCP.
\newblock \BBOQ Fakenewsnet: A data repository with news content, social
  context, and spatiotemporal information for studying fake news on social
  media\BBCQ\
\newblock {\Bem Big Data}, {\Bem 8\/}(3), 171--188.

\bibitem[\protect\BCAY{Shu, Wang,\ \BBA\ Liu}{Shu
  et~al.}{2018}]{shu2018understanding}
Shu, K., Wang, S., \BBA\ Liu, H. \BBOP2018\BBCP.
\newblock \BBOQ Understanding user profiles on social media for fake news
  detection\BBCQ\
\newblock In {\Bem 2018 IEEE MIPR}, \BPGS\ 430--435. IEEE.

\bibitem[\protect\BCAY{Shu, Wang,\ \BBA\ Liu}{Shu et~al.}{2019}]{shu2019beyond}
Shu, K., Wang, S., \BBA\ Liu, H. \BBOP2019\BBCP.
\newblock \BBOQ Beyond news contents: The role of social context for fake news
  detection\BBCQ\
\newblock In {\Bem WSDM}, \BPGS\ 312--320.

\bibitem[\protect\BCAY{Shu, Zheng, Li, Mukherjee, Awadallah, Ruston,\ \BBA\
  Liu}{Shu et~al.}{2020}]{shu2020leveraging}
Shu, K., Zheng, G., Li, Y., Mukherjee, S., Awadallah, A.~H., Ruston, S., \BBA\
  Liu, H. \BBOP2020\BBCP.
\newblock \BBOQ Leveraging multi-source weak social supervision for early
  detection of fake news\BBCQ.

\bibitem[\protect\BCAY{Siegel}{Siegel}{2020}]{social2020}
Siegel, E. \BBOP2020\BBCP\
\newblock
  \url{https://www.kdnuggets.com/how-machine-learning-works-for-social-good.html/}.

\bibitem[\protect\BCAY{Silva, Rich,\ \BBA\ Hall}{Silva
  et~al.}{2016}]{silva2016bullyblocker}
Silva, Y.~N., Rich, C., \BBA\ Hall, D. \BBOP2016\BBCP.
\newblock \BBOQ Bullyblocker: Towards the identification of cyberbullying in
  social networking sites\BBCQ\
\newblock In {\Bem ASONAM}, \BPGS\ 1377--1379. IEEE.

\bibitem[\protect\BCAY{Simonyan, Vedaldi,\ \BBA\ Zisserman}{Simonyan
  et~al.}{2013}]{simonyan2013deep}
Simonyan, K., Vedaldi, A., \BBA\ Zisserman, A. \BBOP2013\BBCP.
\newblock \BBOQ Deep inside convolutional networks: Visualising image
  classification models and saliency maps\BBCQ.

\bibitem[\protect\BCAY{Singh, Vatsa,\ \BBA\ Ratha}{Singh
  et~al.}{2021}]{singh2021trustworthy}
Singh, R., Vatsa, M., \BBA\ Ratha, N. \BBOP2021\BBCP.
\newblock \BBOQ Trustworthy {AI}\BBCQ\
\newblock In {\Bem 8th ACM IKDD CODS and 26th COMAD}, \BPGS\ 449--453.

\bibitem[\protect\BCAY{Slack, Hilgard, Jia, Singh,\ \BBA\ Lakkaraju}{Slack
  et~al.}{2020}]{slack2020fooling}
Slack, D., Hilgard, S., Jia, E., Singh, S., \BBA\ Lakkaraju, H. \BBOP2020\BBCP.
\newblock \BBOQ Fooling lime and shap: Adversarial attacks on post hoc
  explanation methods\BBCQ\
\newblock In {\Bem AIES}, \BPGS\ 180--186.

\bibitem[\protect\BCAY{Smith, Mahdavi, Carvalho, Fisher, Russell,\ \BBA\
  Tippett}{Smith et~al.}{2008}]{smith2008cyberbullying}
Smith, P.~K., Mahdavi, J., Carvalho, M., Fisher, S., Russell, S., \BBA\
  Tippett, N. \BBOP2008\BBCP.
\newblock \BBOQ Cyberbullying: Its nature and impact in secondary school
  pupils\BBCQ\
\newblock {\Bem Journal of Child Psychology and Psychiatry}, {\Bem 49\/}(4),
  376--385.

\bibitem[\protect\BCAY{St{\aa}hl, Falkman, Karlsson,\ \BBA\
  Mathiason}{St{\aa}hl et~al.}{2020}]{staahl2020evaluation}
St{\aa}hl, N., Falkman, G., Karlsson, A., \BBA\ Mathiason, G. \BBOP2020\BBCP.
\newblock \BBOQ Evaluation of uncertainty quantification in deep learning\BBCQ\
\newblock In {\Bem IPMU}, \BPGS\ 556--568. Springer.

\bibitem[\protect\BCAY{Stoica, Riederer,\ \BBA\ Chaintreau}{Stoica
  et~al.}{2018}]{stoica2018algorithmic}
Stoica, A.-A., Riederer, C., \BBA\ Chaintreau, A. \BBOP2018\BBCP.
\newblock \BBOQ Algorithmic glass ceiling in social networks: The effects of
  social recommendations on network diversity\BBCQ\
\newblock In {\Bem The Web Conference}, \BPGS\ 923--932.

\bibitem[\protect\BCAY{S{\"u}hr, Hilgard,\ \BBA\ Lakkaraju}{S{\"u}hr
  et~al.}{2020}]{suhr2020does}
S{\"u}hr, T., Hilgard, S., \BBA\ Lakkaraju, H. \BBOP2020\BBCP.
\newblock \BBOQ Does fair ranking improve minority outcomes? understanding the
  interplay of human and algorithmic biases in online hiring\BBCQ.

\bibitem[\protect\BCAY{Suter, Miladinovic, Sch{\"o}lkopf,\ \BBA\ Bauer}{Suter
  et~al.}{2019}]{suter2019robustly}
Suter, R., Miladinovic, D., Sch{\"o}lkopf, B., \BBA\ Bauer, S. \BBOP2019\BBCP.
\newblock \BBOQ Robustly disentangled causal mechanisms: Validating deep
  representations for interventional robustness\BBCQ\
\newblock In {\Bem ICML}, \BPGS\ 6056--6065. PMLR.

\bibitem[\protect\BCAY{Sweeney}{Sweeney}{2002}]{sweeney2002k}
Sweeney, L. \BBOP2002\BBCP.
\newblock \BBOQ k-anonymity: A model for protecting privacy\BBCQ\
\newblock {\Bem IJUFKS}, {\Bem 10\/}(05), 557--570.

\bibitem[\protect\BCAY{Thiebes, Lins,\ \BBA\ Sunyaev}{Thiebes
  et~al.}{2020}]{thiebes2020trustworthy}
Thiebes, S., Lins, S., \BBA\ Sunyaev, A. \BBOP2020\BBCP.
\newblock \BBOQ Trustworthy artificial intelligence\BBCQ.

\bibitem[\protect\BCAY{Tjoa\ \BBA\ Guan}{Tjoa\ \BBA\ Guan}{2020}]{tjoa2019XAI}
Tjoa, E.\BBACOMMA\  \BBA\ Guan, C. \BBOP2020\BBCP.
\newblock \BBOQ A survey on explainable artificial intelligence (xai): Toward
  medical xai\BBCQ.

\bibitem[\protect\BCAY{Tong, Du,\ \BBA\ Wu}{Tong
  et~al.}{2018}]{tong2018misinformation}
Tong, A., Du, D.-Z., \BBA\ Wu, W. \BBOP2018\BBCP.
\newblock \BBOQ On misinformation containment in online social networks\BBCQ\
\newblock In {\Bem NeurIPS}, \BPGS\ 341--351.

\bibitem[\protect\BCAY{Tram{\`e}r, Zhang, Juels, Reiter,\ \BBA\
  Ristenpart}{Tram{\`e}r et~al.}{2016}]{tramer2016stealing}
Tram{\`e}r, F., Zhang, F., Juels, A., Reiter, M.~K., \BBA\ Ristenpart, T.
  \BBOP2016\BBCP.
\newblock \BBOQ Stealing machine learning models via prediction apis\BBCQ\
\newblock In {\Bem USENIX Security}, \BPGS\ 601--618.

\bibitem[\protect\BCAY{Ungar, De~Veaux,\ \BBA\ Rosengarten}{Ungar
  et~al.}{1996}]{ungar1996estimating}
Ungar, L.~H., De~Veaux, R.~D., \BBA\ Rosengarten, E. \BBOP1996\BBCP.
\newblock \BBOQ Estimating prediction intervals for artificial neural
  networks\BBCQ\
\newblock In {\Bem Proc. of the 9th Yale WALS}.

\bibitem[\protect\BCAY{Varodayan\ \BBA\ Khisti}{Varodayan\ \BBA\
  Khisti}{2011}]{varodayan2011smart}
Varodayan, D.\BBACOMMA\  \BBA\ Khisti, A. \BBOP2011\BBCP.
\newblock \BBOQ Smart meter privacy using a rechargeable battery: Minimizing
  the rate of information leakage\BBCQ\
\newblock In {\Bem ICASSP}, \BPGS\ 1932--1935. IEEE.

\bibitem[\protect\BCAY{Varshney}{Varshney}{2019}]{varshney2019trustworthy}
Varshney, K.~R. \BBOP2019\BBCP.
\newblock \BBOQ Trustworthy machine learning and artificial intelligence\BBCQ\
\newblock {\Bem XRDS: Crossroads, The ACM Magazine for Students}, {\Bem
  25\/}(3), 26--29.

\bibitem[\protect\BCAY{Varshney\ \BBA\ Alemzadeh}{Varshney\ \BBA\
  Alemzadeh}{2017}]{varshney2017safety}
Varshney, K.~R.\BBACOMMA\  \BBA\ Alemzadeh, H. \BBOP2017\BBCP.
\newblock \BBOQ On the safety of machine learning: Cyber-physical systems,
  decision sciences, and data products\BBCQ\
\newblock {\Bem Big Data}, {\Bem 5\/}(3), 246--255.

\bibitem[\protect\BCAY{Vishwamitra, Zhang, Tong, Hu, Luo, Kowalski,\ \BBA\
  Mazer}{Vishwamitra et~al.}{2017}]{vishwamitra2017mcdefender}
Vishwamitra, N., Zhang, X., Tong, J., Hu, H., Luo, F., Kowalski, R., \BBA\
  Mazer, J. \BBOP2017\BBCP.
\newblock \BBOQ Mcdefender: Toward effective cyberbullying defense in mobile
  online social networks\BBCQ\
\newblock In {\Bem CODASPY}, \BPGS\ 37--42.

\bibitem[\protect\BCAY{Vosoughi, Roy,\ \BBA\ Aral}{Vosoughi
  et~al.}{2018}]{vosoughi2018spread}
Vosoughi, S., Roy, D., \BBA\ Aral, S. \BBOP2018\BBCP.
\newblock \BBOQ The spread of true and false news online\BBCQ\
\newblock {\Bem Science}, {\Bem 359\/}(6380), 1146--1151.

\bibitem[\protect\BCAY{Wachter, Mittelstadt,\ \BBA\ Russell}{Wachter
  et~al.}{2017}]{wachter2017counterfactual}
Wachter, S., Mittelstadt, B., \BBA\ Russell, C. \BBOP2017\BBCP.
\newblock \BBOQ Counterfactual explanations without opening the black box:
  Automated decisions and the gdpr\BBCQ\
\newblock {\Bem Harv. JL \& Tech.}, {\Bem 31}, 841.

\bibitem[\protect\BCAY{Wan, Xu, Pinson, Dong,\ \BBA\ Wong}{Wan
  et~al.}{2013}]{wan2013probabilistic}
Wan, C., Xu, Z., Pinson, P., Dong, Z.~Y., \BBA\ Wong, K.~P. \BBOP2013\BBCP.
\newblock \BBOQ Probabilistic forecasting of wind power generation using
  extreme learning machine\BBCQ\
\newblock {\Bem IEEE Transactions on Power Systems}, {\Bem 29\/}(3),
  1033--1044.

\bibitem[\protect\BCAY{Wang, Zhang, Bao, Zhu, Cao,\ \BBA\ Yu}{Wang
  et~al.}{2018a}]{wang2018not}
Wang, J., Zhang, J., Bao, W., Zhu, X., Cao, B., \BBA\ Yu, P.~S.
  \BBOP2018a\BBCP.
\newblock \BBOQ Not just privacy: Improving performance of private deep
  learning in mobile cloud\BBCQ\
\newblock In {\Bem KDD}, \BPGS\ 2407--2416.

\bibitem[\protect\BCAY{Wang, Golbandi, Bendersky, Metzler,\ \BBA\ Najork}{Wang
  et~al.}{2018b}]{wang2018position}
Wang, X., Golbandi, N., Bendersky, M., Metzler, D., \BBA\ Najork, M.
  \BBOP2018b\BBCP.
\newblock \BBOQ Position bias estimation for unbiased learning to rank in
  personal search\BBCQ\
\newblock In {\Bem WSDM}, \BPGS\ 610--618.

\bibitem[\protect\BCAY{Wang, Liang, Charlin,\ \BBA\ Blei}{Wang
  et~al.}{2018c}]{wang2018deconfounded}
Wang, Y., Liang, D., Charlin, L., \BBA\ Blei, D.~M. \BBOP2018c\BBCP.
\newblock \BBOQ The deconfounded recommender: A causal inference approach to
  recommendation\BBCQ.

\bibitem[\protect\BCAY{Wei, Ramamurthy,\ \BBA\ Calmon}{Wei
  et~al.}{2020}]{wei2020optimized}
Wei, D., Ramamurthy, K.~N., \BBA\ Calmon, F. d.~P. \BBOP2020\BBCP.
\newblock \BBOQ Optimized score transformation for fair classification\BBCQ\
\newblock In {\Bem AISTATS}.

\bibitem[\protect\BCAY{Weller}{Weller}{2017}]{weller2017challenges}
Weller, A. \BBOP2017\BBCP.
\newblock \BBOQ Challenges for transparency\BBCQ.

\bibitem[\protect\BCAY{Wikipedia}{Wikipedia}{2021a}]{robust_2021}
Wikipedia \BBOP2021a\BBCP\
\newblock
  \url{https://en.wikipedia.org/w/index.php?title=Robustness_(computer_science)&oldid=1009774103}.
\newblock Page Version ID: 1009774103.

\bibitem[\protect\BCAY{Wikipedia}{Wikipedia}{2021b}]{FacebookCambridge2021}
Wikipedia \BBOP2021b\BBCP\
\newblock
  \url{https://en.wikipedia.org/w/index.php?title=Facebook\%E2\%80\%93Cambridge_Analytica_data_scandal&oldid=1035933869}.
\newblock Page Version ID: 1035933869.

\bibitem[\protect\BCAY{Wikipedia}{Wikipedia}{2021c}]{CambridgeAnalytica2021}
Wikipedia \BBOP2021c\BBCP\
\newblock
  \url{https://en.wikipedia.org/w/index.php?title=Cambridge_Analytica&oldid=1036071985}.
\newblock Page Version ID: 1036071985.

\bibitem[\protect\BCAY{Wold, Esbensen,\ \BBA\ Geladi}{Wold
  et~al.}{1987}]{wold1987principal}
Wold, S., Esbensen, K., \BBA\ Geladi, P. \BBOP1987\BBCP.
\newblock \BBOQ Principal component analysis\BBCQ\
\newblock {\Bem Chemometrics and intelligent laboratory systems}, {\Bem
  2\/}(1-3), 37--52.

\bibitem[\protect\BCAY{Xu, Wu, Yuan, Zhang,\ \BBA\ Wu}{Xu
  et~al.}{2019}]{xu2019achieving}
Xu, D., Wu, Y., Yuan, S., Zhang, L., \BBA\ Wu, X. \BBOP2019\BBCP.
\newblock \BBOQ Achieving causal fairness through generative adversarial
  networks.\BBCQ\
\newblock In {\Bem IJCAI}, \BPGS\ 1452--1458.

\bibitem[\protect\BCAY{Xu, Jun, Zhu,\ \BBA\ Bellmore}{Xu
  et~al.}{2012}]{xu2012learning}
Xu, J.-M., Jun, K.-S., Zhu, X., \BBA\ Bellmore, A. \BBOP2012\BBCP.
\newblock \BBOQ Learning from bullying traces in social media\BBCQ\
\newblock In {\Bem NAACL HLT}, \BPGS\ 656--666. ACL.

\bibitem[\protect\BCAY{Yang\ \BBA\ Feng}{Yang\ \BBA\
  Feng}{2020}]{yang2020causal}
Yang, Z.\BBACOMMA\  \BBA\ Feng, J. \BBOP2020\BBCP.
\newblock \BBOQ A causal inference method for reducing gender bias in word
  embedding relations.\BBCQ\
\newblock In {\Bem AAAI}, \BPGS\ 9434--9441.

\bibitem[\protect\BCAY{Yang, Yang, Dyer, He, Smola,\ \BBA\ Hovy}{Yang
  et~al.}{2016}]{yang2016hierarchical}
Yang, Z., Yang, D., Dyer, C., He, X., Smola, A., \BBA\ Hovy, E. \BBOP2016\BBCP.
\newblock \BBOQ Hierarchical attention networks for document
  classification\BBCQ\
\newblock In {\Bem NAACL HLT}, \BPGS\ 1480--1489.

\bibitem[\protect\BCAY{Yao, Chu, Li, Li, Gao,\ \BBA\ Zhang}{Yao
  et~al.}{2020}]{yao2020causal}
Yao, L., Chu, Z., Li, S., Li, Y., Gao, J., \BBA\ Zhang, A. \BBOP2020\BBCP.
\newblock \BBOQ A survey on causal inference\BBCQ.

\bibitem[\protect\BCAY{Yeo}{Yeo}{2020}]{Yeo_2020}
Yeo, C. \BBOP2020\BBCP.
\newblock \BBOQ What is transparency in {AI}?\BBCQ\
\newblock
  \url{https://medium.com/fair-bytes/what-is-transparency-in-ai-bd08b2e901ac}.

\bibitem[\protect\BCAY{Young, Magassa,\ \BBA\ Friedman}{Young
  et~al.}{2019}]{young2019toward}
Young, M., Magassa, L., \BBA\ Friedman, B. \BBOP2019\BBCP.
\newblock \BBOQ Toward inclusive tech policy design: a method for
  underrepresented voices to strengthen tech policy documents\BBCQ\
\newblock {\Bem Ethics and Information Technology}, {\Bem 21\/}(2), 89--103.

\bibitem[\protect\BCAY{Yuan, He, Zhu,\ \BBA\ Li}{Yuan
  et~al.}{2019}]{yuan2019adversarial}
Yuan, X., He, P., Zhu, Q., \BBA\ Li, X. \BBOP2019\BBCP.
\newblock \BBOQ Adversarial examples: Attacks and defenses for deep
  learning\BBCQ\
\newblock {\Bem IEEE Transactions on Neural Networks and Learning Systems},
  {\Bem 30\/}(9), 2805--2824.

\bibitem[\protect\BCAY{Zhang, Wu, Wong, Xu, Dong,\ \BBA\ Iu}{Zhang
  et~al.}{2014}]{zhang2014advanced}
Zhang, G., Wu, Y., Wong, K.~P., Xu, Z., Dong, Z.~Y., \BBA\ Iu, H. H.-C.
  \BBOP2014\BBCP.
\newblock \BBOQ An advanced approach for construction of optimal wind power
  prediction intervals\BBCQ\
\newblock {\Bem IEEE Transactions on Power Systems}, {\Bem 30\/}(5),
  2706--2715.

\bibitem[\protect\BCAY{Zhang\ \BBA\ Wu}{Zhang\ \BBA\ Wu}{2017}]{zhang2017anti}
Zhang, L.\BBACOMMA\  \BBA\ Wu, X. \BBOP2017\BBCP.
\newblock \BBOQ Anti-discrimination learning: a causal modeling-based
  framework\BBCQ\
\newblock {\Bem JDSA}, {\Bem 4\/}(1), 1--16.

\bibitem[\protect\BCAY{Zhang, Wu,\ \BBA\ Wu}{Zhang
  et~al.}{2016}]{zhang2016causal}
Zhang, L., Wu, Y., \BBA\ Wu, X. \BBOP2016\BBCP.
\newblock \BBOQ A causal framework for discovering and removing direct and
  indirect discrimination\BBCQ.

\bibitem[\protect\BCAY{Zio}{Zio}{2006}]{zio2006study}
Zio, E. \BBOP2006\BBCP.
\newblock \BBOQ A study of the bootstrap method for estimating the accuracy of
  artificial neural networks in predicting nuclear transient processes\BBCQ\
\newblock {\Bem IEEE Transactions on Nuclear Science}, {\Bem 53\/}(3),
  1460--1478.

\end{thebibliography}
\bibliographystyle{theapa}
\end{document}